\documentclass[iop]{emulateapj}
\usepackage{amsmath}
\usepackage{graphicx}

\usepackage{amsmath}
\usepackage{color}
\usepackage{ulem}
\usepackage[safe]{tipa}
\usepackage{multirow}
\usepackage{bold-extra}

\usepackage{natbib}

\usepackage{stackengine}

\newcommand\Msun{\; {\rm M}_{\odot}}
\newcommand\kms{\; {\rm km}\;{\rm s}^{-1}}
\newcommand\pcc{\;{\rm cm}^{-3}}
\newcommand\pc{\;{\rm pc}}

\newcommand\kpc{\;{\rm kpc}}
\newcommand\cm{\;{\rm cm}}
\newcommand\yr{\; {\rm yr}}
\newcommand\K{\; {\rm K}}
\newcommand\Myr{\;{\rm Myr}}
\newcommand\Gyr{\;{\rm Gyr}}
\newcommand\Aunit{\Msun \yr^{-1}}
\newcommand\Surf{\Msun\pc^{-2}}
\newcommand\SFR{{\Sigma_\text{SFR}}}
\newcommand\SFRf{{\Sigma_\text{SFR}}}
\newcommand\SFRunits{{\Msun\pc^{-2}\Myr^{-1}}}
\newcommand\Omgunits{\;\kms\kpc^{-1}}

\newcommand\Rhat{{{\mathbf{\widehat{R}}}}}
\newcommand\phat{{{\boldsymbol{\widehat{\phi}}}}}
\newcommand\vecnabla{{\boldsymbol{\nabla}}}
\newcommand\Omg{{\boldsymbol{\Omega}}}
\newcommand\xhat{{{\mathbf{\hat{x}}}}}
\newcommand\yhat{{{\mathbf{\hat{y}}}}}
\newcommand\zhat{{{\mathbf{\hat{z}}}}}

\usepackage{mathtools}%
\usepackage{xparse, etoolbox}

\shorttitle{}
\shortauthors{}

\begin{document}

\title{Local Simulations of Spiral Galaxies with the TIGRESS Framework: I. Star Formation and Arm Spurs/Feathers}

\author{Woong-Tae Kim}
\affiliation{Department of Physics \& Astronomy, Seoul National
  University, Seoul 08826, Republic of Korea}
\affiliation{Department of Astrophysical Sciences, Princeton University,
  Princeton, NJ 08544, USA}
\author{Chang-Goo Kim}
\affiliation{Department of Astrophysical Sciences, Princeton University,
  Princeton, NJ 08544, USA}
\author{Eve C.\ Ostriker}
\affiliation{Department of Astrophysical Sciences, Princeton University,
  Princeton, NJ 08544, USA}

\email{wkim@astro.snu.ac.kr, cgkim@astro.princeton.edu, eco@astro.princeton.edu}
\slugcomment{Accepted for publication in the ApJ}

\begin{abstract}
Spiral arms greatly affect gas flows and star formation in disk galaxies. We use local three-dimensional simulations of the  vertically-stratified, self-gravitating, differentially-rotating, interstellar medium (ISM) subject to a stellar spiral potential to study the effects of spiral arms on star formation and formation of arm spurs/feathers. We adopt the TIGRESS framework of \citet{cgkim17} to handle radiative heating and cooling, star formation, and ensuing supernova (SN) feedback. We find that more than 90\% of star formation takes place in spiral arms, but the global star formation rate (SFR) in models with spiral arms is enhanced by less than a factor of 2 compared to the no-arm counterpart. This results from a quasi-linear relationship between the SFR surface density $\SFRf$ and the gas surface density $\Sigma$, and supports the picture that spiral arms do not trigger star formation but rather concentrate
star-forming regions. Correlated SN feedback produces gaseous spurs/feathers downstream from arms in both magnetized and unmagnetized models. These spurs/feathers are short-lived and have magnetic fields parallel to their length, in contrast to the longer-lived features with perpendicular magnetic fields induced by gravitational instability.
SN feedback drives the turbulent component of magnetic fields, with the total magnetic field strength sublinearly proportional to $\Sigma$. The total midplane pressure varies by a factor of $\sim 10$ between arm and interarm regions but agrees locally with the total vertical ISM weight, while $\SFRf$ is locally consistent with the prediction of pressure-regulated, feedback-modulated theory.
\end{abstract}


\section{Introduction}

Spiral arms are of paramount importance in dynamical and chemical evolution of
disk galaxies (e.g., \citealt{but96,kor04,sel14}). They not only exert non-axisymmetric torques to drive secular migration of stars and gas clouds (e.g., \citealt{ros08,bov12,dan15,dan18}), but also compress gas to aid the formation of giant molecular clouds (GMCs) in them (e.g., \citealt{elm83,ran93,elm94}; see also \citealt{dob14}). They are also sites of active star formation (e.g., \citealt{elm86,foy10,elm11,sch17,ler17}) and are associated with
spurs or feathers that extend nearly perpendicularly from arm to interarm regions\footnote{More precisely, in the observational literature the term \textit{feathers} has been used to refer to dust features seen as dark against a bright background, while  \textit{spurs} indicates luminous stellar features in optical light (see, e.g., \citealt{lav06,sch17}).  Spurs most likely correspond to dense feathers with recent star formation. In this paper, we use the two terms interchangeably.}  (e.g., \citealt{san61,lyn70,elm80,sco01,lav06,mur09,sch13,sch17,pue14,koo17,elm18,elm19}).

Massive stars formed inside spiral arms -- and in arm spurs/feathers -- profoundly affect the physical properties of the surrounding interstellar medium (ISM) by supplying feedback in the form of stellar winds and radiation during their lifetime and via supernova (SN) explosions at their death (e.g., \citealt{mck07,kru14}).  SN feedback appears strongest among various mechanisms that contribute to driving turbulence in the ISM (e.g., \citealt{mac04}), and is a major agent for regulating galactic star formation rates (SFRs) (e.g., \citealt{ost10,ost11,cgkim11}).

One of the long-standing issues with disk galaxies is whether spiral arms trigger star formation or just organize star-forming regions into an arm shape. The conventional wisdom was that spiral arms enhance SFRs by compressing gas clouds above the threshold density for gravitational collapse (e.g, \citealt{rob69,shu73,rob75}).
\citet{cep90} argued in favor of this notion based on arm \textit{vs.} interarm star formation efficiency (SFE) in atomic gas in M74 and M109 and typical arm/interarm molecular density contrasts (see also \citealt{lor90,kna96,sei02}).
However, there does not in fact seem to be a threshold for star formation in molecular gas \citep{sch11}, and
\citet{elm86} found no obvious correlation between the mean SFRs and the arm strength, indicating that spiral arms are not needed to trigger star formation. A more recent study of \citet{foy10} for galaxies M51 and M74 showed that the SFE within molecular gas is almost similar (within $\sim10\%$) in the arm and interarm regions. Given that the SFR is not correlated with the atomic gas but with the molecular gas \citep{big08,ler08,ler13}, the existing observational results favor the idea of star-formation organization (rather than triggering) by spiral arms.

Another interesting issue regarding spiral arms is what mechanism accounts for the formation of gaseous
spurs/feathers. A number of processes have been put forward to explain spur/feather formation (see, e.g., \citealt{shu16} and references therein). For example, \citet{bal88} argued that swing amplification of hydrodynamic perturbations inside spiral arms is responsible for spurs/feathers, while the presence of magnetic fields appears essential to promote the fast growth of self-gravitating perturbations in spiral arms (e.g., \citealt{elm94,kim02,kim06,she06,she08,lee12,lee14}). In particular, \citet{kim02,kim06} ran magnetodydrodynamic (MHD) simulations of local shearing-box models under an isothermal approximation and showed that elongated spur structures grow via magneto-Jeans instability (MJI), in which magnetic fields remove angular momentum from self-gravitationally contracting regions. In these simulations, gaseous spurs undergo gravitational fragmentation into bound clumps within one or two orbits. When driven by MJI, the mean spacing between spurs along the arm is determined by the Jeans length at the density peak of spiral arms.

There have also been mechanisms proposed for spur/feather formation that do not rely on gaseous self-gravity and magnetic fields. \citet{wad04} ran non-self-gravitating simulations and showed that hydrodynamic spiral shocks are unstable to clump-forming, wiggle instability (WI; see also \citealt{ren13}), which later turns out to arise due to generation of potential vorticity (PV) at perturbed shock fronts that gas passes through periodically \citep{kim14,kim15,sor17}. Using smoothed particle hydrodynamics (SPH) simulations, \citet{dob06} found that changes of angular momenta at shock fronts make the orbits of cold particles crowded, resulting in clumps and feathers behind the shocks.

Although the numerical studies mentioned above were useful to understand physical processes involved in spiral shocks and structure formation inside spiral arms, almost all of them were limited to isothermal models without considering star formation and SN feedback. Realistically, the dominant pressure in the ISM is not thermal pressure but turbulent pressure. Since SN feedback drives turbulence and also creates larger-scale structures including superbubbles \citep[e.g.][]{mcr79,tom81}, it may also control spur/feather formation at least partially. Indeed,  \citet{sch17} found that feathers in M51 are deeply associated with massive star clusters and suggested that their shape and evolution should be influenced by stellar feedback. Therefore, it is necessary to explore how spiral arm substructures form and evolve in a medium continually stirred by SN feedback.

Several recent works have incorporated stellar feedback in global numerical simulations of spiral galaxies and observed spur formation (e.g., \citealt{ren13,bab17,pet17}). From a hydrodynamic simulation of a Milky Way-like galaxy, \citet{ren13} suggested that spurs/feathers originate from velocity shear near spiral arms. Using $N$-body/SPH simulations for galaxies with steady or dynamic spirals, \citet{bab17} found that GMCs formed by cloud-cloud collisions inside arms are sheared into feathers in interarm regions (see also \citealt{dob17}). \citet{pet17} ran models in which spiral arms are driven by a tidal encounter with a companion galaxy (see also \citealt{tre20}). They found that spurs/feathers are much larger and stronger in outer arms (bridges to the companion) that have higher SFRs than inner arms. They also found that feathers are located in between young star clusters in the interarm regions, suggesting that stellar feedback somehow affects the formation and evolution of interarm feathers. These diverse results on spur/feather formation in numerical simulations indicate that there is no consensus as to how star formation and feedback influence these features.

In this paper, we run local three-dimensional simulations of vertically-stratified, self-gravitating, differentially-rotating, magnetized, galactic gaseous disks under the influence of a stellar spiral potential. To handle star formation and SN feedback, we adopt the TIGRESS (Three-phase Interstellar medium in Galaxies Resolving Evolution with Star formation and Supernova feedback) algorithms developed by \citet{cgkim17}. The TIGRESS framework features multi-physics modules including gas accretion to sink/stellar particles and delayed SN feedback in the form of thermal energy and/or momentum, and allows self-consistent simulations of the three-phase star-forming ISM.  The present study  extends the
model setup of \citet{cgkim17} by including a local stellar spiral-arm potential. Our work also extends the isothermal models of \citet{kim02,kim06} by considering radiative cooling and heating, star formation, and stellar feedback (supernovae and FUV heating).

This work has several objectives. First, we want to understand the role of spiral arms on galactic star formation, addressing the issue of star-formation triggering  or organization by spiral arms. Second, we wish to investigate the effects of SN feedback on gaseous  spur formation and evolution, for which it is essential to utilize the TIGRESS algorithms. In addition, spiral structure imposes spatial variations in mean properties that reflect a temporal sequence as gas flows from arm to interarm conditions and back.  Once the simulation reaches a quasi-steady state, it is interesting to calculate profiles of various physical quantities such as the mean gas surface density, SFR, magnetic field strength, and midplane stresses as functions of the distance downstream from the spiral arm peak, and to explore correlations among these physical variables.

The remainder of the paper is organized as follows. In Section \ref{sec:method}, we describe our numerical methods and the models we consider, including briefly summarizing the TIGRESS framework for treating star formation and SN feedback. In Section \ref{sec:evol}, we present the temporal and morphological evolution of our models, with a focus on spur formation. In Section \ref{sec:stat}, we present the
spatial profiles and correlations of various physical quantities
as well as the temporal variations of volume-integrated quantities including SFR and mass fractions. In Section \ref{sec:sumdis}, we summarize our results and discuss their astronomical implications.

\section{Methods}\label{sec:method}

In this paper we study
evolution of the ISM in a vertically-stratified disk under the influence of a stellar spiral potential as well as star formation and SN feedback. Similar work without a spiral potential was reported by \citet{cgkim17}, while an isothermal version without star formation and feedback was studied by \citet{kim02,kim06}. In this section, we present the equations we solve, the numerical methods, and the model parameters we adopt.

\subsection{Basic Equations}

We consider a
Cartesian box that is corotating with a local segment of a stellar spiral potential. The spiral potential is assumed to be tightly wound with a pitch angle $\sin i\ll1$ and rigidly rotating at a constant pattern speed $\Omega_p$ about the center of a galaxy. The local Cartesian frame is centered at a position $(R_0, \phi_0=\Omega_pt, z_0=0)$ in the galactic plane and inclined such that $\bf\hat{x}$ and $\bf\hat{y}$ point toward the directions perpendicular and parallel to the arm, respectively, while $\bf\hat{z}$ corresponds to the direction perpendicular to the galactic plane (e.g., \citealt{rob69,bal88}). This local spiral-arm frame is advantageous for capturing essential physics associated with rotational shear and the spiral potential while affording much higher resolution than global models of spiral galaxies. Our simulation domain is a rectangular parallelepiped with size $L_x\times L_y \times L_z$, where $L_x=2\pi R_0 (\sin i)/m$ is equal to the arm-to-arm distance for an $m$-armed spiral. In this local frame, the background velocity arising from galactic differential rotation is given by
 \begin{equation}\label{eq:v0}
 {\bf v}_0 = R_0(\Omega_0 - \Omega_p) \sin i {\bf\hat{x}}
 + [R_0 (\Omega_0 - \Omega_p) - q\Omega_0 x] {\bf\hat{y}},
 \end{equation}
where $\Omega_0=\Omega(R_0)$ is the angular velocity at $R_0$ in the inertial frame and  $q \equiv -(d\ln\Omega/d\ln R)|_{R_0}=1$  is the shear parameter for a flat rotation curve (e.g., \citealt{kim02}).

The ideal MHD equations expanded in the local frame read
 \begin{gather}
   \frac{\partial \rho}{\partial t} + \boldsymbol\nabla \cdot (\rho \mathbf{v}) =0, \label{eq:con} \\
 \begin{aligned}
   \frac{\partial (\rho\mathbf{v})}{\partial t} + & \boldsymbol\nabla \cdot
   \left[\rho\mathbf{vv} + \left( P + \frac{B^2}{8\pi} \right){\pmb{\mathbb{I}}}
   -\frac{\mathbf{BB}}{4\pi}\right] = \\
   &-2\rho \boldsymbol{\Omega}_0 \times (\mathbf{v}-\mathbf{v}_0)   - \rho\boldsymbol\nabla (\Phi_\text{self} + \Phi_\text{ext}), \label{eq:mom}
\end{aligned} \\
 \begin{aligned}
   \frac{\partial E}{\partial t} + \boldsymbol\nabla\cdot \left[\left( E+ P + \frac{B^2}{8\pi} \right) \mathbf{v} -
\frac{\mathbf{B}(\mathbf{B}\cdot\mathbf{v})}{4\pi}  \right] = \\ -\rho\mathbf{v}\cdot \boldsymbol\nabla (\Phi_\text{self} + \Phi_\text{ext}) -\rho\mathcal{L}, \label{eq:eng}
\end{aligned} \\
\frac{\partial \mathbf{B}}{\partial t} = \boldsymbol\nabla\times (\mathbf{v}\times\mathbf{B}), \label{eq:ind} \\
\boldsymbol\nabla^2\Phi_\text{self} = 4\pi G(\rho +\rho_\text{sp}). \label{eq:pos}
 \end{gather}
Here, $\pmb{\mathbb{I}}$ is the identity matrix,
$E= \rho v^2/2 + {P}/{(\gamma-1)} + {B^2}/{8\pi}$ is the total energy density, $\rho\mathcal{L}$ is the net cooling rate per unit volume per unit time, $\Phi_\text{ext}$ is the external gravitational potential, and $\Phi_\text{self}$ is the self-gravitational potential of gas with density $\rho$ and star particles with density $\rho_\text{sp}$. The other symbols have their usual meaning.
The first two terms in the right-hand side of Equation \eqref{eq:mom} denote, respectively, the Coriolis force and the term counterbalancing the advection of the $y$-velocity along the $x$-direction in the background flow.

The external gravitational potential $\Phi_\text{ext}$ in Equation \eqref{eq:mom} consists of three parts as $\Phi_\text{ext}=\Phi_* + \Phi_\text{dm} + \Phi_\text{arm}$, where
 \begin{align}
   \Phi_* (z) &= 2\pi G\Sigma_*z_*
       \left[\left(1+\frac{z^2}{z_*^2}\right)^{1/2}-1\right],
           \label{eq:exts} \\
  \Phi_\text{dm} (z)&=  2\pi G\rho_\text{dm} R_0^2\ln\left( 1+
                    \frac{z^2}{R_0^2}\right), \label{eq:extdm} \\
 \Phi_\text{arm} (x)&= \Phi_\text{arm}(0)
            \cos\left(\frac{2\pi x}{L_x}\right), \label{eq:extarm}
 \end{align}
respectively representing the fixed gravitational potentials from the stellar disk with surface density $\Sigma_*$ and scale height $z_*$, dark matter halo with mass density $\rho_\text{dm}$, and the stellar spiral arm with amplitude $\Phi_\text{arm}(0)$.

For the net cooling function in Equation \eqref{eq:eng}, we take
 \begin{equation}\label{eq:cooling}
   \rho\mathcal{L} = n_\text{H}^2 \Lambda(T) - n_\text{H}\Gamma,
 \end{equation}
where $n_\text{H}=\rho/(\mu m_\text{H})$ is the number density of hydrogen nuclei, with $\mu(T)$ being the mean molecular weight: we allow $\mu$ to vary
with $T$ from $\mu_\text{ato}=1.295$ (for fully neutral gas) to $\mu_\text{ion}=0.618$ (for fully ionized gas). For the cooling rate $\Lambda(T)$, we adopt the fitting formula of \citet{koy02} for $T<10^{4.2}\rm\,K$  (see also \citealt{cgkim08}), and for $T>10^{4.2}\rm\,K$ we adopt the collisional ionization equilibrium cooling function at solar metallicity  from \citet{sut93}.

The major heating source for warm and cold
gas is the photoelectric effect of dust grains exposed to far ultraviolet (FUV) radiation.
The FUV radiation includes the metagalactic radiation as well as the one emitted from young massive stars.
To allow for the heating by young stars formed in our simulations, we calculate the mean FUV luminosity $\Sigma_{\rm FUV}$ from all star particles averaged over the simulation domain, and use it to evaluate the time-varying heating rate $\Gamma$ as
\begin{equation}\label{eq:Gamma}
    \frac{\Gamma}{\Gamma_0} =\frac{\mu(T)-\mu_\text{ion}}{\mu_\text{ato}-\mu_\text{ion}} \left[0.0024 + \frac{\Sigma_{\rm FUV}}{\Sigma_{\rm FUV,0}}\frac{1-E_2(\tau_\perp/2)}{\tau_{\perp}}\right].
\end{equation}
Here, $\tau_\perp=\kappa_{\rm FUV} \Sigma$ with $\kappa_{\rm FUV}= 10^3\cm^3{\,\rm g^{-1}}$ is the UV optical depth in the global plane-parallel approximation, and $E_2$ is the exponential integral function of order 2; this  form represents global attenuation in the plane-parallel approximation.   For the normalization factors in Equation \eqref{eq:Gamma}, we take the heating rate $\Gamma_0=2\times 10^{-26}\rm\,erg\,s^{-1}$ in the solar neighborhood \citep{koy02}, the interstellar radiation field $\Sigma_{\rm FUV,0}=2.1\times10^{-4}{\rm\,erg\,s^{-1}\,cm^{-2}}$ \citep{dra78}, and the contribution of the metagalactic FUV amounting to $0.24\%$ \citep{ste02}.
The photoelectric heating is completely turned off at high temperature ($T\gtrsim 10^5\K$) where $\mu=\mu_\text{ion}$.  While we do not include heating by photoionization, it would not make a significant difference in $T$ compared to that from the photoelectric heating.

\subsection{Star Formation and Feedback}

We adopt the TIGRESS framework of \citet{cgkim17} to treat star formation and SN feedback. Here we briefly summarize the TIGRESS  algorithms: the reader is referred to \citet{cgkim17} for the complete description. We note from the outset that in the present work we consider SN feedback alone, while ignoring early feedback (e.g., stellar winds, ionizing radiation). This implicitly assumes that the dynamical time in the surrounding clouds is longer than $\sim 4\Myr$,
so that SN feedback dominates early feedback, which is valid for the moderate-density environment under consideration.

In TIGRESS, star formation is modelled by creating a sink/star particle when a cell meets three conditions simultaneously: (1) its density exceeds the Larson-Penston threshold density, (2) it lies at a local potential minimum, and (3) its velocity is converging in all three directions \citep{gon13}. The initial mass of the sink particle is set equal to the difference between the gas mass in the surrounding $3^3$-cell control volume prior to sink formation, and the mass in this region based on the extrapolation  of the surrounding density field. Each sink particle is allowed to accrete mass and momentum from the surrounding gas over time, at a rate based on the fluxes through the surfaces of the  control volume centered at the sink particle. When star particles are actively accreting, physical quantities such as density and velocity (except magnetic fields) in the cells inside the control volume are set as ghost zones based on the surrounding active cells.
Sink particles represent star clusters with stellar populations
that average over the initial mass function (IMF), with an age that reflects the accretion history.
The immediate formation of a sink particle implicitly assumes that the timescale for gravitational collapse and star formation is shorter than the dynamical time of the surrounding gas, and that the star formation efficiency is high.
Particles are merged if their control volumes overlap with each other, with physical quantities set by mass-weighted averages.

Sink/star particles age with time, with the mass-weighted mean age $t_m$ updated to account for accreted gas. Particles are categorized into three species depending on $t_m$ relative to the mean SN onset time $t_\text{SN} \sim 4\Myr$ (determined
stochastically for each particle) and the feedback lifetime $t_\text{life}=40\Myr$: (1) ``growing'' particles if $0<t_m<t_\text{SN}$, (2) ``feedback'' particles if $t_\text{SN}< t_m <t_\text{life}$, and (3)
``passive'' particles if $t_\text{life} < t_m$. Growing particles are treated as sinks
that contribute to the overall FUV radiation radiation field and exert gravity.
Feedback particles do not accrete and merge, but exert gravity and supply FUV radiation and SN feedback. Passive particles neither accrete nor supply feedback, but merely exert gravity on the gas and other particles.

For feedback particles with mass $m_\text{sp}$, the number of SN events expected over the time step $\Delta t$ is drawn from the population synthesis model {\tt STARBURST99} \citep{lei99}, which is roughly $\mathcal{N}_\text{SN}\sim m_\text{sp} \Delta t/(m_*t_\text{life})$, where $m_*=95.5\Msun$ is the mass of a star cluster per SN
for a fully sampled \citet{kro01} IMF. In each time step, we generate a uniform random number $\mathcal{U}_\text{SN}\in (0,1)$ and turn on SN feedback only if $\mathcal{N}_\text{SN}>\mathcal{U}_\text{SN}$.
We also consider SN events from runaway OB stars, by taking a binary fraction $f_{\rm bin}\equiv 2/3$. If the event is determined to occur in a binary, we spawn a massless particle with initial velocity distribution taken from \citet{eld11} and let it explode after a certain delay time (to ensure the entire SN rate the same).

\begin{deluxetable*}{cccccccc}
\tabletypesize{\footnotesize}
\tablewidth{0pt}\tablecaption{Model Parameters and
Selected
Simulation Outcomes}
\tablehead{
\colhead{Model} &
\colhead{$\mathcal{F}$} &
\colhead{$\beta$} &
\colhead{$\log {\Sigma}_{\rm SFR}$ } &
\colhead{$\log P_\text{thm}/k_B$} &
\colhead{$\log P_\text{trb}/k_B$} &
\colhead{$\log \Pi_\text{mag}/k_B$} &
\colhead{$\log P_\text{tot}/k_B$} \\
\colhead{(1)} &
\colhead{(2)} &
\colhead{(3)} &
\colhead{(4)} &
\colhead{(5)} &
\colhead{(6)} &
\colhead{(7)} &
\colhead{(8)} }
\startdata
{\tt F00B10} & $0.0$   & $10$     & $-2.56 \pm  0.13$    & $ 3.77 \pm 0.08$ & $ 3.64 \pm 0.13$ & $ 3.94 \pm 0.07$ & $ 4.29 \pm 0.03$ \\
{\tt F10B10} & $0.1$   & $10$     & $-2.47 \pm  0.13$    & $ 3.74 \pm 0.09$ & $ 3.79 \pm 0.08$ & $ 3.99 \pm 0.05$ & $ 4.34 \pm 0.03$ \\
{\tt F20B10} & $0.2$   & $10$     & $-2.38 \pm  0.20$    & $ 3.75 \pm 0.13$ & $ 3.89 \pm 0.11$ & $ 3.95 \pm 0.10$ & $ 4.36 \pm 0.05$ \\
{\tt F00Binf} & $0.0$  & $\infty$ & $-2.27 \pm  0.19$    & $ 3.85 \pm 0.13$ & $ 3.91 \pm 0.12$ &   \textendash\     &$ 4.19 \pm 0.12$ \\
{\tt F10Binf} & $0.1$  & $\infty$ & $-2.31 \pm  0.17$    & $ 3.78 \pm 0.15$ & $ 3.88 \pm 0.14$ &  \textendash\      & $ 4.15 \pm 0.11$ \\
{\tt F20Binf} & $0.2$  & $\infty$ & $-2.20 \pm  0.11$    & $ 3.80 \pm 0.09$ & $ 4.01 \pm 0.14$ &  \textendash\      & $ 4.22 \pm 0.10$
\enddata
\tablecomments{Column 1: model name. Column 2: arm strength. Column 3: plasma parameter. Column 4: logarithm of the SFR surface density ($\rm M_\odot\,pc^{-2}\,Myr^{-1}$). Columns 5-8: logarithm of the midplane thermal, turbulent, magnetic, and total stresses over $k_B$, respectively ($\rm cm^{-3}\, K$). Quantities in Columns 4-8 are averaged over the simulation domain and $t=200$--$600\Myr$. }\label{t:model}
\end{deluxetable*}

The mode  of SN feedback that is implemented for any given event depends on the local density in the region around the SN.  We start by adopting an initial radius $R_\text{snr}=3\Delta x$ (for grid resolution $\Delta x$) for the feedback region, and compute the ratio of the mass $M_\text{snr}$ within  this region
to the expected shell formation mass $M_\text{sf}=1679\Msun(n_\text{H}/\rm cm^{-3})^{-0.26}$ \citep{cgkim15a}. Here, $M_\text{snr}$ is defined by the sum of the gas mass within the initial SN remnant radius $R_\text{snr}$ and the ejecta mass $M_\text{ej}=10\Msun$. If $\mathcal{R}_M \equiv M_\text{snr}/M_\text{sf}>1$,
we regard that the Sedov-Taylor stage of this SN is unresolved  and inject a radial momentum  $p_\text{snr}=2.8\times10^5\Msun\kms(n_\text{H}/\rm cm^{-3})^{-0.17}$ that represents the final post-cooling value appropriate for the momentum-conserving stage, as calibrated from resolved numerical simulations (e.g., \citealt{cgkim15a,mar15,iff15,wal15}).
If $0.027<\mathcal{R}_M<1$, we regard the SN remnant as being in the Sedov-Taylor phase, and add the SN energy $E_\text{SN}=10^{51}\rm\,erg$ in both thermal $(\sim72\%)$ and kinetic $(\sim28\%)$ forms to the surrounding medium within $R_\text{snr}$. If $\mathcal{R}_M<0.027$, we increase $R_\text{snr}$ until $\mathcal{R}_M\gtrsim 0.027$ and inject $E_\text{SN}$ within the increased $R_\text{snr}$ in both thermal and kinetic forms, as above. When $\mathcal{R}_M$ becomes less than 0.027 even for the prescribed maximum value ($\sim128\pc$) of $R_\text{snr}$, we inject $E_\text{SN}$ in kinetic form only, corresponding to the free expansion of the ejecta.

It turns out that in our simulations, the fraction of SN feedback in the free-expansion, Sedov-Taylor, and momentum-conserving phase are, on average, $\sim13$, 57, and 30\%, respectively, indicating that $\sim$70\% of SN feedback is resolved. The fraction of the resolved feedback is slightly higher for unmagnetized and/or weaker-arm models. SNe in the interarm region tend to explode at lower density and $\sim5\%$ more of the SN events are resolved than those in the arm region. The resolved fraction of SN events is $\sim30\%$ higher for clusters than runaway stars.

We integrate Equations \eqref{eq:con}--\eqref{eq:ind} using a modified version of the {\tt Athena} code that utilizes a directionally unsplit Godunov scheme to solve the ideal MHD equations, including the constrained transport
algorithm to preserve $\boldsymbol\nabla\cdot\mathbf{B}=0$ within machine precision \citep{sto08}. For hydrodynamic variables, we employ shearing-sheet boundary conditions in the horizontal direction \citep{haw95} and outflow
(diode-like) boundary conditions in the vertical direction. The Poisson equation (Equation \eqref{eq:pos}) is solved based on the FFT method with shearing-sheet boundary conditions in the horizontal direction (e.g., \citealt{gam01}) and vacuum boundary conditions in the vertical direction \citep{koy09}.

Appendix \ref{sec:eom} presents the equations of motion that sink/star particles obey in the local spiral-arm coordinates. We integrate Equation \eqref{a:Rob0} using a ``kick-drift-kick'' scheme for a symplectic integrator suggested by \citet{qui10}, with the time $t$ in the last term of Equation \eqref{a:Rob0a} set equal to $t_m$ for each particle. By applying the shearing-sheet boundary conditions in the horizontal plane (e.g., \citealt{hub01,kim07}), we ensure that particles leaving the simulation domain from one $x$-face reenter with shifted $y$-positions through the opposite $x$-face.

\subsection{Model Parameters}

The simulation domain is centered on a location corresponding to a distance $R_0=8\kpc$ from the galactic center, rotating with angular speed of $\Omega_0=30\Omgunits$.
Our simulation domain has a size of $L_x=\pi\kpc$ and $L_y=L_z=2L_x$. The domain is resolved by $256\times512\times512$ cells, corresponding to a physical size $\Delta x=12.3\pc$, which is marginally good enough to obtain converged results for statistical properties such as SFR, velocity dispersions, scale heights, etc.\ \citep{cgkim17}. For models that include spiral-arm forcing, we take the pattern speed $\Omega_p=\Omega_0/2$, pitch angle $\sin i=0.125$, and the arm number $m=2$.

For all models, the initial surface density is spatially uniform and equal to $\Sigma_0=13\Surf$.  To specify $\Phi_*$ and $\Phi_\text{dm}$ in Equations \eqref{eq:exts} and \eqref{eq:extdm}, we adopt the solar neighborhood values of $\Sigma_*=42\Msun\pc^{-2}$, $z_*=245\pc$, and $\rho_\text{dm}=6.4\times10^{-3}\Msun\pc^{-3}$ from \citet{kui89}.

Two key dimensionless parameters that characterize our models are
 \begin{align}\label{eq:Fparam}
  \mathcal{F}&\equiv \frac{m}{\sin i} \left(\frac{|\Phi_\text{arm}(0)|}{R_0^2\Omega_0^2}\right)\,,\\
  \beta &\equiv \frac{8\pi P_0}{B_0^2}, \label{eq:beta}
 \end{align}
where $P_0(z)$ is the initial pressure profile (see below) and ${\bf B}_0=B_0(z){\bf\hat{y}}$ is the initial magnetic field profile, threading the gas parallel to the spiral arm.
The $\mathcal{F}$ parameter represents the ratio of the radial force due to the spiral arm to the mean radial gravitational force \citep{rob69}, while the plasma parameter $\beta$ measures the (inverse of) magnetic pressure  relative to the initial thermal pressure. To study the effect of the arm strength and magnetic fields on the gas evolution and star formation, we run six models that differ in the combination  of $\mathcal{F}$ and $\beta$.

Table \ref{t:model} lists the model parameters and some of the simulation results. Models with $\beta=10$ have $B_0(0)=2.6\mu$G initially at the midplane.
For models with a spiral arm ($\mathcal{F}\neq0$), we turn on the arm potential slowly and make it achieve the full strength at $t=200\Myr$. All simulations are run until $t=700\Myr$.

Initially, the disks are vertically stratified with density $\rho_0(z)$.
For $\rho_0(z)$, we take a double exponential
\begin{equation}\label{eq:inirho}
\begin{split}
 \rho_0(z) &= \rho_1(z) + \rho_2(z) \\
  &= \rho_{10}e^{-\Phi_\text{0,tot}/\sigma_1^2}
  + \rho_{10}e^{-\Phi_\text{0,tot}/\sigma_2^2}\,,
\end{split}
\end{equation}
where $\Phi_\text{0,tot}(z)=\Phi_*(z) + \Phi_\text{dm}(z) + 2\pi G\Sigma_0|z|$ is the total gravitational potential under the assumption that the gaseous disk is razor-thin (without spiral-arm forcing).
With $\rho_{10}=2.85\,m_\text{H}\pcc$, $\rho_{20}=10^{-5}\rho_{10}$, $\sigma_1=7\kms$, and $\sigma_2=10\sigma_1$, $\rho_1$ and $\rho_2$ in Equation \eqref{eq:inirho} represents the warm and hot media, respectively, of the ISM in the Milky Way. We set the initial pressure profile to $P_0 (z)= \sigma_1^2\rho_1 + \sigma_2^2\rho_2$.

\begin{figure*}
\epsscale{1.0} \plotone{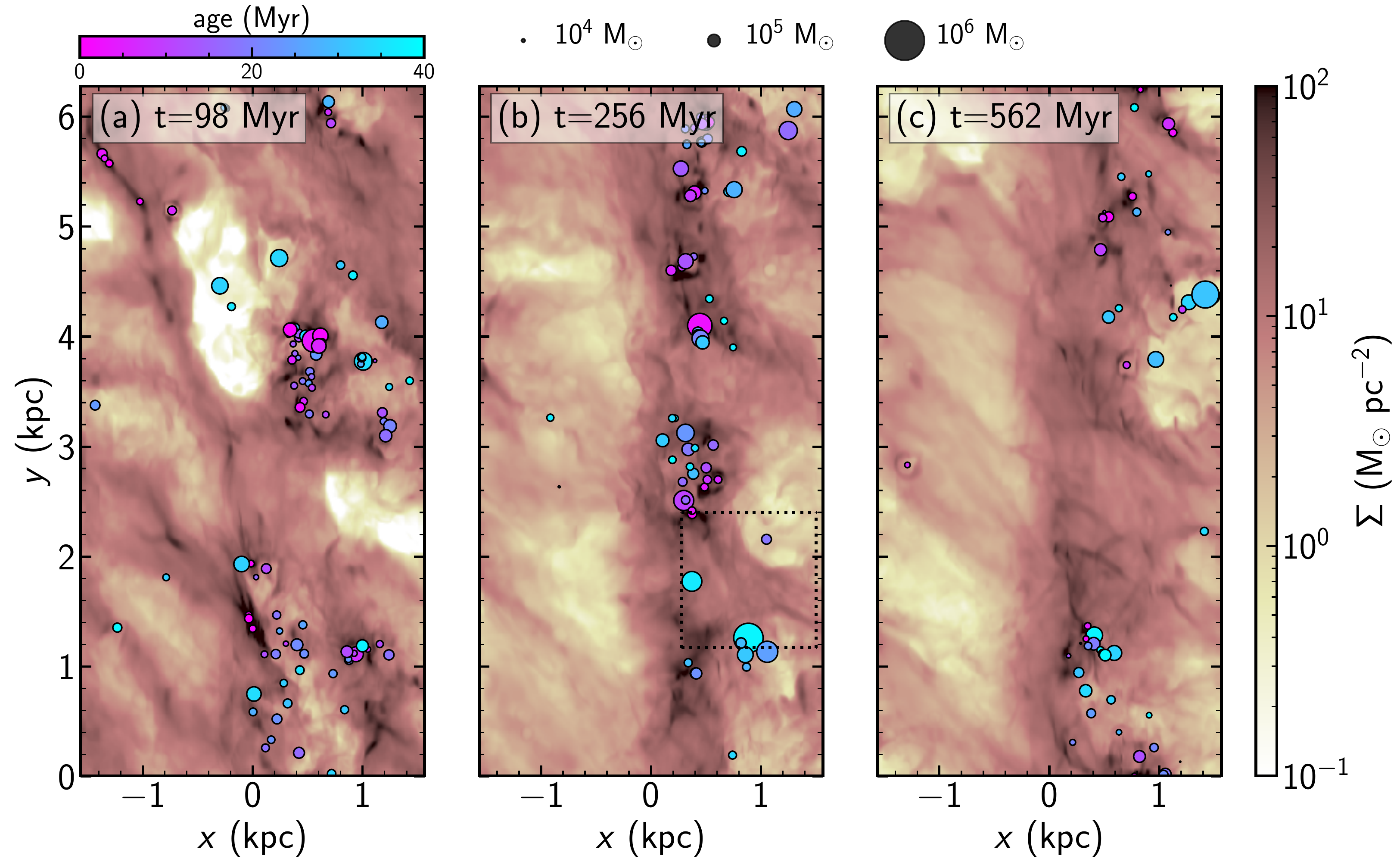}
\caption{Snapshots of the gas surface density of model {\tt F20B10} at (a) $t=98\Myr$, (b) $t=256\Myr$, and (c) $t=562\Myr$. Star particles younger than $40\Myr$ are projected onto the $x$--$y$ plane. The color and size of the circles indicate the age and mass of the star particles, respectively. The spiral potential builds up a dense ridge of gas near the potential minimum where most star formation takes place. SN feedback compresses the gas downstream, forming gaseous spurs jutting out perpendicularly from the arm. The squared section in (b) is zoomed in Figure \ref{fig:spur} to display the configurations of velocity and magnetic fields around a spur.}\label{fig:stdSurf}
\end{figure*}

\begin{figure*}
\epsscale{1.0} \plotone{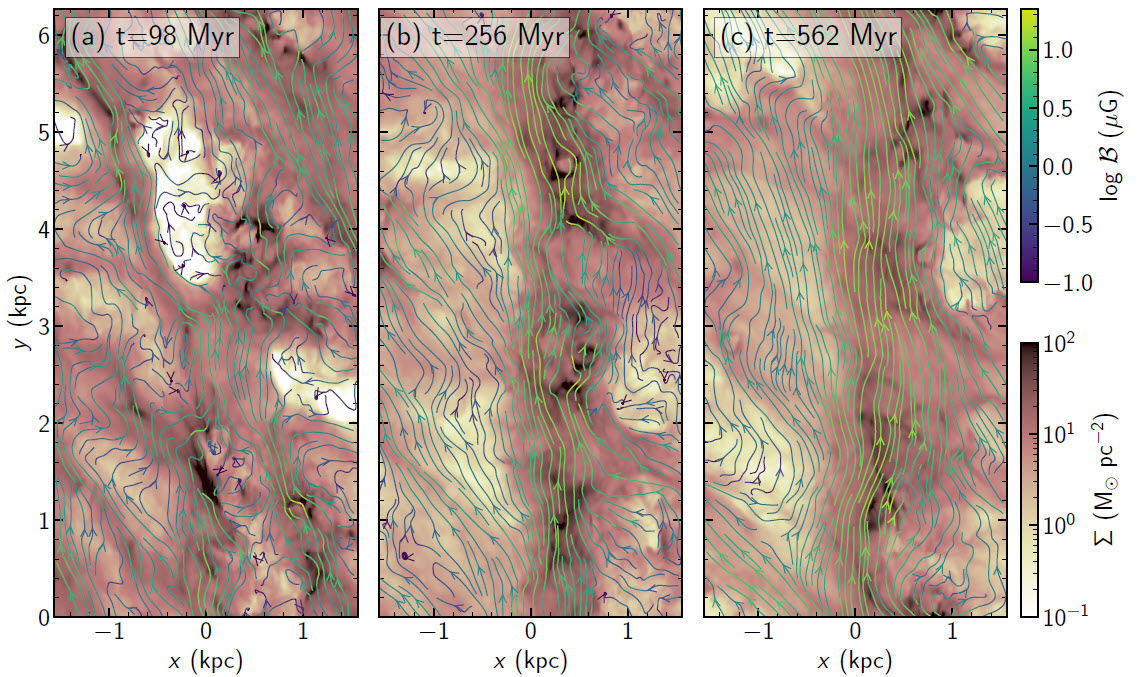}
\caption{Streamline plots of the vertically-integrated magnetic fields $\mathcal{B}_x=\int \rho B_x dz/\Sigma$ and $\mathcal{B}_y=\int \rho B_y dz/\Sigma$ overlaid on the gas surface density $\Sigma$ of model {\tt F20B10} at (a) $t=98\Myr$, (b) $t=256\Myr$, and (c) $t=562\Myr$. The color of the streamlines corresponds to logarithm of the strength $\mathcal{B}=(\mathcal{B}_x^2 + \mathcal{B}_y^2)^{1/2}$. Overall, the magnetic fields are parallel to the arm and stronger in the arm than interarm region.}\label{fig:stdBfield}
\end{figure*}

\begin{figure*}
\epsscale{1.1} \plotone{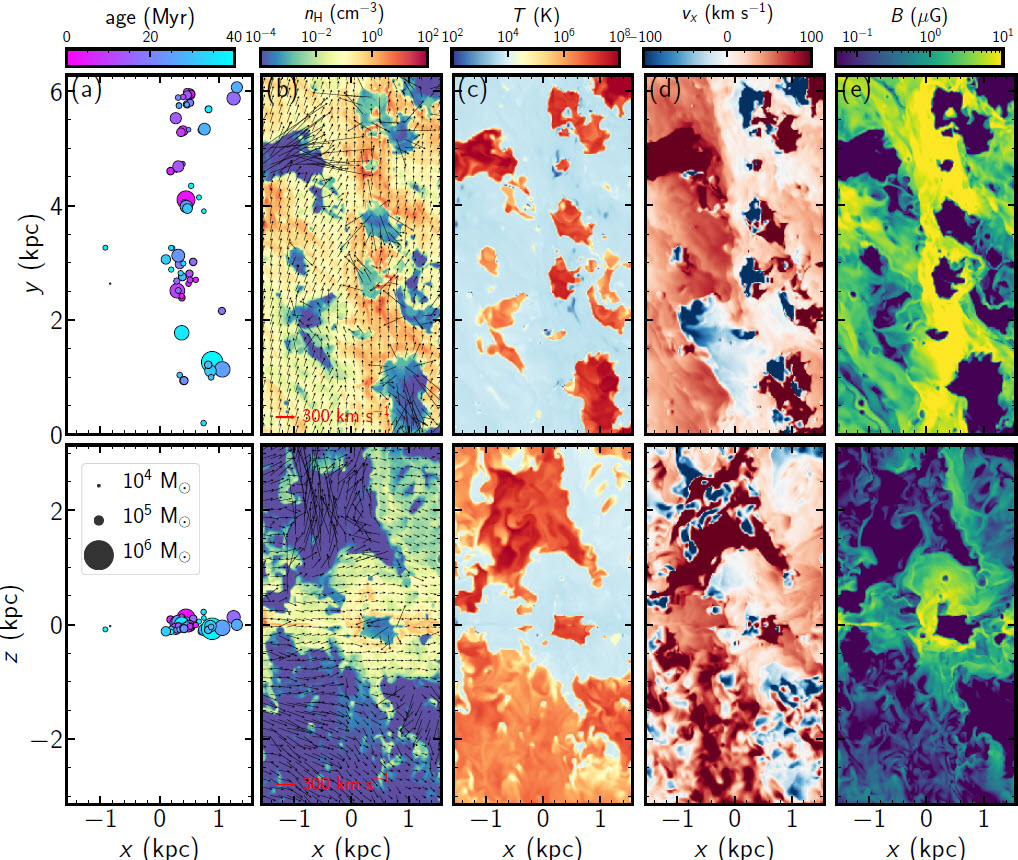}
\caption{Snapshots at $t=256\Myr$ of model {\tt F20B10}.
The top row panels show slices through the midplane, $z=0$, and the bottom row panels show vertical slices that cut through the disk perpendicular to the spiral arm at $y=3.7\kpc$.  Individual panels show
(a) projected positions of sink/star particles, (b) gas number density $n_\text{H}$ together with velocity vectors $(v_x, v_y)$ in the top panel and $(v_x, v_z$) in the bottom panel, (c) temperature $T$, (d) velocity $v_x$ perpendicular to the arm, and (e) magnetic field strength $B$.
In (a), only the sink/star particles with age younger than $40\Myr$ are shown, with their color and size representing age and mass, respectively. In (b), the amplitudes of the velocity vectors are indicated as the red arrows.}\label{fig:slices}
\end{figure*}

Our initial disks described above are initially out of thermal equilibrium, so that they would instantly experience cooling and heating. The cold gas would collapse toward the midplane and form a thin dense layer and undergo an extreme burst of star formation. To prevent this spurious collapse and star formation, we initially generate pseudo-star particles that survive only for an initial $40\Myr$ span,  which stir the disk by injecting SN feedback similarly to the feedback particles.  The creation rate of the pseudo-star particles is set to the value equivalent to the SFR surface density of $\SFR=5\times10^{-3}\SFRunits$, approximately the quasi-steady value of the solar-neighborhood model in the absence of a spiral arm \citep{cgkim17}.
The pseudo-star particle initial locations are chosen randomly in the horizontal direction and  follow an exponential
distribution in the vertical direction with scale height of $10\pc$. These particles are initially assigned background shearing velocity ${\bf v}_0$ as well as random velocities with a one-dimensional dispersion of $10\kms$.

\begin{figure}[t]
\epsscale{1.0} \plotone{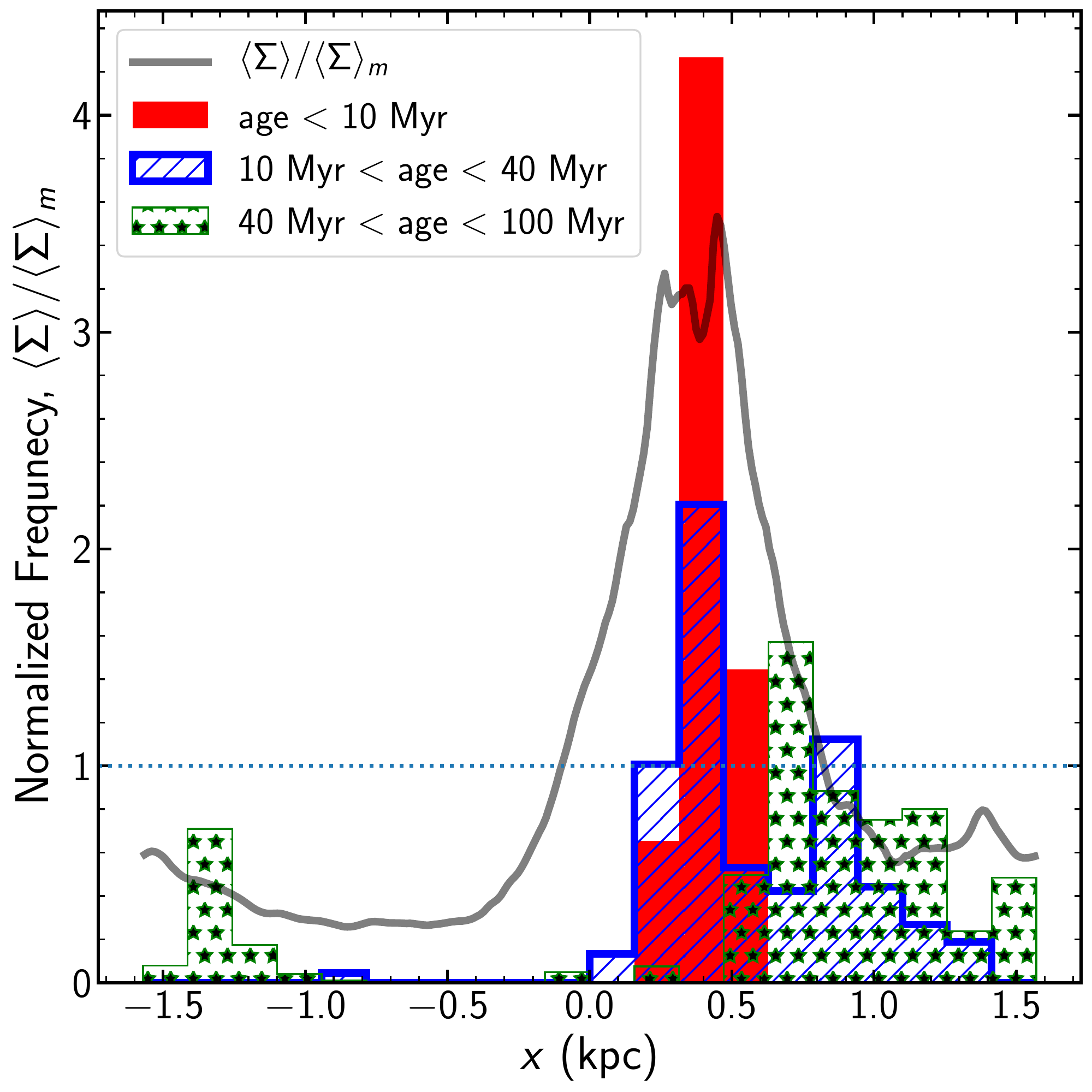}
\caption{Normalized histograms of the mass-weighted positions of star particles in the $x$-direction with ages $t_m<10\Myr$ (red filled), $10\Myr<t_m<40\Myr$ (blue hatched), and $40\Myr<t_m<100\Myr$ (green stars) in model {\tt F20B10} at $t=256\Myr$. For reference we show the normalized surface density $\langle\Sigma\rangle/\langle\Sigma\rangle_m$ averaged along the $y$-direction plotted as a grey solid line. The horizontal dotted line indicates  $\langle\Sigma\rangle/\langle\Sigma\rangle_m=1$.  The  youngest cohort of stars coincides with the peak gas surface density of the arm. }\label{fig:sp_xPDF}
\end{figure}

\begin{figure*}[t]
\epsscale{1.0} \plotone{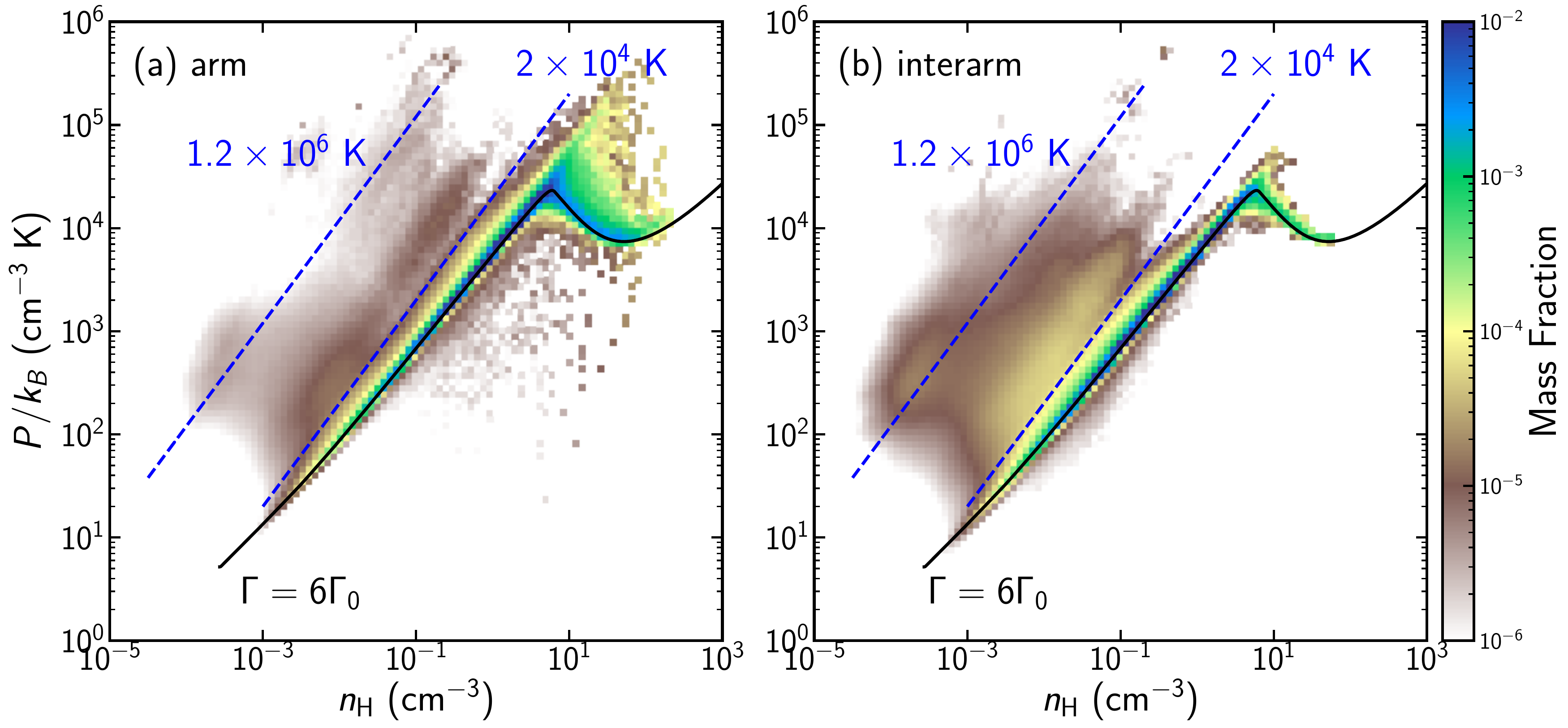}
\caption{Instantaneous distribution in the density-pressure plane of the gas in (a) the arm region at $-0.1\kpc<x<0.8\kpc$ and (b) the interarm region at $x<-0.1\kpc$ or $x>0.8\kpc$ of model {\tt F20B10} at $t=256\Myr$. The color represents the mass fraction of the gas. In each panel, the black solid line draws the thermal equilibrium curve for instantaneous heating rate 6$\Gamma_0$, while blue dashed lines indicate the minimum ($2\times 10^4$ K) and mean ($1.2\times10^6$ K) temperatures of the ionized-hot phase.  }\label{fig:phase}
\end{figure*}

\begin{figure*}
\epsscale{1.0} \plotone{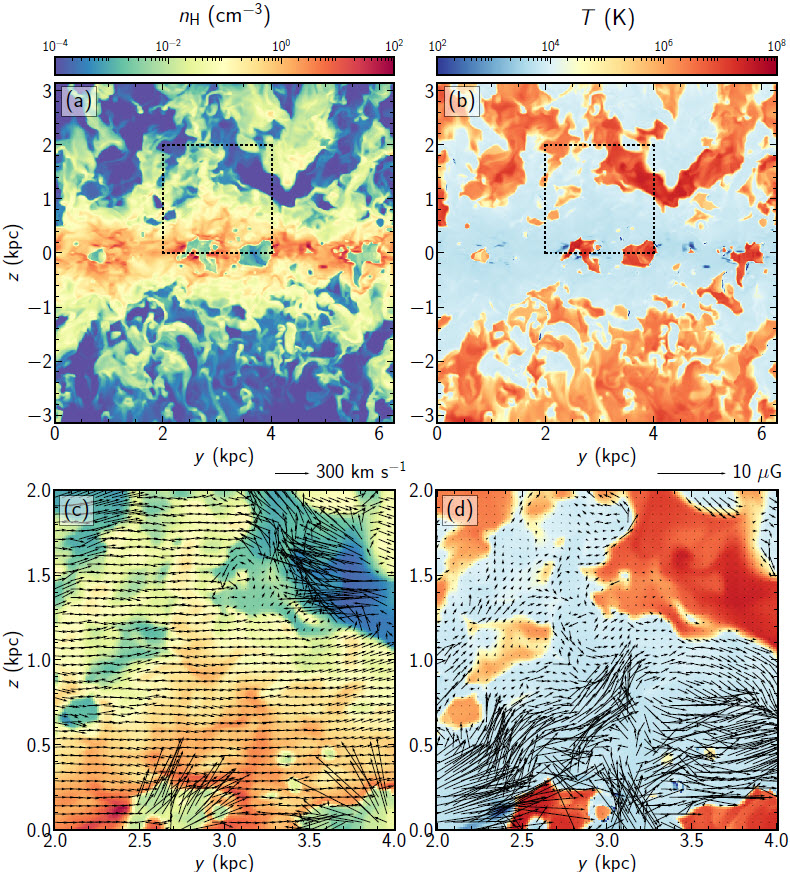}
\caption{Distributions of (a) gas density and (b)  gas temperature for model {\tt F20B10} at $t=256\Myr$. Both panels represent vertical slices that cut through the center of the spiral arm at $x=0.45\kpc$.
The section marked by a dotted square in the upper panels is enlarged in the lower panels to overlay the velocity vectors ($v_y, v_z$) in (c) and the magnetic field vectors ($B_y, B_z$) in (d). The arrows above in (c) and (d) measure the amplitudes of velocity and magnetic-field vectors, respectively.}\label{fig:YZdist}
\end{figure*}

\section{Evolution}\label{sec:evol}

In this section, we focus on the temporal and morphological changes of gaseous structures that form due to spiral compression and star formation feedback.  Physical quantities averaged over space and time and their correlations will be presented in Section \ref{sec:stat}.

\subsection{Overall Evolution of the Fiducial Model}\label{sec:fidu}

We first describe evolution of our fiducial model {\tt F20B10} with  $\mathcal{F}=0.2$ and $\beta=10$. Figure \ref{fig:stdSurf} plots snapshots of the gas surface density $\Sigma=\int \rho dz$, together with the projected distributions of young star particles in the $x$--$y$ plane, at a few selected epoches in model {\tt F20B10}.  Figure \ref{fig:stdBfield} displays streamline plots of the vertically-integrated in-plane magnetic fields $(\mathcal{B}_x, \mathcal{B}_y)=\int \rho (B_x, B_y) dz/\Sigma$ overlaid on the surface density distribution at the same epoches as in Figure \ref{fig:stdSurf}. Figure \ref{fig:slices} plots the distribution of sink/star particles younger than $40\Myr$, gas number density $n_\text{H}$, temperature $T$, velocity $v_x$ perpendicular to the arm, and magnetic field strength $B=|{\bf B}|$ in the (top) $z=0$ and (bottom) $y=3.7\kpc$ plane of model {\tt F20B10} at $t=256\Myr$ when the spiral potential is strong and star formation is active.

Pseudo-star particles introduced initially
provide density and velocity perturbations to the gaseous medium that would otherwise be uniform in the $x$--$y$ plane. Due to the background shear and self-gravity, the perturbations rotate kinematically with time and are amplified as they swing from leading to trailing configurations (e.g., \citealt{kim01}). Dense trailing filaments tend to have stronger magnetic fields than the surrounding underdense regions. In addition to swing amplification, some filaments become denser and colder as they collide with neighbors. Filaments that achieve sufficient density undergo gravitational collapse and spawn star particles (Figure \ref{fig:stdSurf}a). Newly created star particles heat up cold gas to the warm phase. Subsequent SN feedback creates hot gas bounded by expanding shells of the cold/warm gas around the star particles.\footnote{Following the convention of \citet{cgkim17}, we classify the gas into five phases according to its temperature: cold ($T<184\K$), unstable ($184\K<T<5050\K$), warm ($5050\K<T<2\times10^4\K$), ionized ($2\times10^4\K<T<5\times10^5\K$), and hot ($T>5\times 10^5\K$).} Supershells created by clustered SNe
expand to $\sim 1\kpc$ in diameter before being distorted significantly by the background shear or undergoing collisions with neighboring shells. Collisions of expanding, shearing shells produce high-density clumps susceptible to new star formation. Since the strength of the spiral potential is growing slowly, the early phase ($t\lesssim 100\Myr$) of evolution is similar to that of the TIGRESS run without spiral structure presented in \citet{cgkim17}.

\begin{figure*}
\epsscale{1.0} \plotone{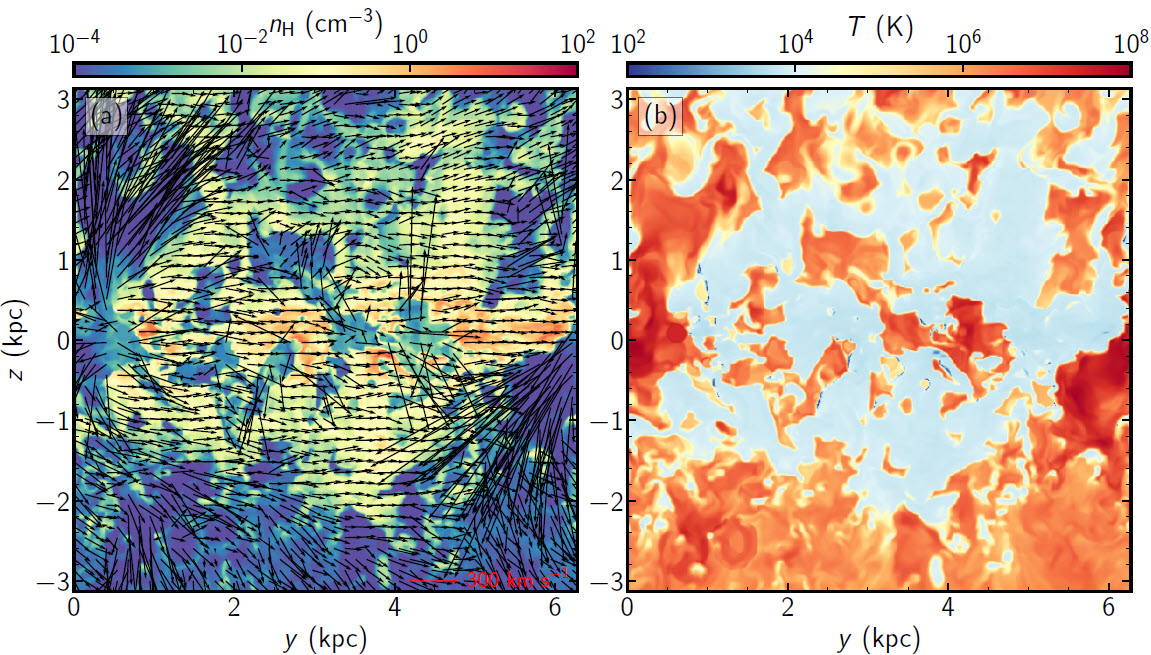}
\caption{Distributions of (a) gas density and (b)  gas temperature
for model {\tt F20B10} at $t=235\Myr$  at $x=0.90\kpc$, downstream from the density peak of the spiral arm. Hot gas is vented to high-$|z|$ regions through chimneys located near $(y,z)=(0.5, 1)\kpc$ and $(6, -0.5)\kpc$. The red arrow at the bottom of (a) measures the amplitudes of velocity vectors.}\label{fig:YZdist_t235}
\end{figure*}

\begin{figure*}
\epsscale{1.1} \plotone{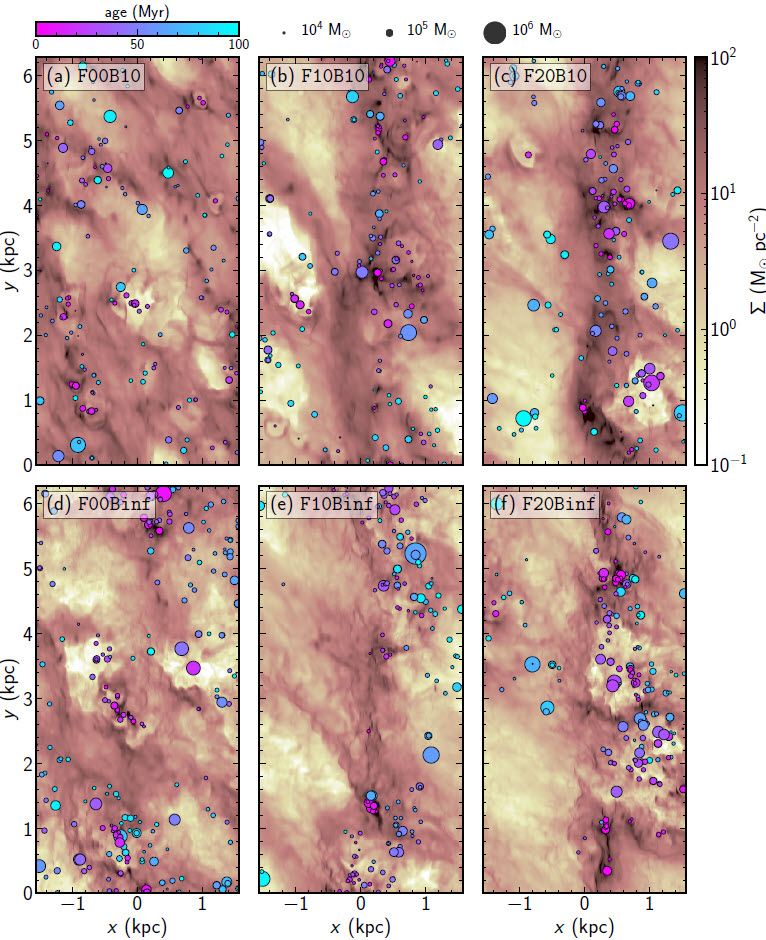}
\caption{Snapshots of gas surface density at $t=300\Myr$ for the (top) magnetized and (bottom) unmagnetized models. Star particles younger than $100\Myr$ are overlaid as circles, with their color and size denoting the age and mass, respectively. In models without spiral potential (left column), star formation occurs in shearing filaments distributed across the simulation domain. In the models with $\mathcal{F}=0.1$ (middle column), the spiral potential produces a star-forming, dense ridge parallel to the arm, and interarm spurs. In the  models with $\mathcal{F}=0.2$ (right column), dense ridges and spurs are more pronounced than in the $\mathcal{F}=0.1$ counterpart.}\label{fig:allsnap}
\end{figure*}

As the strength of the spiral potential increases further, material is preferentially gathered
toward the potential minimum. These converging motions of the gas inevitably result in the formation of a dense ridge of gas located at $x\sim0.2\kpc$ roughly parallel to the $y$-direction, which occurs at $t\sim162\Myr$ in model {\tt F20B10}.
Magnetic fields are overall parallel to the ridge and stronger in the regions with larger gas density, while being slightly inclined in the interarm region (Figure \ref{fig:stdBfield}). The arm density is naturally inhomogeneous along its length due to collisions of shearing filaments and shells.
It is in the high-density  regions along the ridge where most star formation takes place, although some clusters also form in downstream gaseous spurs after they develop.

Star formation
is usually clustered, as indicated by the distributions of star particles shown in Figure \ref{fig:stdSurf}(b). Clustered star formation forms supperbubbles filled with weakly-magnetized, rarefied hot gas bounded by strongly-magnetized, dense shells (Figure \ref{fig:slices}). Strong SN feedback from the star particles
expels the gas away from the star-forming sites.
Gas motions induced by feedback are limited, however, by the high-density ridge slightly to the right (downstream) from the potential minimum.
Gas pushed by feedback toward the upstream side of the galaxy rotation collides with the gas that is flowing into the arm, reinforcing the high-density ridge.
Large-scale spiral shocks produced by these gas collisions, as delineated by sharp discontinuities of $v_x$ at $x\sim-0.6$--$0\kpc$, are evident in Figure \ref{fig:slices}(d). Apparently, the shock fronts are not straight in the $y$-direction and consist of numerous small-scale curved shocks. The shocked gas is soon pushed back by the ram pressure of the interarm gas toward the potential minimum, suggesting that the spiral shocks oscillate with large amplitudes along the $x$-direction. On the other hand, gas pushed downstream by supernova explosions can easily leave the arm region, supplying
the interarm region.

New star particles produced in the arm climb out of the arm potential well as they follow their epicycle orbits downstream (see Appendix \ref{sec:eom}). They inject SN feedback in both arm and interarm regions. The mean speed of star particles perpendicular to the arm is $v_{0x}\sim 15\kms$, so that the SN feedback is concentrated mostly in a zone within $\sim v_{0x}t_\text{life}=0.6\kpc$ from the star-forming sites. Feedback from spatially uncorrelated star particles tends to simply increase random velocity dispersions of the surrounding gas.

The boundaries of distorted superbubbles typically take the form of dense structures attached to the arm in the surface density maps, as shown in Figure \ref{fig:stdSurf}.
Due to the velocity acquired during the superbubble expansion, swept-up gas in the midplane moves in positive or negative $y$-direction, with velocities $\sim 20$--$30\kms$ relative to the local background.
Relatively weak filaments are quickly swept/washed out by expansions of newly created bubbles nearby. Filaments strong enough to survive interactions with small bubbles undergo a collision with neighboring one, forming even denser, longer structures that extend almost over the whole length of the interarm region. Figure \ref{fig:stdSurf}(b) displays three large-scale gas structures located at $y \sim 1.6$, 4.0, and $5.7\kpc$, with the peak surface densities of $\sim15$, 22, $20\Surf$ measured at $x=1\kpc$, respectively. Such large-scale structures may be observed as gas and dust spurs/feathers in real spiral galaxies. In our simulations, strong spurs can produce star formation at their densest parts usually close to the arm, but sometimes deep in the interarm region.

Since star particles form mostly in the dense ridge and age as they move downstream, one can naturally expect an age gradient along the direction perpendicular to the arm. Figure \ref{fig:sp_xPDF} plots the normalized histograms of the mass-weighted $x$-positions of star particles with age $t_m<10\Myr$ (red filled), $10\Myr<t_m<40\Myr$ (blue hatched), and $40\Myr<t_m<100\Myr$ (green stars) in model {\tt F20B10} at $t=256\Myr$. For comparison, we also plot the $y$-averaged surface density $\langle\Sigma\rangle =\int \Sigma dy/L_y$ normalized to the mean value $\langle\Sigma\rangle_m \equiv \int \langle\Sigma\rangle dx/L_x$ as a grey solid line. As expected, younger star particles tend to be crowded closer to the density peak located at $x\sim0.3$--$0.5\kpc$: the mass-weighted mean locations of the star particles with $t_m<10\Myr$, $10\Myr<t_m<40\Myr$, and $40\Myr<t_m<100\Myr$ are at  $x_m=0.42$, 0.59, and $1.1\kpc$, respectively, with the standard deviations of $\sigma(x_m) =0.47, 0.87, 1.4\kpc$, respectively.\footnote{In calculating $x_m$ and $\sigma(x_m)$, we add $L_x$ to the $x$-coordinate for the star particles located at $x<0$, consistent with the periodic boundary condition.} These
correspond roughly to $dx_m/dt_m \sim 10\pc\Myr^{-1}$ and
$d\sigma(x_m)/dt_m \sim 14\pc\Myr^{-1}$ at $t=256\Myr$.
When averaged over $t=200$--$600\Myr$, we find the mean age gradient $dx_m/dt_m \sim 8.7\pm 3.9\pc\Myr^{-1}$ and the dispersion $d\sigma(x_m)/dt_m \sim 15.2\pm9.3\pc\Myr^{-1}$,
insensitive to $\mathcal{F}$ and $\beta$. This shows that older stars tend to be found farther away from the ridge and spread out more widely than younger stars.

We note, however, that older star particles are not always found downstream from younger star particles. This is because star formation sometimes occurs away from the ridge, and because some particles move in the negative $x$-direction on their epicycle orbits, mixing particles with different ages.

Based on the mean surface density, we define ``arm'' and ``interarm'' regions as the regions with $\langle\Sigma\rangle/\langle\Sigma\rangle_m$ greater or smaller than unity, respectively (cf. Fig. \ref{fig:sp_xPDF}). In model {\tt F20B10}, the arm region is at $-0.1\kpc<x<0.8\kpc$, varying slightly with time, and the rest of the domain is regarded as the interarm region.

Figure \ref{fig:phase} plots the mass-weighted probability distribution functions (PDFs) in the pressure--density plane for the gas in the (a) arm and (b) interarm regions at $t=256\Myr$ of model {\tt F20B10}. Two dashed diagonal lines mark the minimum ($2\times 10^4$ K) and mean ($1.2\times10^6$ K) temperatures for the ionized--hot phase. At this time, the majority of the gas is close to the instantaneous thermal equilibrium curve indicated by the black solid line with $\Gamma=6\Gamma_0$.

Including gas at all heights, the mass fractions of the cold-unstable, warm, and ionized-hot phases in the arm region are $0.16$, $0.83$, and $7.7\times10^{-3}$, respectively, which change to $2.6\times10^{-2}$, $0.95$, $2.4\times10^{-2}$ in the interarm region. The corresponding volume fractions of the cold-unstable, warm, and ionized-hot phases are $2.6\times 10^{-3}$, 0.44, and 0.56 in the arm, and $8.7\times10^{-4}$, 0.30, and 0.70 in the interarm region, respectively. For gas within $|z|\leq 1\kpc$, the volume fraction of the warm phase is increased to 0.78 and 0.50 in the arm and intearm regions, respectively. For gas at the midplane, the area fractions of the warm and ionized-hot phases are 0.71 and 0.26 in the arm, and 0.74 and 0.25 in the interarm regions, respectively. This indicates that gas in both arm and interarm regions is predominantly in the warm phase by mass (also by volume near the midplane), and most of the cold gas is in the arm region. Due to the deeper gravitational potential, the gas in the arm region tends to be denser and colder, and has higher pressure to support its weight against gravity, than in the interarm region.

Cold and warm gas is lifted to high-$|z|$ regions from the midplane due to the vertical momenta of superbubbles which drift with the background flows as they expand.
Figure \ref{fig:YZdist} plots the $y$--$z$ distributions of the gas density and temperature at $x=0.45\kpc$ of model {\tt F20B10} at $t=256\Myr$ in the upper panels as well as their zoomed-in views together with velocity and magnetic field structures in the lower panels. Apparently, the gas distribution near the midplane in spiral arms is highly clumpy, characterized by expanding low-density bubbles surrounded by distorted high-density shells. Expansion speeds of hot gas in superbubbles easily exceed the speed of galaxy rotation,  pushing on the surrounding dense  shell to drive turbulence in the warm ISM. Expanding superbubbles transport the midplane gas to high-$|z|$, causing the height of the regions with largest density (hence highest SFR) to occur at $|z|\sim40\pc$ rather than at $z=0$ (see below).

Superbubbles usually break out from the cold-warm midplane layer in the interarm region downstream from the densest region of the arm, leading to the rise of hot gas to high-$|z|$ regions.
Figure \ref{fig:YZdist_t235} plots a slice through the density and temperature distributions in the $y$--$z$ plane at $x=0.90\kpc$ for model {\tt F20B10} at $t=235\Myr$. A superbubble centered at  $(y,z)=(0.5\kpc, 0)$ breaks out vertically and produces two slightly inclined chimneys though which hot, tenuous gas is vented to high-$|z|$ regions. Because of our periodicity in $y$, the vent towards positive $z$ appears at small $y$ and the vent towards negative $z$ appears mostly at large $y$ in  the figure.
As previously discussed by \citet{cgkim18}, superbubbles can drive warm gas to large $|z| \sim 1-3 \kpc$.  Since this warm gas does not achieve the escape velocity ($\sim200\kms$), it falls back toward the midplane, creating a warm galactic fountain. In contrast, the hot gas in chimneys has high velocities, and in our simulations is lost through the vertical boundaries located at $z=\pm\pi\kpc$.  In real galaxies, this venting produces a hot galactic wind.

Supernova/superbubble shell expansions, mutual collisions, and interactions with the background shear and gravitational potential drive turbulent motions in the gas. While SN feedback is concentrated primarily in zones immediately downstream from star-forming regions,
the injected energies and momenta are advected by the mean flows to make all of the gas turbulent across the simulation domain. The related gas motions also stretch, twist, and amplify magnetic fields that tend to be strong in high-density regions. Figure \ref{fig:YZdist}(d) shows that  although magnetic fields point on average in the $y$-direction (i.e., parallel to the spiral arm), they change their directions quite rapidly in the $y$--$z$ plane, indicating that the turbulent component of magnetic fields is as strong as the regular component. We defer more detailed discussion of the magnetic fields to Section \ref{sec:mag}.

After the spiral potential is fully turned on, the system
reaches a quasi-steady state in the sense that feedback from star formation
in the arm region drives turbulence and heating
that balance
dissipation and radiative losses of turbulent and thermal energies. Still, Figure \ref{fig:stdSurf} shows that the spiral arm at $t=562\Myr$ is less concentrated, harbors less active star formation, and is associated with less pronounced spurs, compared to the arm at $t=256\Myr$. This is caused primarily by the secular reduction in the gas mass over time in our simulations.  With no gas inflow from outside, continued star formation as well as mass loss at the vertical boundaries keep decreasing the gas mass over time.  Reduced self-gravity further lowers the gas density and hence SFR in the arm region.

\begin{figure*}
\epsscale{1.0} \plotone{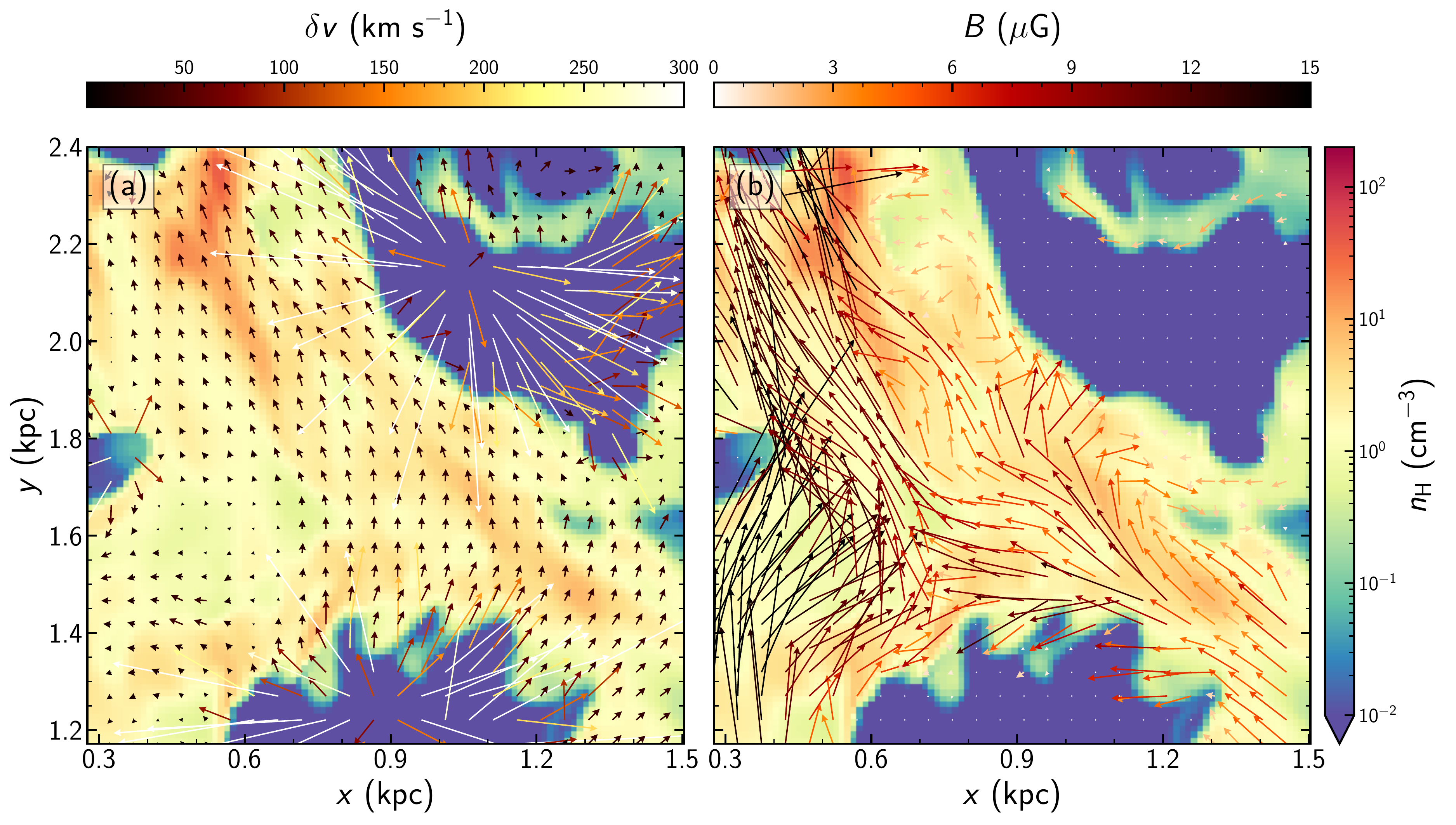}
\caption{Zoomed-in view of the spur in the region marked by a square in Figure \ref{fig:stdSurf}(b), from model {\tt B10F20} at $t=256\Myr$.
Vector fields of the perturbed velocity $(\delta v_x, \delta v_y)\equiv (v_x-v_{0x}, v_y-v_{0y})$ and the magnetic field $(B_x, B_y)$ are  overlaid on the midplane gas number density (color scale). The color (and length) of the vectors represents the amplitude of (a) the in-plane perturbed velocity $\delta v=(\delta v_x^2+\delta v_y^2)^{1/2}$ and (b) the in-plane magnetic field $B=(B_x^2+B_y^2)^{1/2}$.
The spur is confined by two superbubbles centered at $(x,y)=(0.9, 1.2)\kpc$ and $(1.1, 2.1)\kpc$. Magnetic fields are parallel to the arm at $x\lesssim0.7\kpc$, and to the spur at $x\gtrsim0.7\kpc$. }\label{fig:spur}
\end{figure*}

\subsection{Other Models}

The evolution in models with no spiral potential is similar to that of the ``Solar neighborhood'' TIGRESS run presented in \citet{cgkim17}, in that most star formation takes place in sheared filaments distributed across the simulation domain. There is neither formation of dense ridges resembling spiral arms nor gaseous spurs perpendicular to the arms.
In contrast,
models with a spiral potential all form dense arm ridges and gaseous spurs, regardless of the presence of magnetic fields, similarly to the fiducial model.   The ridge density and the number of spurs depends on $\mathcal{F}$ and $\beta$.

Figure \ref{fig:allsnap} plots the distributions of the gas surface density in the $x$--$y$ plane for all models at $t=300\Myr$. Star particles with age less than $100\Myr$ are overlaid as circles. While star particles are widely distributed in models without a spiral potential (left column), they are concentrated along the dense ridges in models with spiral potential (middle and right columns). Note that all models with $\mathcal{F}\neq0$ possess spurs that jut out approximately perpendicularly from the arm and turn to trailing configurations in the interarm region. Weaker spiral-arm forcing in model {\tt F10B10} produces a less pronounced ridge and induces more distributed star formation in the $x$-direction compared to model {\tt F20B10}. Without magnetic support, star formation in unmagnetized models is more active, which in turn makes the ridges less dense and broader in the $x$-direction compared to the magnetized counterpart.

\subsection{Spur/Feather Formation}

\begin{figure}
\epsscale{1.0} \plotone{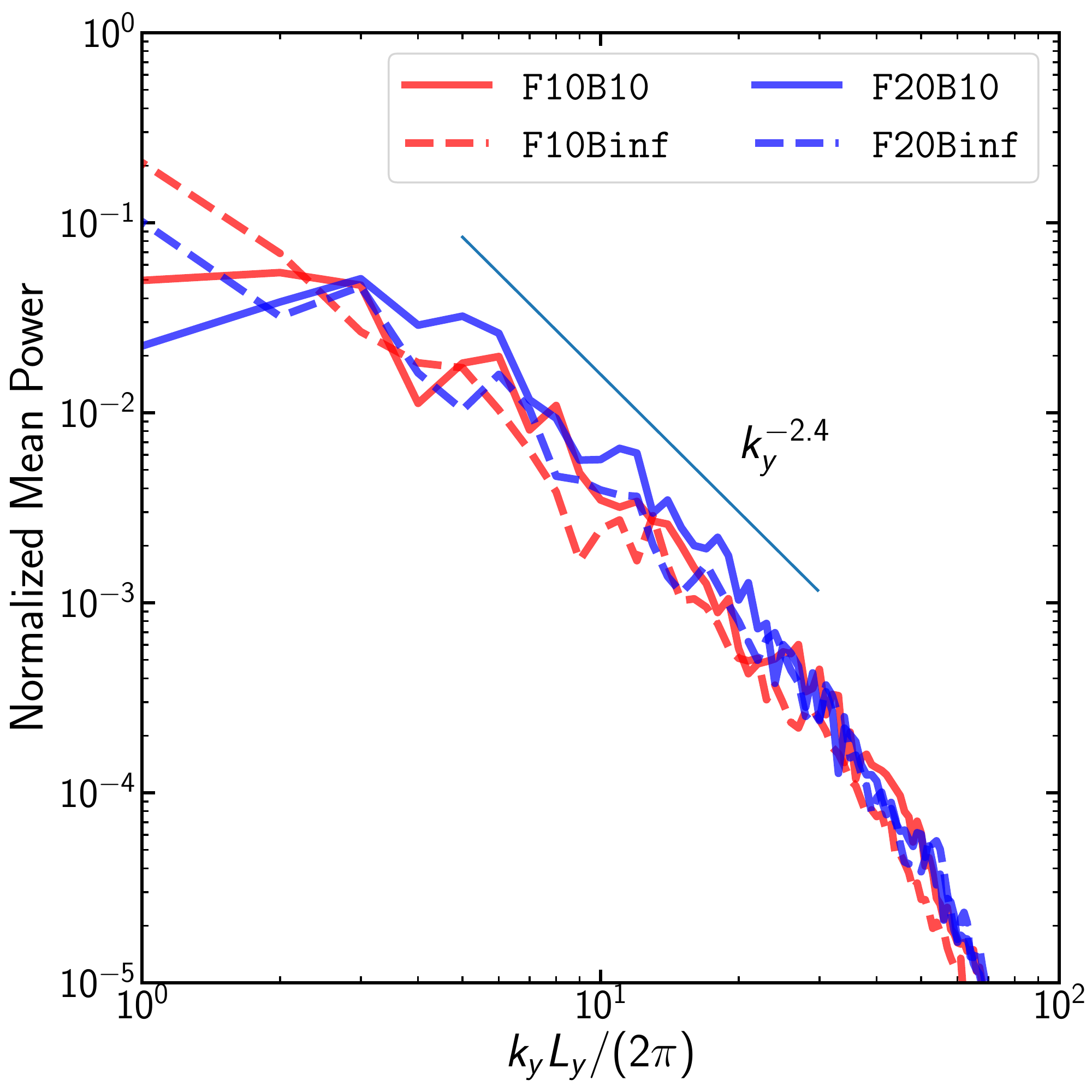}
\caption{Mean power spectra of the integrated surface density $\int_{0.8\kpc}^{1.2\kpc}\Sigma(x,y,t) dx$ averaged over $200\Myr \leq t \leq 300\Myr$ as a function of the dimensionless wavenumber $n_y=k_yL_y/(2\pi)$ for models with a spiral arm. Each power spectrum is normalized to unity at $n_y=0$. The short line segment draws the slope of $-2.4$, which best describes the surface-density power in the range with $5\leq n_y \leq 30$.}\label{fig:fpower}
\end{figure}

As mentioned above, gaseous spurs/feathers form in both magnetized and unmagnetized models as long as the spiral potential is present. To illustrate the structure of large-scale spurs formed in our simulations, Figure \ref{fig:spur} enlarges the square sector marked in Figure \ref{fig:stdSurf}(b) and plots the in-plane velocity and magnetic fields overlaid on the gas density in the $z=0$ plane. Clearly, the spur at $y\sim1.6\kpc$ is bounded by the SN feedback regions created by two groups of star particles located at $(x,y)=(0.9, 1.2)\kpc$ and $(1.1, 2.1)\kpc$. The material in the spur moves roughly parallel to the arm with velocities $\delta v_y = v_y - v_{0y}\sim 30\kms$ relative to the background flows.
In the arm region, magnetic fields have strength $\sim 10$--$14\,\mu$G,  and are oriented overall parallel to the arm. In the interarm region, on the other hand,  compression
by the expanding superbubbles orients magnetic fields
so that they  approximately follow the direction of the spurs (see also Figure \ref{fig:stdBfield}), with strength  $\sim4$--$8\,\mu$G. The lifetime of feedback-induced spurs is rather short, of order of $\sim 30\Myr$.  After cessation of the correlated feedback that created a given spur structure,  its destruction is accomplished primarily by feedback from neighboring regions (and sometimes the spur itself).

Large-scale spurs in our simulations are created by the collisions of sheared supershells that in turn form due to clustered star formation in the arm region. This implies that the presence of spurs requires strong star formation in the arm and spatially correlated feedback. Our fiducial model possesses prominent spurs mostly during the period $200\Myr \lesssim t \lesssim 300\Myr$ when the spiral potential is fully turned on and the arm has a plenty of gas available for active star formation. To estimate the typical spur spacing in our simulations, we calculate the mean surface density averaged over a narrow strip with $x\in [0.8, 1.2]\kpc$ where spurs are strong, and take its Fourier transform along the $y$-direction.  Figure \ref{fig:fpower} plots the mean power spectra averaged over $t=200$--$300\Myr$ as a function of the dimensionless $y$-wavenumber $n_y = k_yL_y/(2\pi)$, normalized to unity at $n_y=0$, for models with a spiral potential. The power spectra in models {\tt F10B10} and {\tt F20B10} are dominated by the modes with $n_y=2$ and $3$,  corresponding to two and three spurs, respectively, while $n_y=1$ and 2 modes are significant in models  {\tt F10Binf} and {\tt F20Binf}. This indicates that the typical spur spacing in our magnetized simulations is $\sim2$--$3\kpc$, with the smaller value corresponding to a stronger arm. In all models, the power spectra have a slope of $-2.4$ over $n_y \in [5, 30]$, which is slightly shallower than the two-dimensional (2D) Kolmogorov spectrum with a slope of $-8/3$.

\begin{figure*}
\epsscale{1.0} \plotone{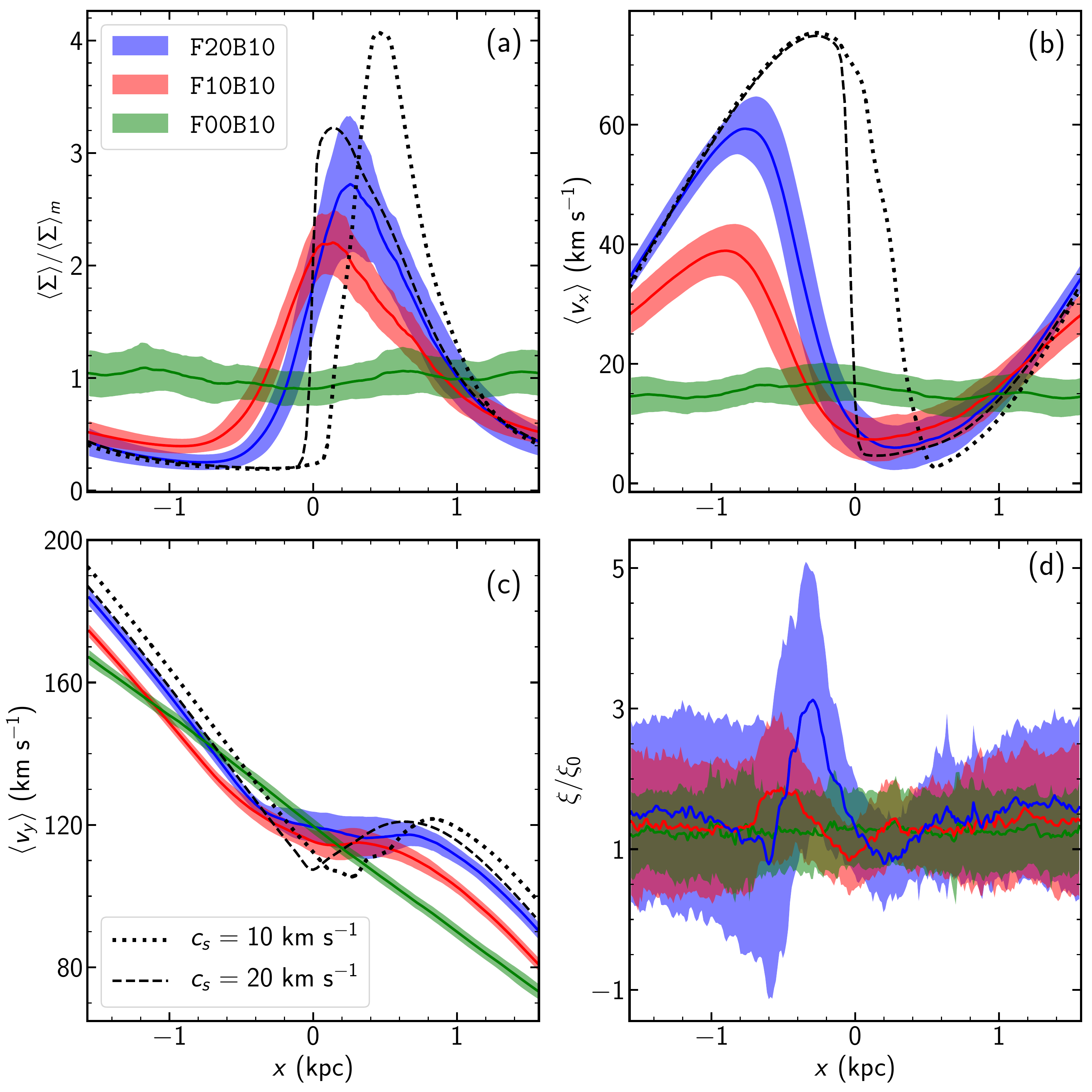}
\caption{
Profiles perpendicular to the arm
of (a) normalized surface density $\langle\Sigma\rangle/\langle\Sigma\rangle_m$, (b)
velocity perpendicular  to the arm
$\langle v_x\rangle$, (c)
velocity parallel to the arm $\langle v_y\rangle$, and (d) normalized PV $\xi/\xi_0$. All magnetized models are shown, with  $\mathcal{F}$ for each model indicated in the key.  The solid lines give the ($y$-averaged) mean values over $t=200$--$600\Myr$, while the shades indicate the standard deviations. For reference, the dotted and dashed lines in (a)--(c) plot the arm profiles in a vertically-stratified, isothermal disk without star formation, with a sound speed $c_s=10$ and $20\kms$, respectively,
and other parameters as in model {\tt F20B10}.
}\label{fig:SigVxVy}
\end{figure*}

\begin{figure*}
\epsscale{1.0} \plotone{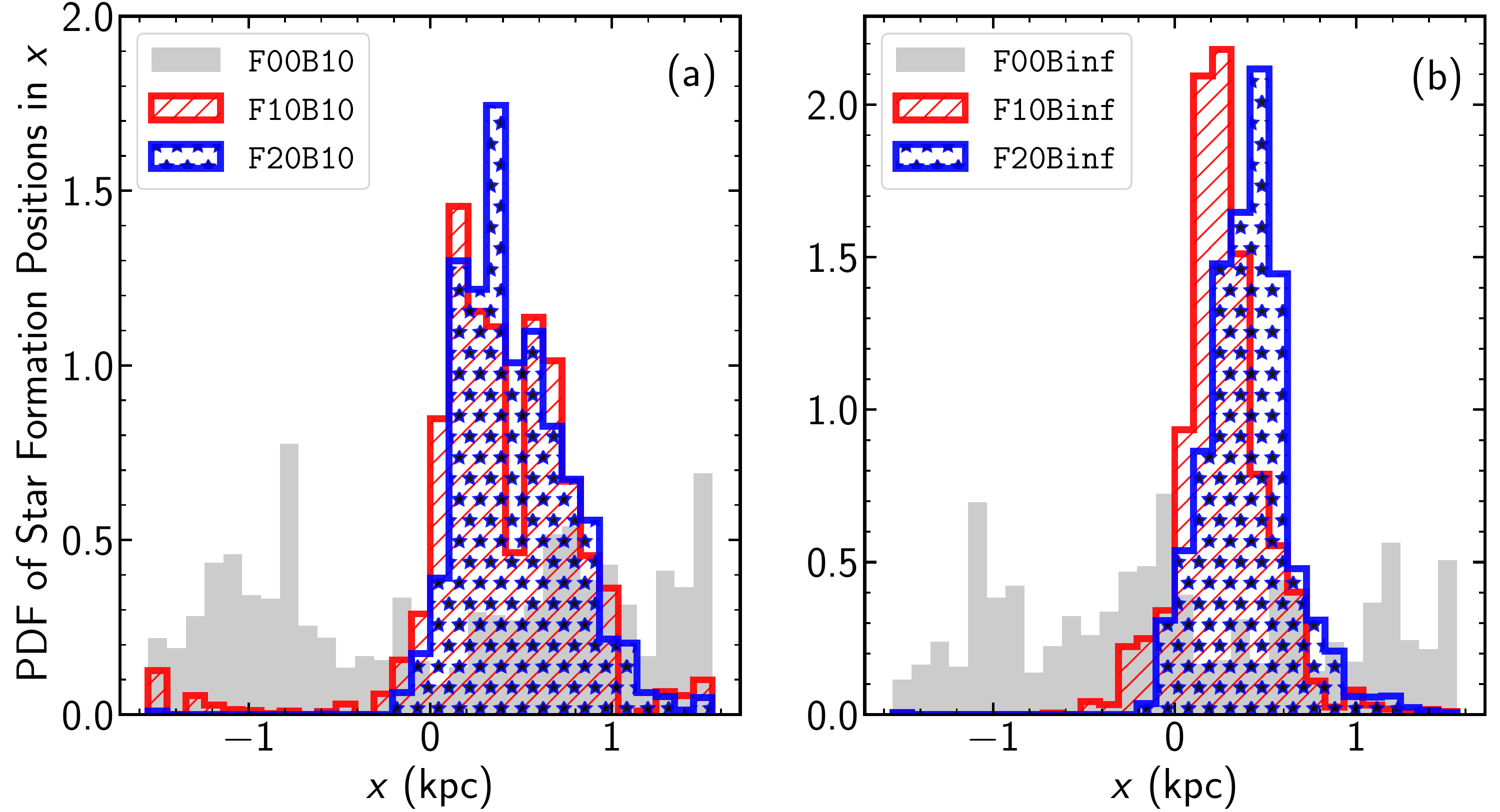}
\caption{PDFs of the mass-weighted positions of star-forming regions in the $x$-direction over $t=200$--$600\Myr$ for (a) magnetized and (b) unmagnetized models. }\label{fig:SF_XPDF}
\end{figure*}

\begin{figure*}
\epsscale{1.0} \plotone{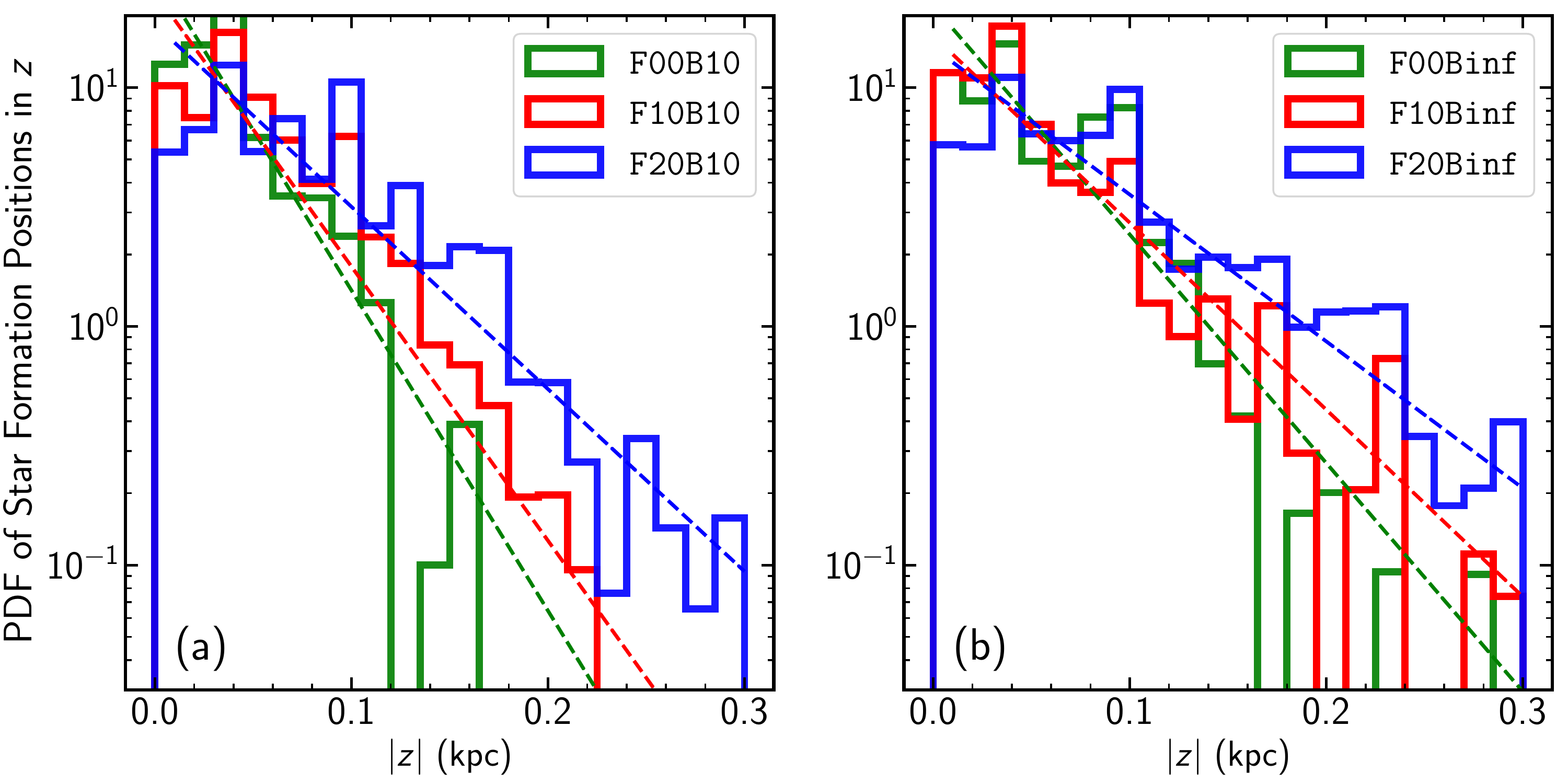}
\caption{PDFs of the mass-weighted positions of star-forming regions in the $z$-direction over $t=200$--$600\Myr$ for (a) magnetized and (b) unmagnetized models. The dashes lines are fits to the PDFs. See text for details.}
\label{fig:SF_ZPDF}
\end{figure*}

With reduction in the arm gas density, star formation at $t\gtrsim300\Myr$ becomes distributed and spatially uncorrelated, and the associated feedback occurs almost randomly in the $y$-direction. Bubbles and shells that form are usually small (less than $\sim0.5\kpc$) in size, and collide with each other to create small filaments, some of which undergo star formation in the interarm region. We note that arm star formation can still be clustered and correlated, but this occurs relatively rarely during the late time evolution of our simulations. When highly clustered star formation events do occur at late times, they are followed by formation of large-scale spurs.

Finally, we note that smaller-scale spurs can form temporarily as some of the material collected into ``superclouds'' in the arm moves out into the interam region, where it is sheared out into trailing filaments.  This process is somewhat reminiscent of the spur-formation mechanism originally identified by \citet{kim02}.  There, self-gravitating perturbations grew in the dense, magnetized arm regions, with the overdensities advected into the interarm region then forming spurs.  In those earlier models, the only physical source driving large-scale perturbations was self gravity (aided by magnetic stresses).  In contrast, when correlated supernovae are present as in the current work, they are able to drive large-scale velocity flows that create (and also destroy) spiral arm spurs.

\section{Temporal and Spatial Variations of Physical Quantities}\label{sec:stat}

We now investigate the
profiles of various physical quantities such as surface densities, velocities, SFR, magnetic fields etc.,
as well as correlations among them.  For these profiles, the independent variable is the quasi-radial coordinate, i.e., the distance perpendicular to the spiral arm.  We also present the temporal evolution of volume-integrated quantities including the SFR and mass fractions.

\subsection{Spiral Arm Profiles}

\begin{figure*}
\epsscale{1.0} \plotone{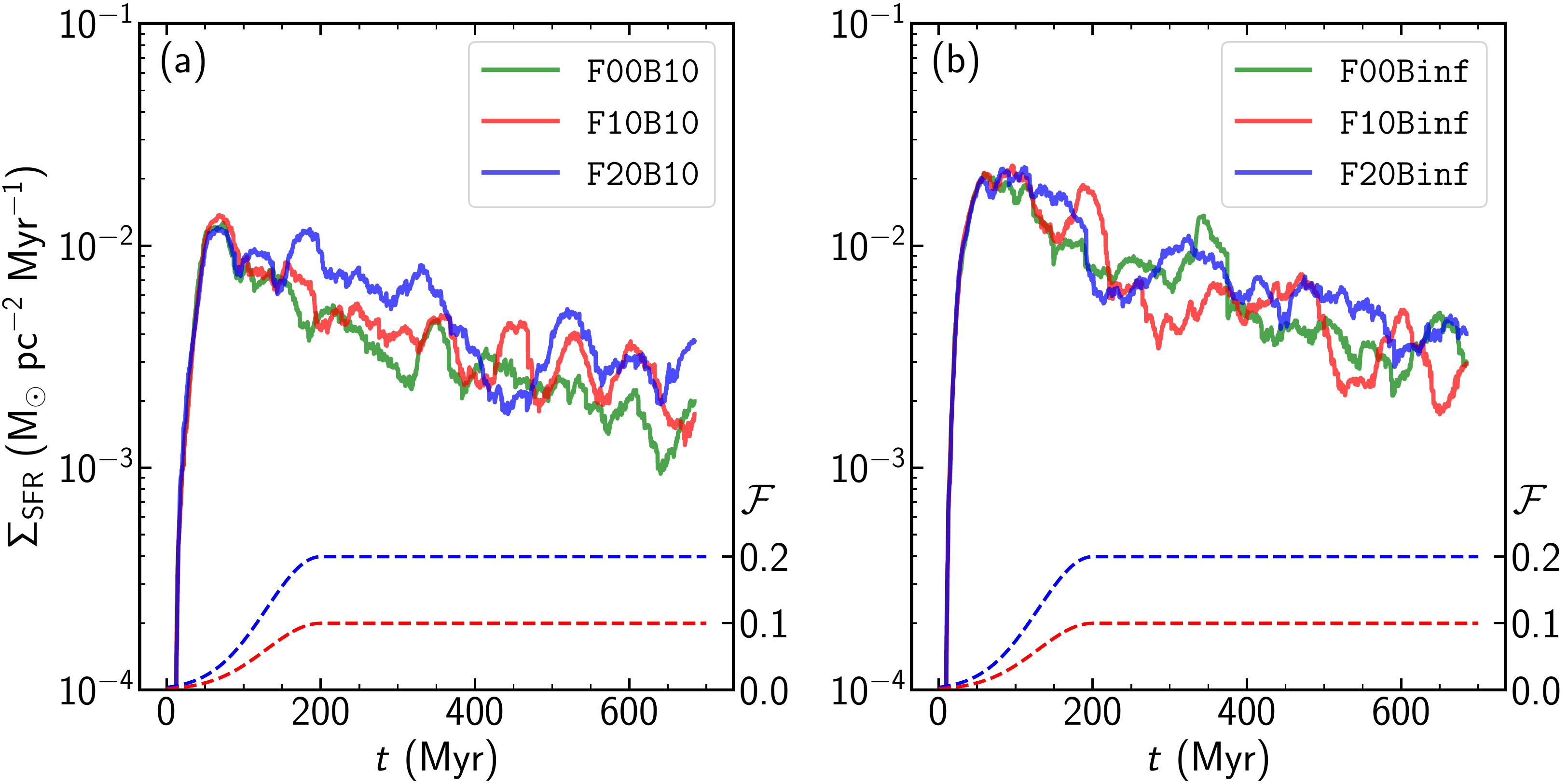}
\caption{Temporal evolution of the global SFR surface density (solid lines; left axis), $\SFRf$, measured using star particles younger than $40\Myr$, for (a) magnetized and (b) unmagnetized models. Dashed lines (right axis) plot the time variation of the spiral-arm forcing, $\mathcal{F}$, in our spiral-arm models.
$\SFRf$ rapidly rises initially,
achieves a peak at $t\sim60$--$70\Myr$, and then decreases secularly with time. Unmagnetized models have higher $\SFRf$ than their  magnetized counterparts.  The spiral-arm forcing enhances $\SFRf$, but by less than a factor 1.6 on average. }\label{fig:SFR_evol}
\end{figure*}

It is interesting to compare the averaged arm profiles from our simulations with the much simpler case of isothermal spiral shocks without star formation (theory for the latter dates back to \citealt{rob69}).\footnote{\citet{rob69} took a quasi-axisymmetric approximation to obtain one-dimensional periodic solutions of spiral shocks.  While the full velocity $\langle v_y \rangle$ shown in Figure \ref{fig:SigVxVy}(c) is not periodic in $x$, its perturbed part, $\langle v_y \rangle -v_{0y}$, is periodic.
Although the background velocity $v_{0y}$ varies linearly in $x$ (Equation \ref{eq:v0}), this does not alter the form of the perturbation equations that control the (linear or nonlinear) arm profile \citep[see also][]{shu73,kim14}.} Figure \ref{fig:SigVxVy} plots for the magnetized models the dependence on $x$ of (a) the normalized gas surface density $\langle\Sigma\rangle/\langle\Sigma\rangle_m$, (b) the density-weighted quasi-radial velocity $\langle v_x \rangle \equiv \int \rho v_x dydz /\int \rho dydz$, (c) the density-weighted quasi-azimuthal velocity $\langle v_y \rangle \equiv \int \rho v_y dydz /\int \rho dydz$, and (d) the normalized PV $ \xi \equiv  (d \langle v_y \rangle/dx + 2\Omega_0)/\langle\Sigma\rangle$ relative to the initial value $\xi_0 = (2-q)\Omega_0/\Sigma_0\rightarrow \Omega_0/\Sigma_0$ (e.g., \citealt{bal88,kim02,kim06}). The solid lines draw the temporal averages over $t=200$--$600\Myr$, while the shades represent the standard deviations. For comparison, we also plot examples of  profiles in  $x$ for quasi-steady solutions in the  isothermal case, including self-gravity (but with no star formation and feedback), based on a vertical  average of an $x-z$ spiral shock solution.  The cases shown as black dotted and dashed lines have sound speed $c_s=10$ and $20\kms$, respectively, with the other arm and disk parameters set equal to those of model {\tt F20B10}.

In model {\tt F00B10} with no arm, the averaged profiles are very close to their initial distributions. The mean PV is enhanced by about $\sim20\%$ most likely due to vorticity generation across numerous curved shocks produced by SN feedback. Also, the presence of magnetic fields contributes to the changes of PV (e.g., \citealt{web15}).

In models with spiral arms, the spiral potential compresses the gas toward the potential minimum, increasing (decreasing) gas surface density in the arm (interarm) region.
For a steady flow that varies only with $x$, mass flux $\Sigma v_x$ and PV $\xi$  are conserved, which leads to the characteristic profiles shown for the isothermal models: (1) with the rapid increase of  $\Sigma$  moving into the shock, $v_x$ rapidly decreases, while in the post-shock and interarm region where  $\langle\Sigma\rangle$ steadily decreases,  $v_x$ steadily increases; (2) in the arm region where $\Sigma/\Sigma_0 > 2/(2-q)\rightarrow 2$, the azimuthal shear $dv_y/dx\approx R_0 d\Omega/dR$ becomes positive (e.g., \citealt{rob69,shu73,bal85}).
Simple isothermal models with stronger spiral-arm forcing and lower sound speed also tend to have stronger shocks with density peaks further downstream (e.g., \citealt{kim02,kim14}).
Some of these features are recovered in the present (much more physically complex) simulations, while others are not.

Evidently, stronger spiral-arm forcing in the {\tt F20B10} model compared to {\tt F10B10} does result in larger changes in $\Sigma$ and $v_x$, and places the density peak farther downstream.
In terms of the peak density and its location, the spiral-arm profiles in model {\tt F20B10} are similar to those in the isothermal counterpart with $c_s=20\kms$.
We find, however, that the oscillations of spiral shocks in the $x$-direction (both in time and along $y$) mentioned in Section \ref{sec:fidu} smear out the averaged profiles considerably, especially near $x\sim0$ where the spiral shocks are found.
This smearing makes the $y$-averaged increase in $\Sigma$ and decrease in $v_x$ on the upstream side of the arm less steep than the isothermal models.
The smearing, in combination with magnetic and Reynolds stresses that break conservation of angular momentum,
removes the region of shear reversal (i.e., $dv_y/dx>0$)
that is a characteristic feature of simple, one-dimensional isothermal spiral shocks.
Curved shocks along the dense ridges (see Figures \ref{fig:stdSurf} and \ref{fig:allsnap}) are responsible for the local peak in PV near $x\sim -0.3\kpc$ in model {\tt F20B10}.

\subsection{Star Formation}\label{sec:SF}

Figures \ref{fig:SF_XPDF} and \ref{fig:SF_ZPDF} plot the PDFs of the mass-weighted positions of star formation sites in the $x$- and $z$-directions, respectively, during $t=200$--$600\Myr$, for the (a) magnetized and (b) unmagnetized models. Star formation in models with no arm is widely distributed in the $x$-direction, with large fluctuations due to stochasticity. The spiral potential gathers the star-forming regions close to the arm where the gas is also concentrated. When the arm is defined as the region where the $y$-averaged  gas surface density exceeding the domain-averaged value (see Section \ref{sec:fidu} and Figure \ref{fig:SigVxVy}(a)), for example, $\sim 95\%$ and $\sim90\%$ of all star formation occurs inside the arm in models {\tt F20B10} and {\tt F10B10}, respectively. The fraction of the arm star formation is increased to $\sim 98\%$ and $\sim97\%$ in models {\tt F20Binf} and {\tt F10Binf} without magnetic fields, respectively. Most of the interarm star formation occurs in strong spurs/feathers.

Figure \ref{fig:SF_ZPDF} shows that star formation is concentrated toward the midplane, although the most probable locations are at $|z|\sim 40\pc$ rather than at $z=0$ for all models. The PDFs in $|z|$ are overall described by an exponential function, $\propto\exp(-|z|/H_\text{SF})$, with a scale height of $H_\text{SF}=75, 87, 131\pc$ in models {\tt F00B10}, {\tt F10B10}, {\tt F20B10} and $H_\text{SF}=100, 128, 163\pc$ in models {\tt F00Binf}, {\tt F10Binf}, {\tt F20Binf}, respectively. Models with stronger arms have higher $H_\text{SF}$, and the presence of magnetic fields tends to reduce $H_\text{SF}$ by about $30\%$. This is because a higher SFR in the unmagnetized models can lift the dense gas to higher-$|z|$ regions than in the magnetized models.  Due to the spiral arm compression, gas in models with higher $\mathcal{F}$ is denser in the arm and can achieve the threshold density for star formation at higher $|z|$.

To define profiles of the star formation rate surface density, $\Sigma_\mathrm{SFR}$, in the direction perpendicular to the arm,
at any given time we select young star particles with age less than $40\Myr$ and bin their $x$-positions into $n_\text{bin}=15$ partitions with width $w_\text{bin}=L_x/n_\text{bin}\simeq0.21\kpc$. Let $M_{\text{sp},i}$ denote the total mass of the young star particles in the $i$-th bin, with integer $i=1,2,\cdots,n_\text{bin}$.
At a given time, the local SFR surface density in the $i$-th bin can then be calculated as
\begin{equation}\label{eq:lSFR40}
   \SFRf (x_i) \equiv \frac{M_{\text{sp},i}}
   {40\Myr\cdot w_\text{bin}L_y},
\end{equation}
where $x_i = -L_x/2 + (i+0.5)w_\text{bin}$.
The global SFR surface density across the simulation domain is given by $\SFRf =n_\text{bin}^{-1}\sum_{i=1}^{n_\text{bin}} \SFRf (x_i)$.

Figure \ref{fig:SFR_evol} plots temporal variations of the global SFR surface density $\SFRf$ for (a) magnetized and (b) unmagnetized models. For reference we also indicate (dashed lines) the temporal changes of $\mathcal{F}$ in models with spiral-arm forcing. In all models, $\SFRf$ rapidly increases with time initially, as the system adjusts from its initial transient state.
The value of $\SFRf$ tops out at
around $t\sim60$--$70\Myr$, independent of $\mathcal{F}$ and $\beta$, when SN feedback and radiative heating self-consistently balance turbulence dissipation and cooling.  At this time, the spiral forcing shown as the dashed lines remains weak.
The peak value of $\SFRf\sim 1.2\times 10^{-2}\SFRunits$ in the magnetized models (Figure \ref{fig:SFR_evol}a) is similar to $\SFRf$ in the (magnetized) TIGRESS run of \citet{cgkim17} which has the same surface density, external potential, etc.  The peak $\SFRf$ in magnetized models is about half that of the unmagnetized models shown in Figure \ref{fig:SFR_evol}b, and as expected the overall $\SFRf$ is higher for unmagnetized models than their magnetized counterparts (see \citealt{cgkim15b}). Subsequently, in all models $\SFRf$ exhibits a secular decay as well as large-amplitude quasi-periodic variations with period $\sim 50\Myr$. The secular decay is caused by the decrease in the gas mass in the simulation domain (see below), while the quasi-periodic fluctuations reflect self-regulation cycles of star formation: SN feedback and heating puff up the disk vertically and thus reduce the SFR; lower feedback in turn cools down the disks and promotes a new round  of  star formation and feedback \citep{cgkim17}.

Inclusion of the spiral potential enhances $\SFRf$, but the effect is relatively modest overall, and at some times $\SFRf$ can even be larger in models without a spiral potential.
Column 4 of Table \ref{t:model} lists the mean SFR surface density and its standard deviation averaged over $200\Myr\leq t\leq 600\Myr$ for all models. The mean SFR of model {\tt F20B10} is larger, but only by a factor of $1.6$, compared to model {\tt F00B10}. The SFR enhancement factor is reduced to $1.2$ in the unmagnetized models. This suggests that triggering of star formation by spiral arms is only moderate and the main effect of arms is rather to collect
star-forming regions into narrow ridges.

\begin{deluxetable}{c|c|ccc}
\tabletypesize{\footnotesize}
\tablewidth{0pt}\tablecaption{Mass fractions, Velocity Dispersions, and Scale Heights of Various Phases}
\tablehead{
\colhead{Model} &
\colhead{Phase} &
\colhead{$\log f_m$} &
\colhead{$\log \sigma_z$} &
\colhead{$\log H$} \\
\colhead{(1)} &
\colhead{(2)} &
\colhead{(3)} &
\colhead{(4)} &
\colhead{(5)}
}
\startdata
\multirow{4}{*}  {\tt F00B10}  &
total          & $0$               & $ 1.05 \pm 0.07$ & $ 2.52 \pm 0.03$ \\
\cline{2-5}
&cold-unstable & $ -0.66 \pm 0.13$ & $ 0.72 \pm 0.04$ & $ 1.86 \pm 0.05$ \\
&warm          & $ -0.12 \pm 0.04$ & $ 1.07 \pm 0.05$ & $ 2.55 \pm 0.04$ \\
&ionized-hot   & $ -2.26 \pm 0.11$ & $ 1.90 \pm 0.08$ & $ 3.17 \pm 0.02$ \\
\hline
\multirow{4}{*}  {\tt F10B10}  &
total          & $0$               & $ 1.12 \pm 0.05$ & $ 2.59 \pm 0.04$ \\
\cline{2-5}
&cold-unstable & $ -0.72 \pm 0.14$ & $ 0.84 \pm 0.03$ & $ 2.02 \pm 0.06$ \\
&warm          & $ -0.10 \pm 0.03$ & $ 1.12 \pm 0.04$ & $ 2.62 \pm 0.05$ \\
&ionized-hot   & $ -2.23 \pm 0.10$ & $ 1.93 \pm 0.08$ & $ 3.18 \pm 0.03$ \\
\hline
\multirow{4}{*}  {\tt F20B10}  &
total          & $0$               & $ 1.16 \pm 0.06$ & $ 2.63 \pm 0.04$ \\
\cline{2-5}
&cold-unstable & $ -0.78 \pm 0.14$ & $ 0.92 \pm 0.04$ & $ 2.12 \pm 0.07$ \\
&warm          & $ -0.09 \pm 0.04$ & $ 1.15 \pm 0.05$ & $ 2.65 \pm 0.04$ \\
&ionized-hot   & $ -2.15 \pm 0.12$ & $ 1.92 \pm 0.07$ & $ 3.16 \pm 0.02$ \\
\hline
\multirow{4}{*}  {\tt F00Binf}  &
total          & $0$               & $ 1.40 \pm 0.07$ & $ 2.80 \pm 0.05$ \\
\cline{2-5}
&cold-unstable & $ -0.93 \pm 0.14$ & $ 0.94 \pm 0.05$ & $ 2.02 \pm 0.08$ \\
&warm          & $ -0.07 \pm 0.02$ & $ 1.31 \pm 0.03$ & $ 2.81 \pm 0.05$ \\
&ionized-hot   & $ -1.72 \pm 0.08$ & $ 2.08 \pm 0.07$ & $ 3.19 \pm 0.02$ \\
\hline
\multirow{4}{*}  {\tt F10Binf}  &
total          & $0$               & $ 1.43 \pm 0.06$ & $ 2.82 \pm 0.08$ \\
\cline{2-5}
&cold-unstable & $ -0.97 \pm 0.14$ & $ 1.01 \pm 0.07$ & $ 2.08 \pm 0.10$ \\
&warm          & $ -0.06 \pm 0.02$ & $ 1.36 \pm 0.06$ & $ 2.83 \pm 0.09$ \\
&ionized-hot   & $ -1.70 \pm 0.12$ & $ 2.06 \pm 0.07$ & $ 3.18 \pm 0.02$ \\
\hline
\multirow{4}{*}  {\tt F20Binf}  &
total          & $0$               & $ 1.44 \pm 0.03$ & $ 2.83 \pm 0.07$ \\
\cline{2-5}
&cold-unstable & $ -1.00 \pm 0.14$ & $ 1.08 \pm 0.07$ & $ 2.16 \pm 0.09$ \\
&warm          & $ -0.06 \pm 0.02$ & $ 1.37 \pm 0.05$ & $ 2.84 \pm 0.07$ \\
&ionized-hot   & $ -1.71 \pm 0.12$ & $ 2.04 \pm 0.06$ & $ 3.16 \pm 0.03$
\enddata
\tablecomments{The mean values and standard deviations are taken over $t=200$--$600\Myr$. Column 3: logarithm of the mass fraction. Column 4: logarithm of the vertical velocity dispersion ($\rm km\,s^{-1}$). Column 5: logarithm of the vertical scale height (pc). }\label{t:Sig_H}
\end{deluxetable}

The very weak dependence of $\SFRf$ on $\mathcal{F}$ is, of course, due to the regulation of star formation by radiative heating and SN feedback: strong star formation in the arm is accompanied by correspondingly strong
feedback that temporarily reduces the SFR by injecting energy and momentum, leading to $\SFRf\propto \Sigma$ for a fixed external gravity (see below; see also \citealt{ost10,ost11}). This is more significant in the unmagnetized models where stronger feedback makes the arm gas more turbulent and thus less prone to the spiral-arm forcing.

\begin{figure*}
\epsscale{1.1} \plotone{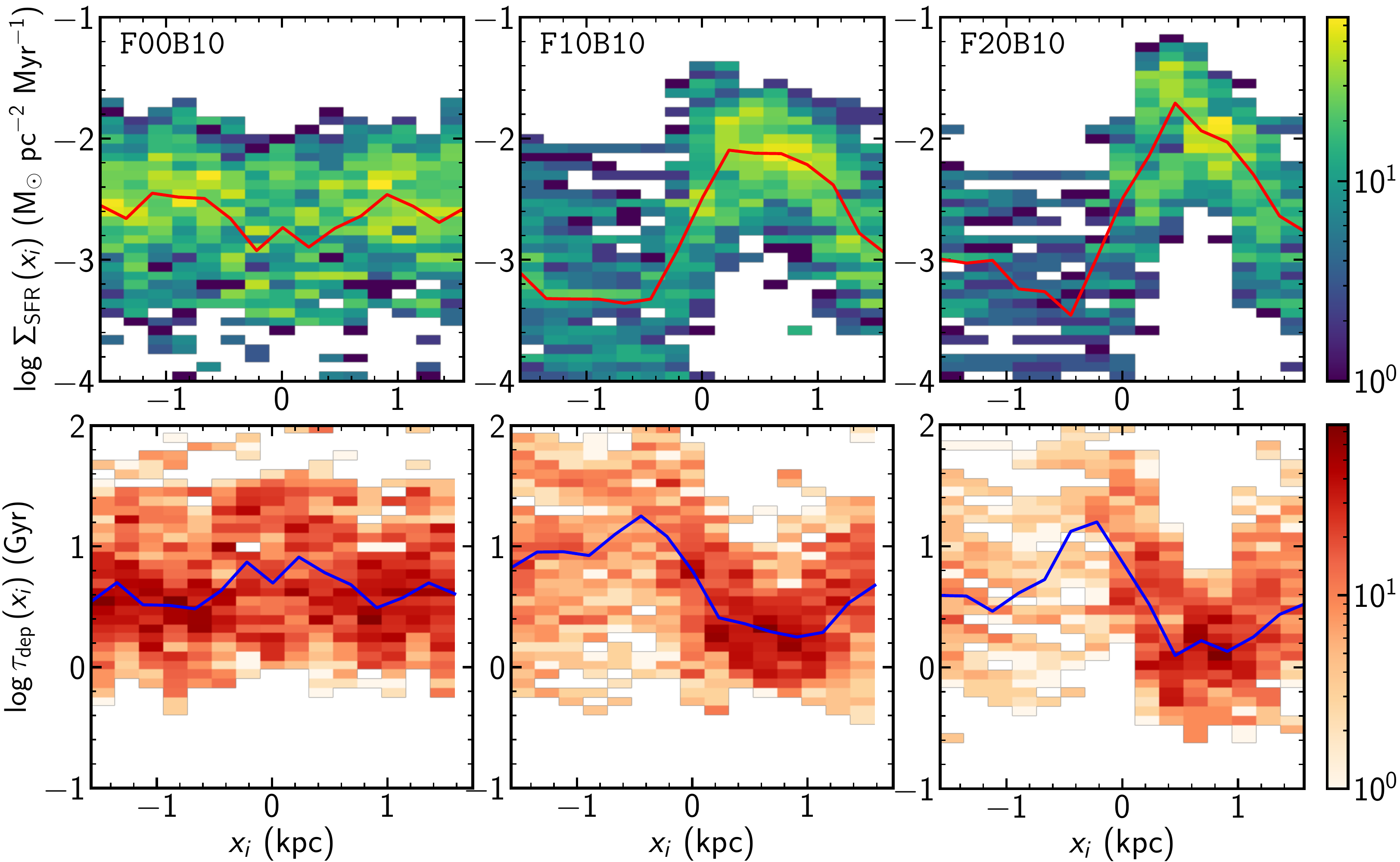}
\caption{Histograms over $t=200$--$600\Myr$  of (top) the local SFR surface density $\SFRf$ and
(bottom) the gas depletion time $\tau_\text{dep}$ for (left) model $\tt F00B10$, (middle) model $\tt F10B10$, and (right) model $\tt F20B10$. The colorbars represent frequency of occurrence. The solid lines draw the median values. Star formation rates are based on star particles younger than $40\Myr$.}\label{fig:SFR40_pos}
\end{figure*}

\begin{figure}
\epsscale{1.0} \plotone{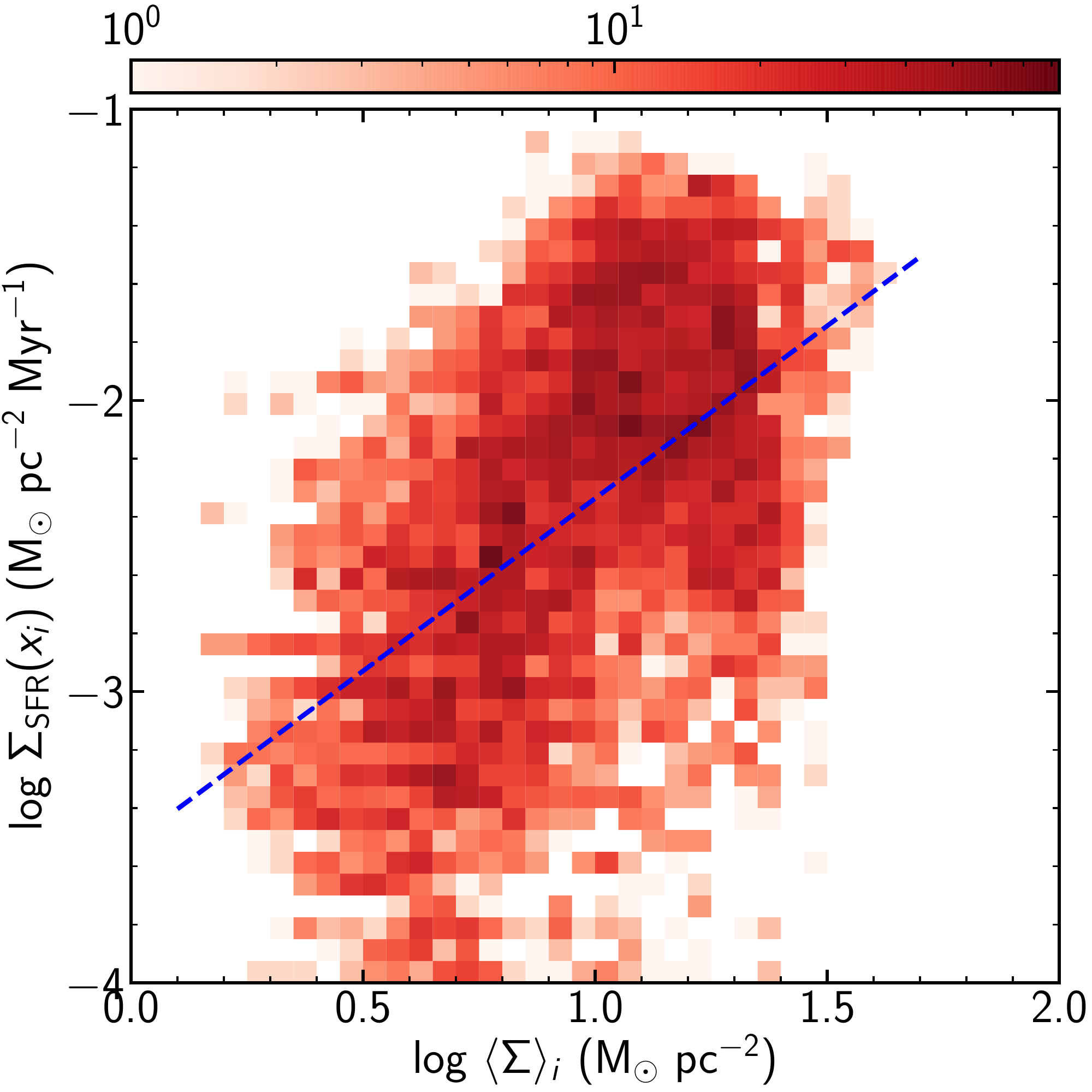}
\caption{2D histogram of the local SFR surface density $\SFRf (x_i)$ and the local surface density $\langle \Sigma \rangle_i$ over $t=200$--$600\Myr$ for all models with spiral-arm forcing ($\mathcal{F}\neq0$). The dashed line draws the best fit, Equation \eqref{eq:SFRSigma}.} \label{fig:SFR_Sigma}
\end{figure}

\begin{figure*}
\epsscale{1.0} \plotone{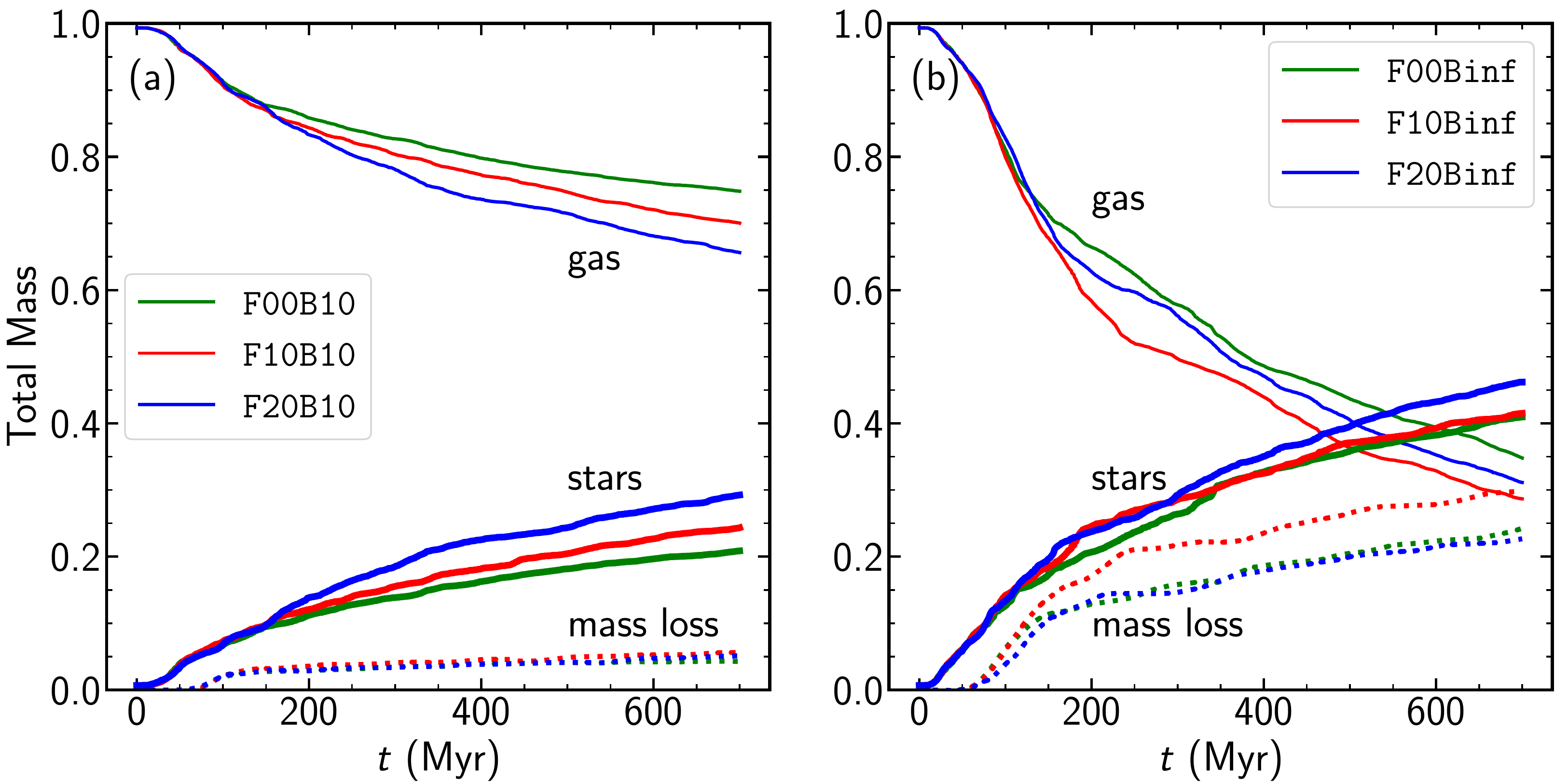}
\caption{Temporal evolution of the mass fractions of gas (thin solid), star particles (thick solid), and gas lost as outflows through the vertical boundaries (dotted) for (a) magnetized and (b) unmagnetized models.}\label{fig:masstime}
\end{figure*}

\begin{figure*}
\epsscale{1.1} \plotone{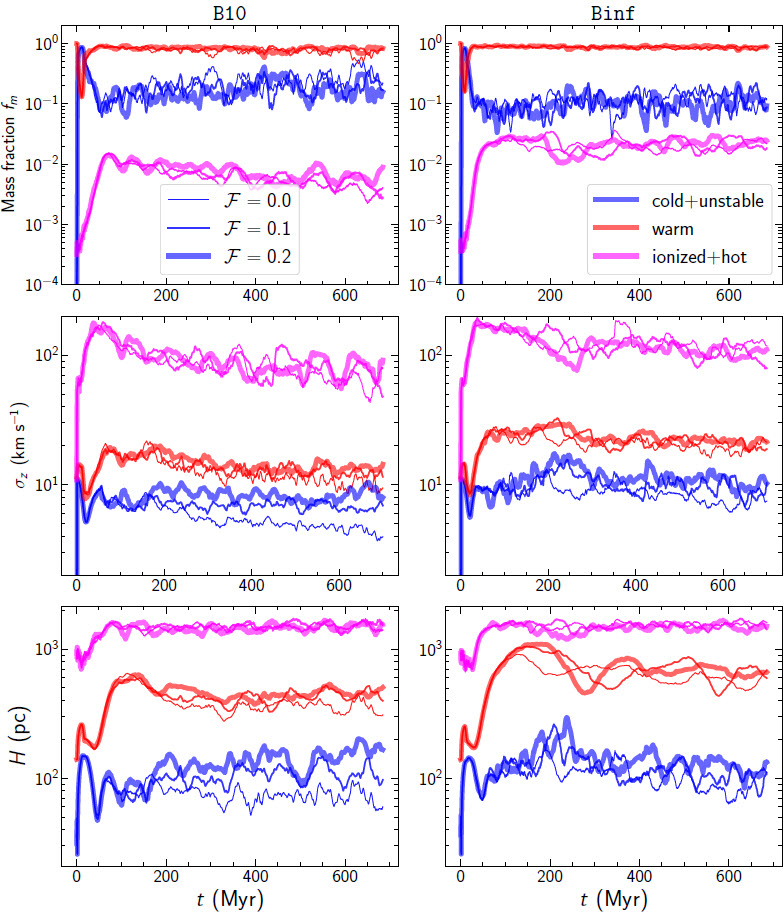}
\caption{Comparison of temporal evolution of gas properties, divided by thermal phase: (top) the mass fractions $f_m$, (middle) the density-weighted vertical velocity dispersions $\sigma_z$, and (bottom) the density-weighted scale heights $H$, for (left) magnetized and (right) unmagnetized models.    In each  panel, we separate cold plus unstable ($T<5050\K$; blue), warm ($5050\K<T<2\times 10^4\K$; red), ionized plus hot phases ($T>2\times10^4\K$; pink). Thick, intermediate, and thin lines correspond to the models with $\mathcal{F}=0.2$, 0.1, and 0, respectively.}\label{fig:sigzH_time}
\end{figure*}

Equation~\eqref{eq:lSFR40} gives $\SFRf (x_i)$ at any given time, and for each radial position bin $x_i$ we can compute a histogram of these values.
The top panels of Figure \ref{fig:SFR40_pos} plots these histograms over $t=200$--$600\Myr$ for the magnetized models. In the lower panels of Figure \ref{fig:SFR40_pos}, we show histograms of
the corresponding gas depletion times $\tau_{\text{dep}} (x_i) = {M_{\text{gas},i}}/[w_\text{bin}L_y{\SFRf (x_i)}]$ as a function of $x_i$. Here, $M_{\text{gas},i}= L_y \int_{x_i - w_\text{bin}/2}^{x_i+w_\text{bin}/2} \langle \Sigma \rangle dx$ is the gas mass in the $i$-th bin.  The solid lines draw the median values (excluding points with zero $\SFRf$ or infinite $\tau_{\text{dep}}$), while the color represents frequency of occurrence. The corresponding distributions for the unmagnetized models (not shown) are qualitatively similar, although they have, on average, twice larger $\SFRf$ and three times smaller $\tau_{\text{dep}}$ than the magnetized counterparts. In model {\tt F00B10} with no arm, $\SFRf (x) \sim 2\times 10^{-3}\SFRunits$ and $\tau_\text{dep} (x) \sim4.7\Gyr$, roughly constant in $x$. In model {\tt F20B10}, star formation is concentrated in the arm with a rate $\SFRf \gtrsim 10^{-2}\SFRunits$, which is more than an order of magnitude larger than in the interarm region where star formation is scarce. Correspondingly, the median value of the gas depletion time is $\sim 1\Gyr$ in the arm,  about an order of magnitude shorter than in the interarm region.

Comparison of Figures \ref{fig:SigVxVy} and \ref{fig:SFR40_pos} shows that the SFR is higher in regions with higher density. To describe the relationship
between the SFR surface density and the gas surface density based on profiles, we bin  $\langle \Sigma \rangle$ into $n_\text{bin}=15$ partitions to calculate the local mean density $\langle \Sigma \rangle_i = w_\text{bin}^{-1}\int_{x_i - w_\text{bin}/2}^{x_i+w_\text{bin}/2} \langle \Sigma \rangle dx$ in the $i$-th bin.
Figure \ref{fig:SFR_Sigma} plots 2D histograms of $\SFRf(x_i)$ and $\langle \Sigma \rangle_i$ for all spiral-arm models over $t=200$--$600\Myr$,  analogous to a Schmidt-Kennicutt plot \citep{sch59,ken89,ken98}.
Since the current models have only a single value of $\Sigma_0$ and since averaging along $y$ reduces large excursions in $\langle  \Sigma\rangle_i$ relative to the mean (while simultaneously increasing scatter in $\SFRf(x_i)$),
the ranges of $\SFRf$ and $\langle \Sigma \rangle$ are quite narrow in each of our simulations. Therefore, we have combined all the data from models with spiral-arm forcing.
Despite the large scatter, there is a rough
correlation between $\SFRf(x_i)$ and  $\langle \Sigma \rangle_i$. The dashed line is our best fit:
\begin{equation}\label{eq:SFRSigma}
 \SFRf(x_i) =4.61\times10^{-3} \SFRunits \left( \frac{\langle \Sigma \rangle_i }{10\,\rm M_\odot\,pc^{-2}}\right)^{1.19},
\end{equation}
for models with $\mathcal{F}\neq0$. We remark that Equation \eqref{eq:SFRSigma} is based on $y$-averaged binning approach that can encompass quite diverse conditions, resulting in a lot of scatter.

\subsection{Mass Fractions, Velocity Dispersion, and Scale Height}

Figure \ref{fig:masstime} plots temporal changes of the mass fractions in gas (thin solid) and star particles (thick solid) in the simulation domain together with the fraction of the gas lost  through the vertical boundaries (dotted), relative to the initial gas mass, for the (a) magnetized and (b) unmagnetized models. Due to the larger SFRs, the increase in the stellar mass and the lost gas mass is larger in the unmagnetized models. The effect of the spiral potential is  insignificant (tens of percent) compared to the effect of the magnetic field (factor of two). For instance, model {\tt F20Binf} has converted $46\%$ of its initial gas mass to stars at $t=700\Myr$, which is only $13\%$ larger than in model {\tt F00Binf}. In comparison, the respective stellar conversion proportions at the same time are 29\% and 21\% for models {\tt F20B10} and {\tt F00B10}.  The amount of the mass lost in the unmagnetized models is $\sim20$--$30\%$, which is decreased to $\sim 5\%$ in the magnetized models. Model {\tt F10Binf} experiences massive mass loss around $t\sim200$--$220\Myr$, which is caused by an explosive star formation event occurred at $t\sim190\Myr$.

Figure \ref{fig:sigzH_time} plots temporal evolution of key properties of the ISM gas, separated by thermal phase. We show the mass fractions $f_m$, the density-weighted vertical velocity dispersions $\sigma_z$, and the density-weighted scale heights $H$ of the gas in different phases for (left) magnetized and (right) unmagnetized models.  Here, $\sigma_z$ and $H$ of each phase are calculated as
\begin{equation}\label{eq:sigH}
\begin{split}
  \sigma_z &= \left( \frac{\int \rho v_z^2 \Theta(T) dxdydz}{\int \rho \Theta(T) dxdydz}\right)^{1/2}, \\
  H &= \left( \frac{\int \rho z^2 \Theta(T) dxdydz}{\int \rho \Theta(T) dxdydz}\right)^{1/2},
\end{split}
\end{equation}
where $\Theta(T)$ is an on/off function such that $\Theta=1$ if the gas temperature is within the range of the phase, and $\Theta=0$ otherwise.
All quantities reach roughly a quasi-steady state after $t=100\Myr$, although they all  fluctuate and a few quantities show modest secular changes associated with the decrease in the gas mass and SFR.

In all models, most of the gas mass is in the warm phase. The unmagnetized models have less cold-unstable phase and more ionized-hot phase than the magnetized model due to more active star formation. The unmagnetized models also have larger velocity dispersions and scale heights for all phases. Overall, the vertical velocity dispersion and the scale height are smallest ($\sigma_z\sim10\kms$ and $H\sim0.1\kpc$) for the cold-unstable phase and largest ($\sigma_z\sim100\kms$ and $H\sim1.5\kpc$) for the ionized-hot phase. Table \ref{t:Sig_H} lists the time-averaged values and standard deviations over $t=200$--$600\Myr$ for the mass factions, vertical velocity dispersions, and scale heights of each phase, as well as the total gas.

\subsection{Magnetic Fields}\label{sec:mag}

\begin{figure*}
\epsscale{1.0} \plotone{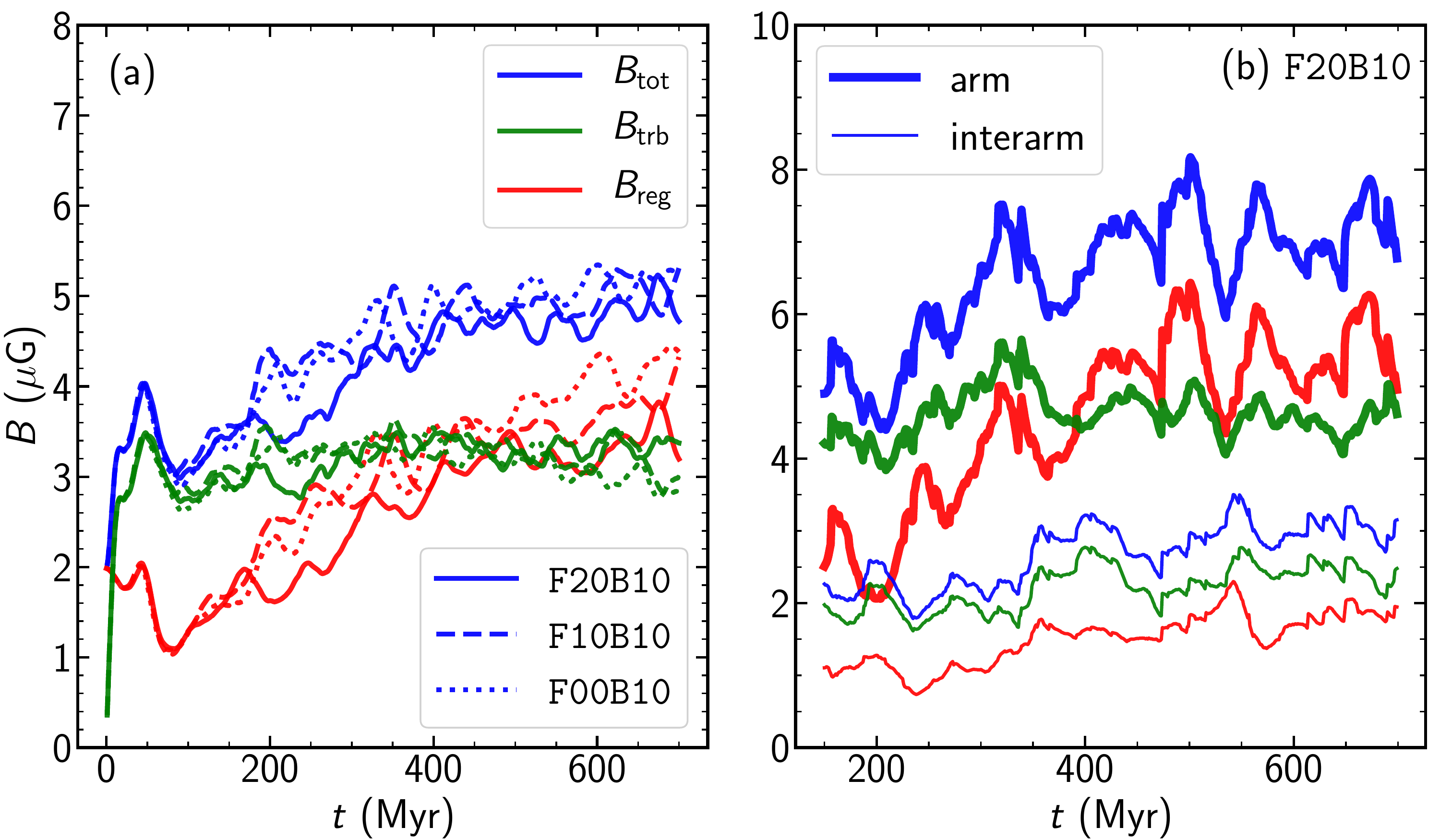}
\caption{Temporal variations of the density-weighted, total (${B}_{\rm tot}$),
turbulent (${B}_{\rm trb}$), and regular (${B}_{\rm reg}$) components of magnetic fields averaged over (a) the whole simulation domain for all magnetized models, and (b) the arm and interarm regions, separately, for model {\tt F20B10}. While the turbulent fields saturate early, the regular and total components exhibit a secular growth.
The magnetic fields are stronger in the arm, by about a factor of 2.5, than in the interarm region.}\label{fig:Bfields_time}
\end{figure*}

\begin{figure*}
\epsscale{1.0} \plotone{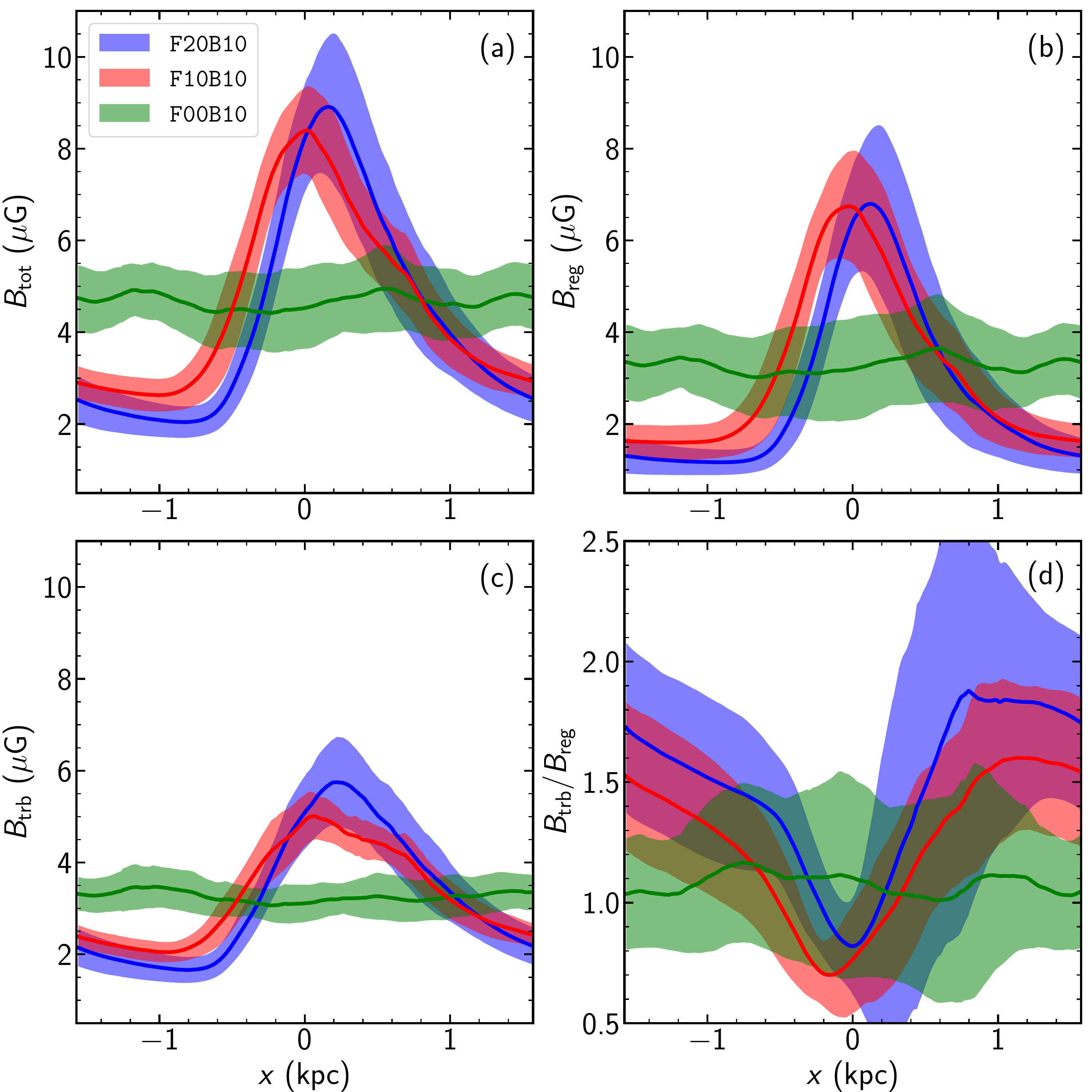}
\caption{
Profiles perpendicular to the arm
of (a) total, (b) regular, (c) turbulent components of magnetic fields, and (d) the ratio of the turbulent to regular component in the magnetized models. The solid lines and colored shades give the mean and standard deviations over $t=200$--$600\Myr$, respectively. }\label{fig:Bfield_B10}
\end{figure*}

\begin{deluxetable*}{c|c|cc c cc c cc}
\tabletypesize{\footnotesize}
\tablewidth{0pt}\tablecaption{Regular, Turbulent, and Total Magnetic Fields}
\tablehead{
Model &
Component &
\multicolumn{2}{c}{$B_\text{reg}$} &&
\multicolumn{2}{c}{$B_\text{trb}$} &&
\multicolumn{2}{c}{$B_\text{tot}$}  \\
(1) & (2) & \multicolumn{2}{c}{(3)}
&& \multicolumn{2}{c}{(4)}
&& \multicolumn{2}{c}{(5)} \\
\cline{3-4} \cline{6-7} \cline{9-10}
 &  & Arm & Interarm && Arm & Interarm && Arm & Interarm
}
\startdata
\multirow{4}{*}  {\tt F00B10}
&total & \multicolumn{2}{c}{$3.29\pm0.94$} && \multicolumn{2}{c}{$3.26\pm0.43$} && \multicolumn{2}{c}{$4.70\pm0.78$} \\
\cline{2-10}
&$x$-comp. & \multicolumn{2}{c}{$-0.58\pm0.13\;\;\;$} && \multicolumn{2}{c}{$1.77\pm0.27$} && \multicolumn{2}{c}{$1.89\pm0.27$} \\
&$y$-comp. & \multicolumn{2}{c}{$3.22\pm0.95$} && \multicolumn{2}{c}{$2.33\pm0.37$} && \multicolumn{2}{c}{$4.03\pm0.81$} \\
&$z$-comp. & \multicolumn{2}{c}{$0.00\pm0.16$} && \multicolumn{2}{c}{$1.41\pm0.23$} && \multicolumn{2}{c}{$1.43\pm0.24$} \\
\hline
\multirow{4}{*}  {\tt F10B10}
&total & $4.85\pm1.80$ & $2.01\pm0.88$ && $4.41\pm0.62$ & $2.53\pm0.53$ && $6.68\pm1.56$ & $2.53\pm0.53$ \\ \cline{2-10}
&$x$-comp. & $-0.60\pm0.15\;\;\;$ & $-0.60\pm0.14\;\;\;$ && $2.41\pm0.33$ & $1.49\pm0.26$ && $2.49\pm0.33$ & $1.49\pm0.26$ \\
&$y$-comp. & $4.79\pm1.82$ & $1.89\pm0.90$ && $2.90\pm0.51$ & $1.74\pm0.42$ && $5.69\pm1.60$ & $1.74\pm0.42$ \\
&$z$-comp. & $0.00\pm0.28$ & $0.00\pm0.13$ && $2.24\pm0.46$ & $1.04\pm0.30$ && $2.25\pm0.46$ & $1.04\pm0.30$ \\
\hline
\multirow{4}{*}  {\tt F20B10}
&total & $4.62\pm2.17$ & $1.63\pm0.89$ && $4.85\pm1.06$ & $2.27\pm0.70$ && $6.86\pm2.09$ & $2.27\pm0.70$ \\
\cline{2-10}
&$x$-comp. & $-0.57\pm0.17\;\;\;$ & $-0.57\pm0.12\;\;\;$ && $2.74\pm0.57$ & $1.44\pm0.38$ && $2.81\pm0.56$ & $1.44\pm0.38$ \\
&$y$-comp. & $4.56\pm2.21$ & $1.49\pm0.93$ && $3.08\pm0.73$ & $1.47\pm0.52$ && $5.61\pm2.07$ & $1.47\pm0.52$ \\
&$z$-comp. & $0.03\pm0.30$ & $-0.02\pm0.12$ && $2.51\pm0.65$ & $0.92\pm0.39$ && $2.53\pm0.66$ & $0.92\pm0.39$ 
\enddata
\tablecomments{The mean values and standard deviations are taken over $t=200$--$600\Myr$. Columns 3-5: density-weighted regular, turbulent, and total magnetic fields, respectively ($\mu$G). The arm and interarm regions are defined as the regions with $\langle \Sigma\rangle /\langle \Sigma\rangle_m$ larger and smaller than unity, respectively.}
\label{t:Mag}
\end{deluxetable*}

As Figures \ref{fig:stdBfield}, \ref{fig:YZdist}, and \ref{fig:spur} show, magnetic fields in our simulations possess a regular component as well as an irregular, turbulent component. Following \citet{cgkim15b}, we define the $y$-average of a physical quantity $q$ as
\begin{equation}
  \overline{q}(x,z) \equiv \int q dy /L_y,
\end{equation}
and calculate the regular field $\overline{\bf B}(x,z)$ and turbulent field $\delta {\bf B}\equiv {\bf B} - \overline{\bf B}$.
The density-weighted regular, turbulent, and total components in the $j\,(=x,y,z)$ direction are then calculated as
\begin{subequations}\label{eq:Bmean}
\begin{align}
   {B}_{\text{reg},j} (x) &= \frac{\int\overline{\rho} \overline{B}_j dz}{\int \overline{\rho} dz}, \\
   {B}_{\text{trb},j} (x) &= \frac{\int\overline{\rho} \overline{\delta{B_j}^2}^{1/2} dz} {\int \overline{\rho} dz},\\
   {B}_{\text{tot},j} (x) &= \frac{\int\overline{\rho} \overline{B_j^2}^{1/2} dz} {\int \overline{\rho} dz},
\end{align}
\end{subequations}
respectively. Note that $\overline{B_j^2}=\overline{B}_j^2+\overline{\delta{B_j}^2}$ by definition. If the magnetic fields change their directions rapidly in space, $|{\bf B}_\text{reg}|$ would underestimate the total strength significantly.

Figure \ref{fig:Bfields_time}(a) plots the temporal variations of $B_\text{tot}=|{\bf B}_\text{tot}|$, $B_\text{trb}=|{\bf B}_\text{trb}|$, and $B_\text{reg}=|{\bf B}_\text{reg}|$ averaged over the whole computational domain for all magnetized models.  SN feedback creates the turbulent component from the regular component, while background shear tends to increase the regular component at the expense of the turbulent component. Note that the turbulent component saturates at $B_\text{trb}\sim3\mu$G, while the regular (and thus total) component grows secularly with time.
The domain-averaged field strength is largely independent of the spiral-arm forcing. Figure \ref{fig:Bfields_time}(b) plots the temporal changes of the density-weighted magnetic fields in the arm region (with $\langle \Sigma \rangle$ larger than the mean value) and interarm region (with $\langle \Sigma \rangle$ smaller than the mean value) at $t=150$--$700\Myr$ for model {\tt F20B10}. Overall, the magnetic fields in the arm are about 2.5 time stronger and thus grow more strongly than in the interarm regions.

Table \ref{t:Mag} gives the mean values of the regular, turbulent, and total magnetic fields averaged over $t=200$-$600\Myr$. The spatial averages are taken over the whole domain for model {\tt F00B10} and over the arm and interarm regions, separately, for models {\tt F10B10} and {\tt F20B10}. The regular component has $|{B}_{\text{reg},y}| > |{B}_{\text{reg},x}| \gg
|{B}_{\text{reg},z}|$. Although the regular fields are dominated by the $y$-component, the turbulent fields are comparable in all directions. Note that ${B}_{\text{reg},x}<0$ and ${B}_{\text{reg},y}>0$ for all models, implying that the mean fields are trailing and inclined relative to the arm (or $y$-direction).  The inclination angle of the regular magnetic fields relative to the arm is $\theta_B\equiv -\tan^{-1}({B}_{\text{reg},x} /{B}_{\text{reg},y} )\sim 10^\circ$ in model {\tt F00B10}, while $\theta_B\sim7^\circ$ in the arm region and $\theta_B\sim20^\circ$ in the interarm regions, insensitive to the arm strength, in models {\tt F10B10} and {\tt F20B10}. The mean magnetic fields follow trailing gaseous spurs or filaments in the interarm region, while being roughly parallel to the arm in the arm region (see Figure \ref{fig:stdBfield}).

Figure \ref{fig:Bfield_B10} plots the profiles in $x$ (i.e. vs. offset from the arm)
of the strength of the total, regular, and turbulent fields, as well as the ratio $B_\text{trb}/B_\text{reg}$ for all magnetized models. The solid lines give the mean values over $t=200$--$600\Myr$, while the colored shades represent the standard deviations. In model {\tt F00B10}, $B_\text{tot}\sim 4.7\mu$G and $B_\text{reg}\sim B_\text{trb}\sim3.3\mu$G, almost constant over $x$. The spiral-arm forcing compresses magnetic fields to $B_\text{reg}\sim 6.8\mu$G at the arm density peak, insensitive to the arm strength, which is $\sim3$ times stronger than the interarm regular fields. In the spiral-arm models,  the turbulent component is stronger than the regular component almost everywhere, except near the upstream side of the arm region. This is because the turbulent magnetic fields are generated by the random gas motions driven by SN feedback that is most active in the feedback zone downstream from the arm. In model {\tt F20B10}, $B_\text{trb}/B_\text{reg}\sim1.4$ in the interarm region where the velocity dispersion is high due to low gas density.

\begin{figure*}
\epsscale{1.1} \plotone{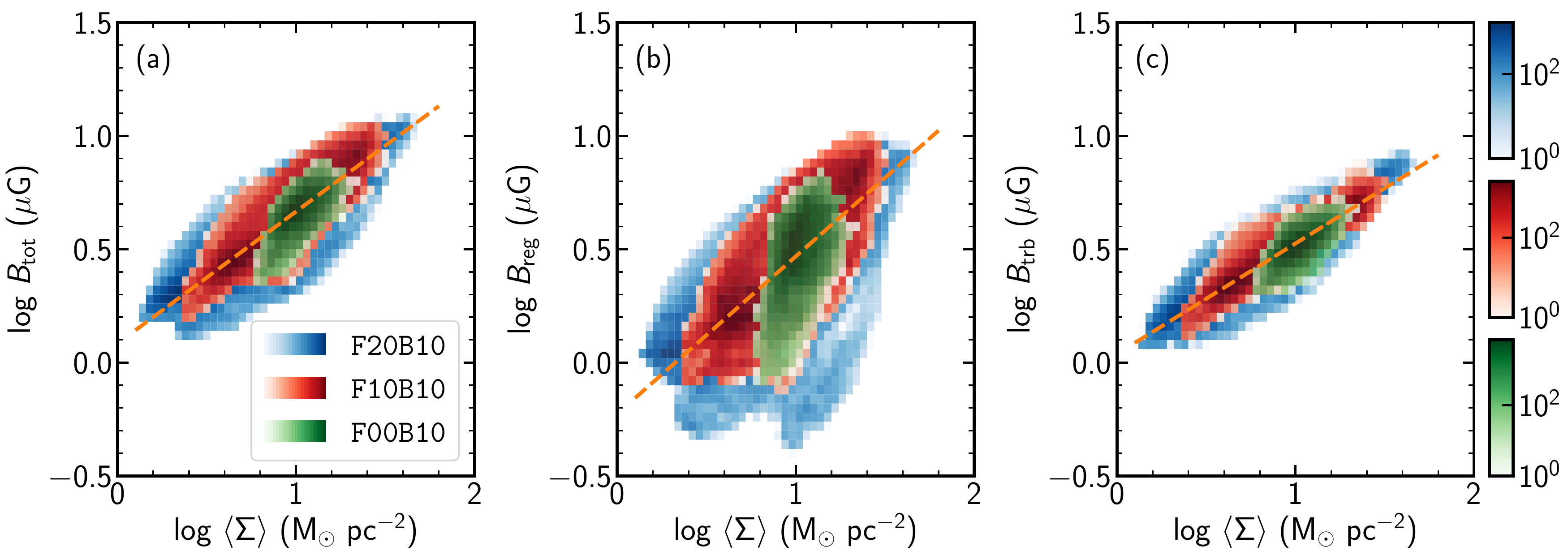}
\caption{2D histograms of the (a) total, (b) regular, and (c) turbulent components of magnetic fields and the gas surface density. Colorbars in blue, red, and green indicate the frequency of occurrence over $t=200$--$600\Myr$ for magnetized models with $\mathcal{F}=0.2$, 0.1, and 0, respectively. The dashed line in each panel draws the best fits, Equation \eqref{eq:mag}, for the combined data.}\label{fig:Mag_Sigma}
\end{figure*}

Like the SFR, the strength of magnetic fields is correlated with the gas surface density. Figure \ref{fig:Mag_Sigma} plots the relationships between the gas surface density $\langle\Sigma\rangle$ and the (a) total, (b) regular, and (c) turbulent components of magnetic fields over $t=200$--$600\Myr$ for magnetized models. Clearly, magnetic fields are stronger in regions with higher gas density. The dashed lines are our best fits
\begin{subequations}\label{eq:mag}
\begin{align}\label{eq:mag00}
 B_\text{tot}=4.63\mu\text{G} \left(\frac{\langle\Sigma\rangle}{10\rm\,M_\odot\,pc^{-1}}\right)^{0.58}, \\
 \label{eq:mag10}
 B_\text{reg}=2.94\mu \text{G} \left(\frac{\langle\Sigma\rangle}{10\rm\,M_\odot\,pc^{-1}}\right)^{0.69}, \\
 \label{eq:mag20}
 B_\text{trb}=3.35\mu \text{G} \left(\frac{\langle\Sigma\rangle}{10\rm\,M_\odot\,pc^{-1}}\right)^{0.49},
\end{align}
\end{subequations}
for the combined data of all magnetized models.

We note that even though $B_{\rm reg}$ is dominated by the $y$-component, the scaling of $B_{\rm reg}$ with $\langle \Sigma \rangle$ is sublinear, where a linear relation would apply for one-dimensional compression in $x$ of a uniform magnetic field $B_y$.  This sublinear scaling is because magnetic fields can expand vertically when the gas is compressed horizontally, and indeed the enhancement of thermal and turbulent pressure from feedback in the arm aids in this vertical expansion.

\subsection{Midplane Stress}

\begin{figure*}
\epsscale{1.0} \plotone{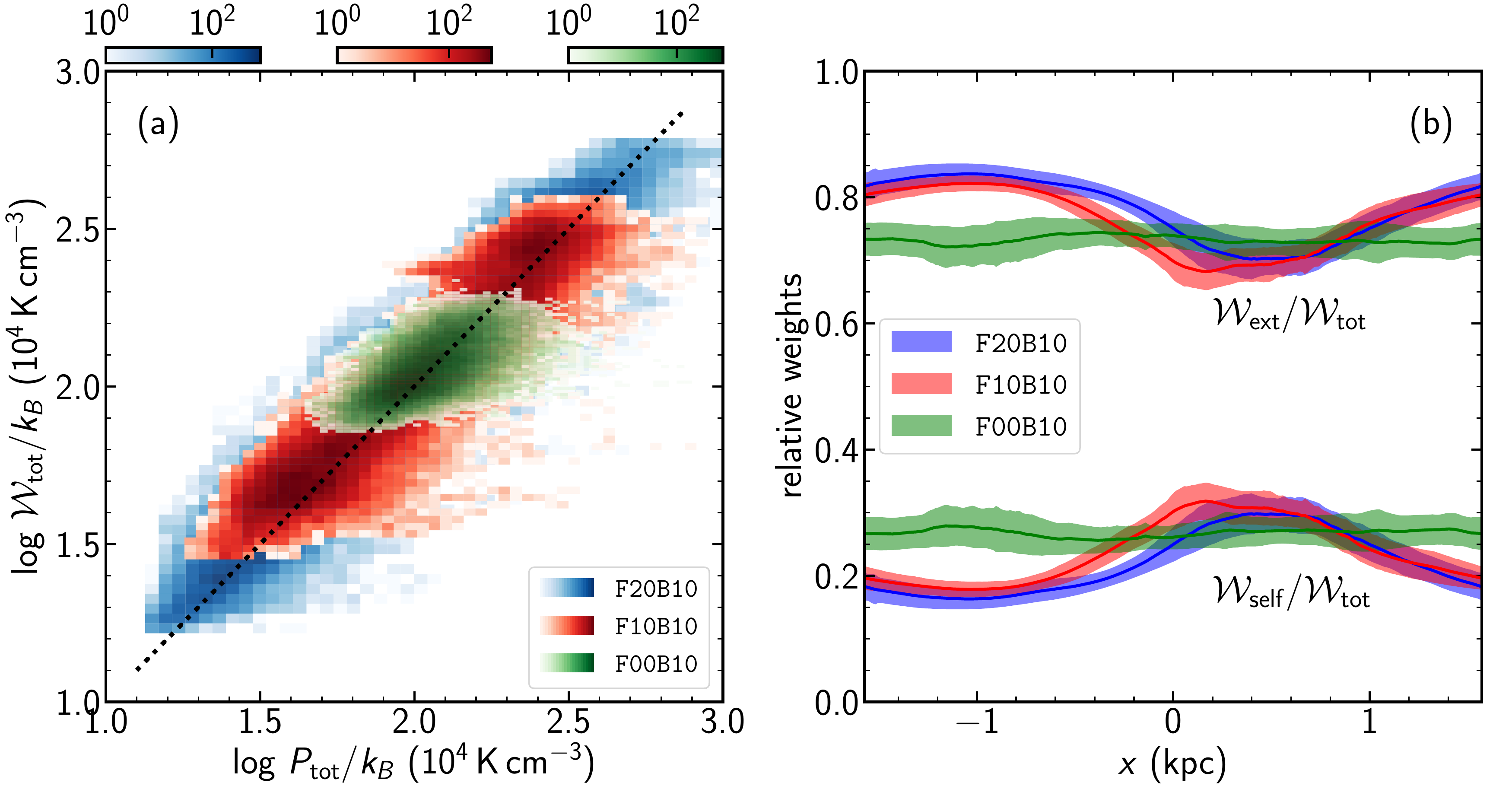}
\caption{Gas weight and midplane pressure for all magnetized models.
(a) 2D histograms of the total midplane stress $P_\text{tot}$ and the total gas weight $\mathcal{W}_\text{tot}$, and (b)  profiles of the
relative contributions to the total gas weight  from
external gravity ($\mathcal{W}_\text{ext}$) and self-gravity ($\mathcal{W}_\text{self}$).
The dashed line in (a) corresponds to $\mathcal{W}_\text{tot}/P_\text{tot}=1$, and colorbars indicate the frequency of occurrence over $t=200$--$600\Myr$. In (b), the solid lines and colored shades give the mean and standard deviations over $t=200$--$600\Myr$, respectively.} \label{fig:Weights}
\end{figure*}

\begin{figure*}
\epsscale{1.0} \plotone{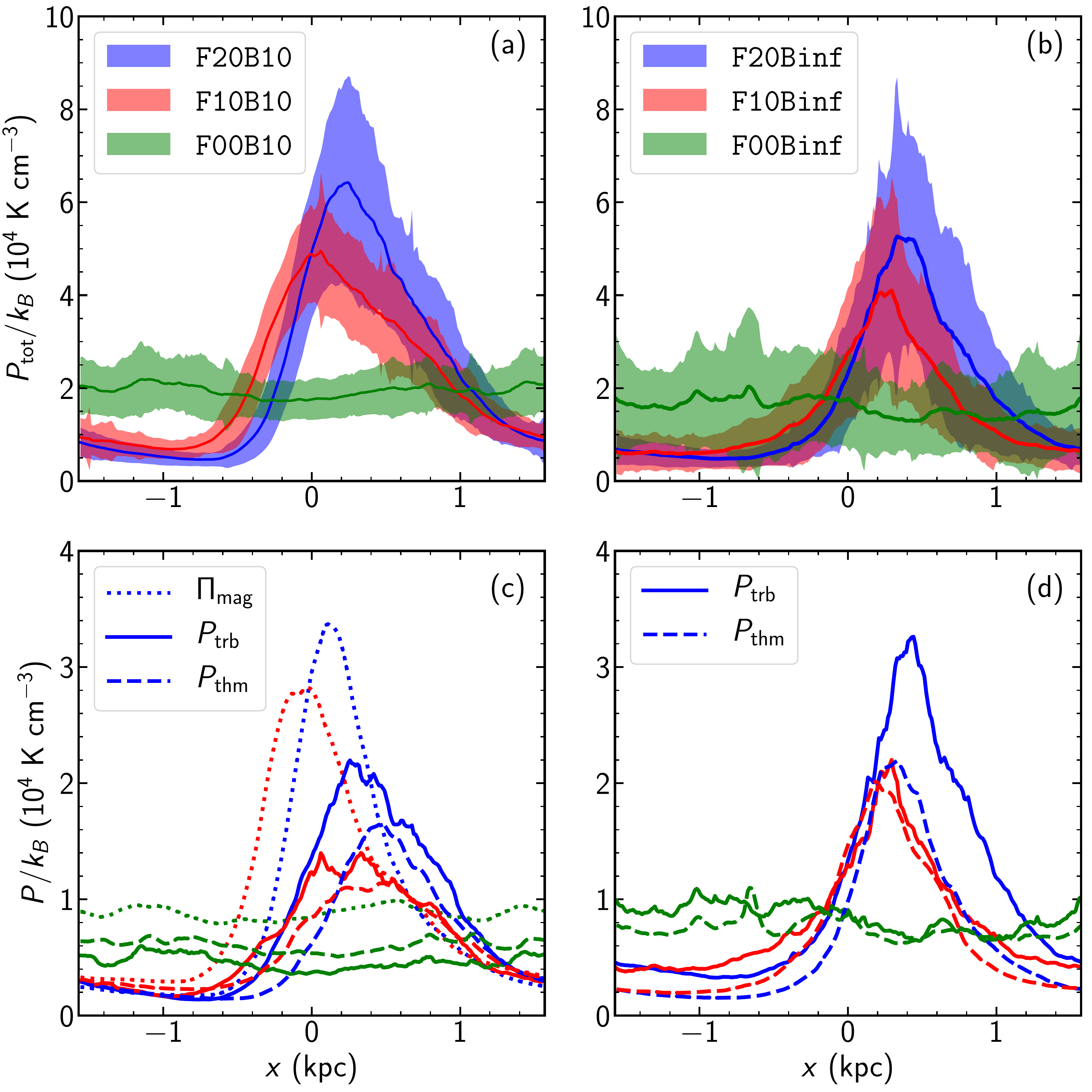}
\caption{
Profiles perpendicular to the arm
of (a,b) the total midplane stress and (c,d) various midplane stresses for the (left) magnetized and (right) unmagnetized models. Blue, red, and green correspond to models with $\mathcal{F}=0.2$, 0.2, and 0, respectively. The lines and colored shades represent the mean values and standard deviations over $t=200$--$600\Myr$. For clarity, only mean values are shown in (c) and (d).
} \label{fig:Pressures}
\end{figure*}

When a disk is in a quasi-steady state, the force balance in the vertical direction requires that the weight of the gas under the total gravitational field should be supported by the total midplane stress. In the presence of magnetic fields, not only the magnetic pressure but also the vertical tension force $\partial (B_z^2/4\pi) /\partial z$ contribute to the magnetic forces, so that the relevant magnetic stress is $B^2/(8\pi) -  B_z^2/(4\pi)$ \citep{bou90,pio07,ost10}.
We measure the thermal, turbulent, and magnetic stresses at the midplane as
\begin{subequations}\label{eq:Press}
\begin{align}\label{eq:Pressa}
P_\text{thm} &=\frac{1}{2\Delta zL_y} \iint_{z=-\Delta z}^{z=\Delta z} P dzdy, \\
P_\text{trb} &=\frac{1}{2\Delta zL_y} \iint_{z=-\Delta z}^{z=\Delta z} \rho v_z^2 dz dy, \label{eq:Pressb}\\
\Pi_\text{mag} &=\frac{1}{2\Delta zL_y} \iint_{z=-\Delta z}^{z=\Delta z} \frac{B^2-2B_z^2}{8\pi}  dz dy, \label{eq:Pressc}
\end{align}
\end{subequations}
respectively.  Here, $\Delta z=12.3\pc$ refers to the grid spacing in the $z$-direction. In $\Pi_\mathrm{mag}$, the tension term is about 17\% of the magnetic pressure term in our simulations.  The total midplane stress is $P_\text{ tot}=P_\text{thm}+P_\text{trb}+\Pi_\text{mag}$.

We also measure the total weight of the gas as $\mathcal{W}_\text{tot} = \mathcal{W}_\text{ext} + \mathcal{W}_\text{self}$, where
\begin{subequations}\label{eq:Weight}
 \begin{align}
   \mathcal{W}_\text{ext} &= \frac{1}{2}\iint_{z=-L_z/2}^{z=L_z/2} \rho \left| \frac{d\Phi_\text{ext}}{dz}\right| dzdy, \\
   \mathcal{W}_\text{self} &= \frac{1}{2}\iint_{z=-L_z/2}^{z=L_z/2} \rho \left| \frac{d\Phi_\text{self}}{dz}\right| dzdy,
 \end{align}
\end{subequations}
representing the gas weight under the external gravity or self-gravity alone, respectively.

Figure \ref{fig:Weights} plots the relationship between $P_\text{tot}$ and $\mathcal{W}_\text{tot}$ as well as the ratios $\mathcal{W}_\text{ext}/\mathcal{W}_\text{tot}$ and $\mathcal{W}_\text{self}/\mathcal{W}_\text{tot}$ as functions of $x$ for the magnetized models over $t=200$--$600\Myr$: the distributions are similar for the unmagnetized models. As expected, $P_\text{tot}\approx \mathcal{W}_\text{tot}$ within 12\%, demonstrating that the disks are overall in dynamical equilibrium in the vertical direction.  On average,  $\mathcal{W}_\text{self}\approx 0.24\mathcal{W}_\text{tot}$ for the magnetized models and $\mathcal{W}_\text{self}\approx 0.20\mathcal{W}_\text{tot}$ for the unmagnetized models since the latter have higher $\SFRf$ and thus lower gas mass.  Inside the arm, $\mathcal{W}_\text{self}/\mathcal{W}_\text{tot}$ increases up to 0.30 and 0.27 in models {\tt F20B10} and {\tt F20Binf}, respectively. This demonstrates that the gas weight in our models is dominated by the external gravity rather than self-gravity even inside the spiral arms.

Figure \ref{fig:Pressures} plots the profiles in $x$ of the mean values and standard deviations of (upper panels) the total midplane stress  $P_\text{tot}$ as well as (lower panels) the thermal, turbulent, and magnetic stresses, over $t=200$--$600\Myr$. The left and right panels correspond to the magnetized and unmagnetized models, respectively.  Columns 5--8 in Table \ref{t:model} give the mean values and standard deviations of the various midplane stresses averaged spatially in the $x$-direction and temporally over $t=200$--$600\Myr$.

The magnetized models have a higher total midplane stress, by a factor of $\sim 1.3$ on average, than the unmagnetized counterpart. This is because they contain more gas (Figure \ref{fig:masstime}) with a larger scale height (Table \ref{t:Sig_H}), and thus have higher gas weight than the unmagnetized counterparts. In the magnetized models, the magnetic stress closely follows the $\langle \Sigma\rangle$ distribution shown in Figure \ref{fig:SigVxVy}(a) and has  $\Pi_\text{mag}\sim(0.39-0.45)P_\text{tot}$ on average, while the thermal and turbulent pressures are larger in the region with stronger star formation (and SN feedback; see Figure \ref{fig:SF_XPDF}). The thermal and turbulent pressures are individually higher in the unmagnetized models, by a factor of $\sim1.1$ and 1.5, respectively, than in the magnetized models, although the addition of the magnetic stress makes the total larger. All the stresses are almost flat in the $x$-direction in the no-arm models, and higher in the arm region owing to stronger self-gravity.

\begin{figure}
\epsscale{1.0} \plotone{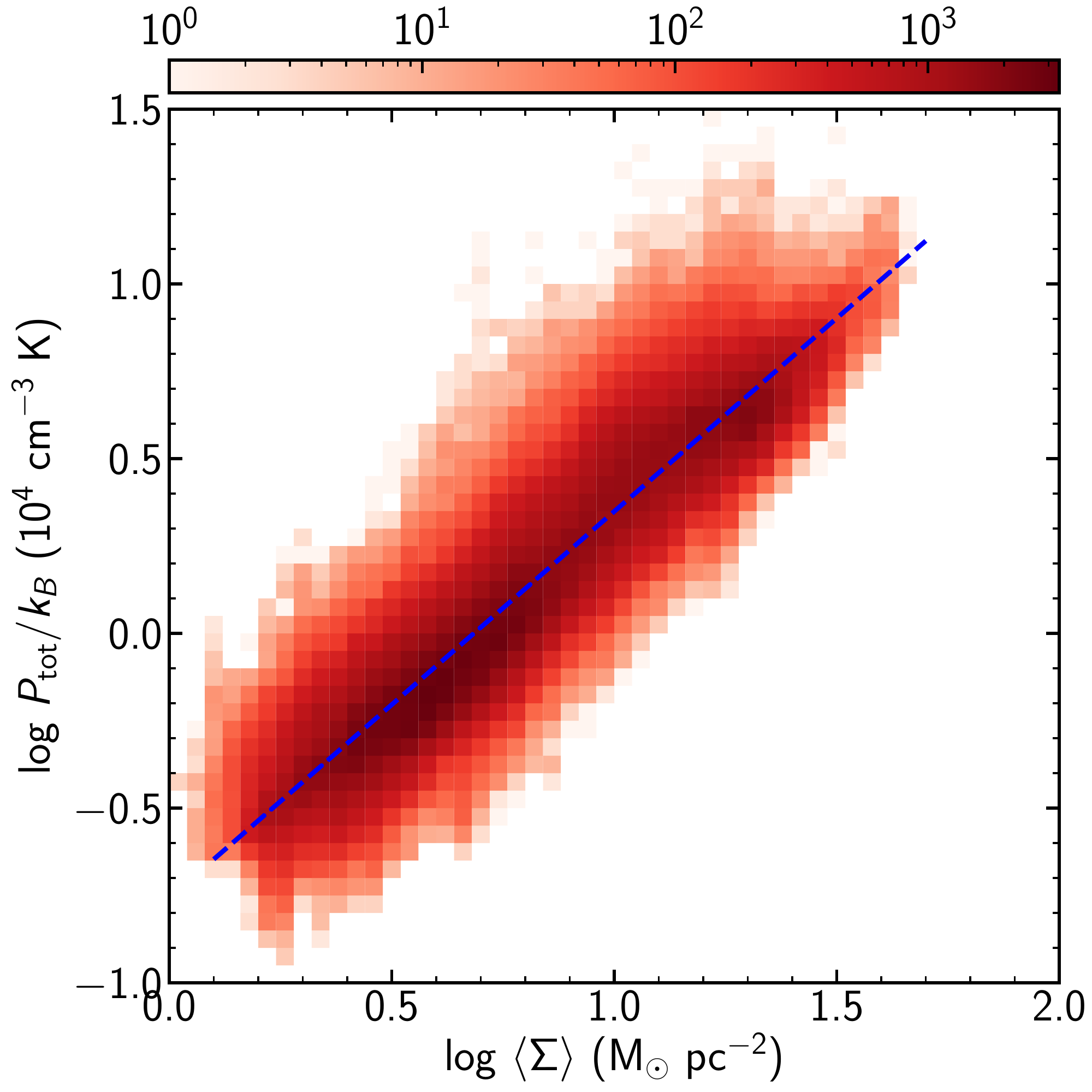}
\caption{2D histogram of the total midplane stress $P_\text{tot}$ and the gas surface density $\langle \Sigma\rangle$ averaged along the $y$-direction for all models with spiral forcing ($\mathcal{F}\neq0$). The colorbar indicates the frequency of occurrence over $t=200$--$600\Myr$. The dashed line draws the best fit, Equation \eqref{eq:Ptot}.}\label{fig:Ptot_Sigma}
\end{figure}

Figure \ref{fig:Ptot_Sigma} plots a 2D histogram of the total midplane stress $P_\text{tot}$ and the gas surface density $\langle \Sigma\rangle$ for all models with spiral forcing. The colorbar represents the frequency of occurrence over $t=200$--$600\Myr$. The dashed line is the best fit
\begin{equation}\label{eq:Ptot}
  P_\text{tot} =
 2.23\times 10^4\, k_B\rm\,cm^{-3}\,K \left(\frac{\langle\Sigma\rangle}{\rm 10\, M_\odot\,pc^{-2}}\right)^{1.11}.
\end{equation}
In spite of the large variations of both surface density and pressure as the gas flows from interarm to arm and back (Figures \ref{fig:SigVxVy} and \ref{fig:Pressures}), there is a  nearly-linear relationship between $P_\text{tot}$ and $\langle \Sigma\rangle$.  The basic reason for this is that the vertical ISM weight $\mathcal{W}_\text{tot}$ is dominated by $\mathcal{W}_\text{ext} \propto \langle\Sigma\rangle$, and since dynamical equilibrium is satisfied,  the total stress (which must balance $\mathcal{W}_\text{tot}$) must vary as $P_\text{tot} \propto \langle \Sigma\rangle$.  We note that the proportionality coefficient in Equation \eqref{eq:Ptot} is set by the vertical gravitational field of the background stars (see Equation \eqref{eq:Weight}), which varies only tens of percent between arm and interarm.  The coefficient would increase (decrease) in a galactic region where the vertical gravity from the stellar disk is higher (lower).

It is interesting to explore the relationship between the SFR surface density and the midplane stresses. For this purpose, we first calculate the average values $P_c (x_i)=w_\text{bin}^{-1} \int_{x_i-w_\text{bin}/2}^{x_i+w_\text{bin}/2} P_c dx$ in the 15 bins, as we did for the SFR surface density in Section \ref{sec:SF}. Here, the subscipt ``$c$'' denotes ``tot'', ``thm'', ``trb'', and ``mag'' for the respective midplane stress. We then measure the feedback ``yields''
 \begin{equation}
     \eta_c \equiv \frac{P_{c}(x_i)}{\Sigma_{\text{SFR}(x_i)}};
 \end{equation}
these yields $\eta$ have velocity units, but can be converted to the scaled values adopted in \citet{cgkim13} and \citet{cgkim15b} by  multiplying by $4.8\times 10^{-3}$. When the numerical data for models with  $\mathcal{F}=0.1$ and 0.2 are combined, and defining
$\Sigma_{\text{SFR},-3}=\SFRf (x_i)/(10^{-3} {\rm\,M_\odot\,pc^{-2}\,Myr^{-1}})$,
our magnetized models yield
$\eta_\text{tot}=1880\kms\Sigma_{\text{SFR},-3}^{-0.17}$,
$\eta_\text{thm}=693\kms\Sigma_{\text{SFR},-3}^{-0.43}$,
$\eta_\text{trb}=527\kms\Sigma_{\text{SFR},-3}^{-0.10}$,
$\eta_\text{mag}=187\kms\Sigma_{\text{SFR},-3}^{1.00}$, while the unmagnetized models give
$\eta_\text{tot}=1380\kms\Sigma_{\text{SFR},-3}^{-0.36}$,
$\eta_\text{thm}=558\kms\Sigma_{\text{SFR},-3}^{-0.36}$,
$\eta_\text{trb}=531\kms\Sigma_{\text{SFR},-3}^{-0.14}$.
For a reference value of $\Sigma_{\text{SFR},-3}=3.5$ (see Table \ref{t:model}), these correspond to
$\eta_\text{tot}\approx 1519\kms$,
$\eta_\text{thm}\approx 404\kms$,
$\eta_\text{trb}\approx 464\kms$, and
$\eta_\text{mag}\approx 654\kms$
for the magnetized models, showing that the magnetic yield is about 45\% of the total.

Figure \ref{fig:SFR_Ptot} plots the 2D histogram of $\SFRf(x_i)$ and $P_\text{tot} (x_i)$ for all models with spiral forcing. The colorbar represents the frequency of occurrence over $t=200$--$600\Myr$. Notwithstanding the large scatter caused by the $y$-averaging, $P_\text{tot} (x_i)$ has a good correlation with $\SFRf (x_i)$.
The dashed line draws our best fit,
\begin{equation}\label{eq:SFR}
 \SFRf  =1.64\times 10^{-3} {\rm\,M_\odot\,pc^{-2}\,Myr^{-1}} \left(\frac{P_\text{tot}/k_B}{10^4\rm\,cm^{-3}\,K}\right)^{1.24}.
\end{equation}
This is very close to Equation (26) from \citet{cgkim13}, plotted as a solid line, which was obtained from unmagnetized models with no spiral forcing, and a two-phase ISM model. This suggests that star formation in our spiral-arm simulations is self-regulated, similarly to \citet{cgkim13}, in such a way that
it keeps the disk in thermal, turbulent, and dynamical equilibrium  \citep[see also][]{ost10,cgkim11}.
With $P_\mathrm{tot}$ nearly linearly dependent on $\langle \Sigma \rangle$ from Equation \eqref{eq:Ptot}, Equation \eqref{eq:SFR} is consistent with the relationship between $\SFRf$ and $\Sigma$ indicated by Equation \eqref{eq:SFRSigma}.

\section{Summary and Discussion}\label{sec:sumdis}

\subsection{Summary}\label{sec:sum}

Spiral arms greatly affect gas flows,  magnetic fields, and star formation in disk galaxies. Two unsolved problems for spiral galaxies  are (1) whether the spiral arms enhance the SFR in the disks or not and (2) how gaseous spurs/feathers perpendicular to arms form and evolve. To address these issues, in this paper we have extended the TIGRESS simulations of \citet{cgkim17} to include the effect of a stellar spiral-arm potential. TIGRESS is a numerical framework that accurately handles the formation and evolution of star clusters (represented as star particles in the simulation) and the heating plus SN explosions they create. The TIGRESS framework allows us to model the turbulent, magnetized, multiphase (cold-warm-hot), differentially-rotating, self-gravitating, vertically-stratified ISM  self-consistently with star formation and key feedback effects produced by massive stars.  We adopt the local spiral-arm coordinates of \citet{rob69} where the two orthogonal axes in the galactic plane correspond to the directions perpendicular ($\xhat$) and parallel $(\yhat)$ to a local segment of a spiral arm, respectively, while the third coordinate ($\zhat$) is perpendicular to the galactic plane. We derive the equations of motions for star particles in the spiral-arm coordinates (Appendix \ref{sec:eom}), and use them in the simulations.

Our simulation domain is a rectangular box corotating with the arm. The box  size $L_x$ in the $x$-direction is set equal to the arm-to-arm distance.  We represent the spiral arm using a fixed gravitational potential of a simple sinusoidal shape (Equation \eqref{eq:extarm}), with its minimum occurring at the middle of the box ($x=0$).
All models have initial gas surface density $\Sigma_0=13\Surf$, similar to that in the Solar neighborhood and other ``mid-disk'' galactic environments.
For magnetized models, the disks are initially threaded by
magnetic fields parallel to the spiral arm. We consider 6 models that differ in the arm strength and magnetic field strength which are characterized by the dimensionless parameters $\mathcal{F}$ and $\beta$, respectively (Equations \eqref{eq:Fparam} and \eqref{eq:beta}). In order to avoid transients caused by a sudden introduction of the arm potential, we increase its amplitude slowly to the full strength achieved at $t=200\Myr$.  All models are run up to $t=700\Myr$, long enough for the disk to reach a quasi-steady state.

The main results can be summarized as follows.

\begin{enumerate}

\item{\textit{Overall Evolution} -- }
As the amplitude of the spiral potential grows with time,
a dense ridge forms in the region slightly downstream from the potential minimum.  Gathering of individual filaments of sheared gas and shells created by expanding superbubbles contributes to the formation of this ridge.  Compared to the interarm region, the gas in the arm is denser and colder, has stronger magnetic fields and midplane pressure, and is the locus of
most star formation in the simulation domain. Star particles formed in the arm exhibit a spatial gradient with respect to age amounting to $dx_m/dt_m\sim 9\pm4 \pc\Myr^{-1}$ on average, with $x_m$ and $t_m$ denoting the mass-weighted mean position and age of star particles, respectively. Arm star formation is usually clustered, with SNe injecting thermal and kinetic energy in the downstream feedback zones to create superbubbles. These interact and are advected by the background flow to produce turbulence throughout the simulation domain, with the larger superbubbles breaking out of the disk to vent hot gas through chimneys. The system rapidly reaches a quasi-steady state in which turbulence driven by SN feedback balances dissipation, and the  heating (radiation plus SN energy input) balances cooling. Without gas inflows from outside the domain, continued star formation and mass loss through the vertical boundaries make the gas mass and SFR decrease with time in our simulations.

\begin{figure}
\epsscale{1.0} \plotone{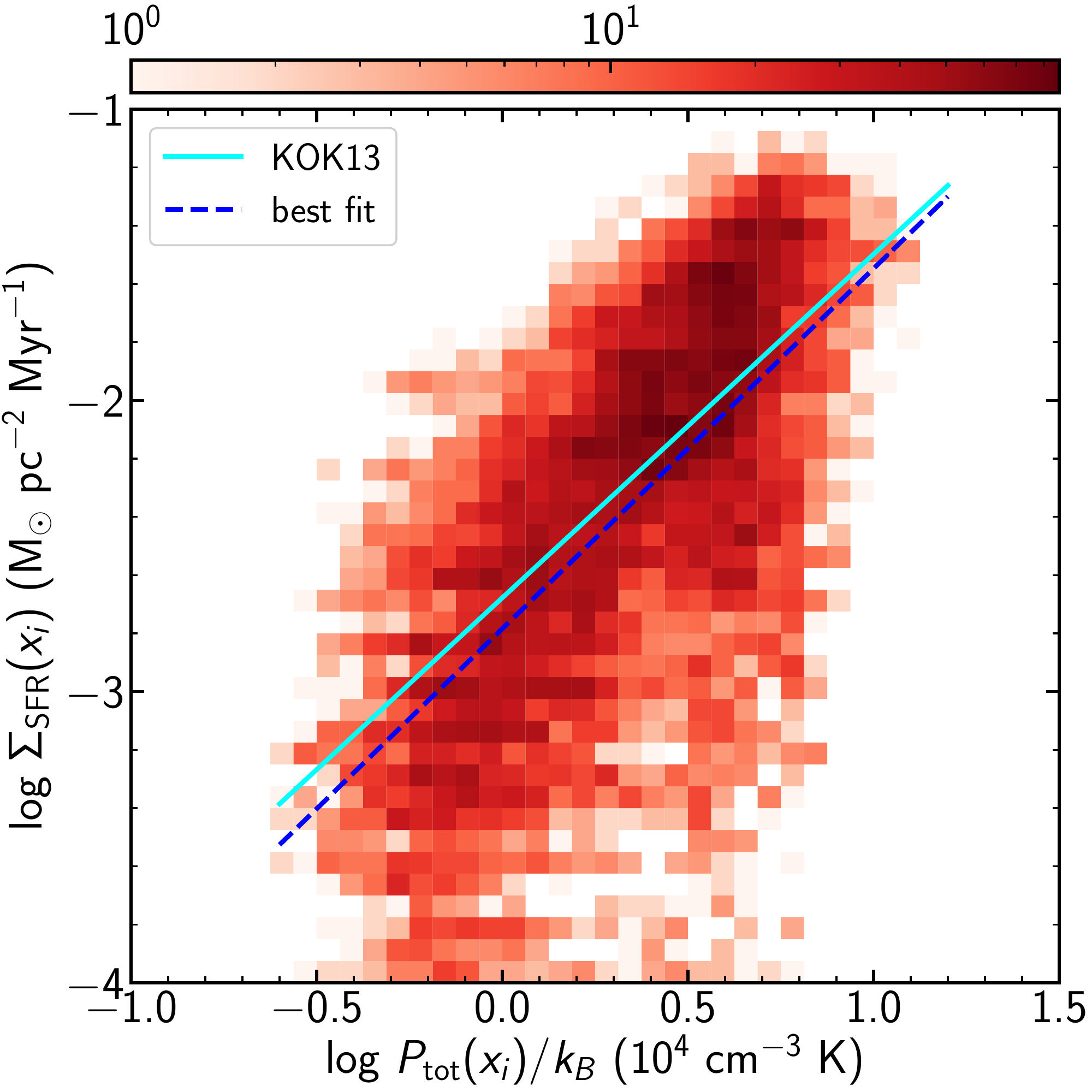}
\caption{2D histogram of the local SFR surface density $\SFRf (x_i)$ and the local midplane stress $P_\text{tot} (x_i)$ for all models with spiral-arm forcing ($\mathcal{F}\neq0$). The colorbar indicates the frequency of occurrence over $t=200$--$600\Myr$. The solid line draws Equation (26) of \citet{cgkim13}, while the dashed line is our best fit, Equation \eqref{eq:SFR}.}\label{fig:SFR_Ptot}
\end{figure}

\item{\textit{Spurs/Feathers} -- }
Clustered and correlated star formation in the arm produces expanding superbubbles that are distorted as they move downstream, with their boundaries turning to dense filaments attached to the spiral arm. While weak filaments are readily destroyed by nearby feedback, strong filaments survive and undergo collisions with neighboring ones, forming even denser, large-scale structures.
These dense structures, bounded on both sides by regions where feedback has cleared out gas (see Figures~\ref{fig:stdSurf} and \ref{fig:spur}), resemble gaseous spurs/feathers seen in real disk galaxies with strong spirals. These feedback-induced spurs form regardless of the presence of magnetic fields. Spurs in our models protrude nearly perpendicularly from the arm and turn to a trailing configuration in the interarm region.
In magnetized models, the magnetic field is generally parallel to the ``long axis'' of the spur.
They are transient and last for $\sim30\Myr$ before being destroyed by nearby feedback and turbulence.  Their mean spacing along the arm is $\sim2$--$3\kpc$, tending to be larger in unmagnetized models with a weaker spiral-arm forcing.

\item{\textit{Star Formation} --}
In our models, more than 90\% of star formation is concentrated in the arm, defined as the region where the gas surface density is above the mean value.  The vertical distribution of star-forming positions are well described by an exponential function, with a scale height proportional to the SFR. The local SFR surface density $\SFRf$ averaged along the $y$-direction is proportional roughly linearly to the local gas surface density $\Sigma$ averaged along the $y$-direction (Equation \eqref{eq:SFRSigma}) and to the total midplane stress $P_\text{tot}$ (Equation \eqref{eq:SFR}), where the latter is also equal to the ISM vertical weight $\mathcal{W}_\text{tot}$. Compared to the no-arm counterpart, the global SFR surface density averaged over the entire domain is only modestly enhanced by the spiral-arm forcing, by respective factors of 1.6 and 1.2 in models {\tt F20B10} and {\tt F20Binf} (which have $\mathcal{F}=0.2$).  This suggests that spiral arms do not trigger star formation much, but rather concentrate star-forming regions into narrow ridges. The local $\SFRf$ in the arm is about an order of magnitude or more higher than in the interarm region. The weak dependence of $\SFRf$ on $\mathcal{F}$ is consistent with the quasi-linear dependence of local $\SFRf$  on the local $\Sigma$.  The near-linear relationship we find between $\SFRf$ and $P_\text{tot}$ is quantitatively in agreement with results from our previous simulations (Figure \ref{fig:SFR_Ptot}).

\item{\textit{Magnetic Fields and Midplane Stress} -- }
Turbulence driven by SN feedback stretches and twists magnetic fields to generate an irregular, turbulent component. Due to background shear, the regular magnetic fields are trailing, with a mean inclination angle of $\sim7^\circ$ in the arm and $\sim20^\circ$ in the interarm regions, relative to the spiral arm. Both regular and turbulent components of the magnetic field are strong inside the arm and weak in the interarm region, with their strength sublinearly proportional to the gas surface density (Equation \eqref{eq:mag}). In models with spiral-arm forcing, the turbulent magnetic  component is stronger than the regular components especially in the interarm region, up to a factor of 1.5.  Our disks are overall in local dynamical equilibrium in which the total midplane stress $P_\text{tot}$ (with  thermal, turbulent, and magnetic contributions) balances the gas weight $\cal{W}_\mathrm{tot}$ along the vertical direction (Figure \ref{fig:Weights}). The contribution of self-gravity to the total gas weight $\mathcal{W}_\text{self}/\mathcal{W}_\text{tot}$ is $20$--$24\%$ on average and $27$--$30\%$ in the spiral-arm density peaks. Because the gas weight is dominated by ${\mathcal{W}}_\text{ext} \propto\Sigma $, the midplane stress $P_\text{tot}$ is approximately linearly proportional to $\langle \Sigma\rangle$ (Equation \eqref{eq:Ptot}).   The magnetic tension term in the vertical force balance is about $\sim17\%$ of the magnetic pressure term in the midplane, while the magnetic stress amounts to about $\sim40\%$ of $P_\text{tot}$.

\end{enumerate}

\subsection{Discussion}\label{sec:dis}

In our models, spiral-arm spurs form due to SN feedback in the interarm regions immediately downstream from arms, as a result of clustered and correlated arm star formation. This mechanism is different from the MJI of quasi-steady spiral arms proposed for spur formation by \citet{kim02,kim06}. In the MJI, spurs are nonlinear waves that grow via self-gravity. They collect material along the direction parallel to the arm, which is also parallel to the mean magnetic field. Thus, MJI-induced spurs are predicted to have magnetic fields perpendicular to their ``long axis'' and to be associated with converging velocity fields with amplitudes of order $\sim10\kms$ (e.g., \citealt{kim02}). This is in stark contrast to feedback-induced spurs, which would have magnetic fields parallel to the length of the spur, and could have larger converging velocities up to $\sim50$--$100\kms$. In addition, spurs resulting from the MJI endure for a long time ($\sim2\pi/\Omega_0 = 200\Myr$)  until they undergo gravitational fragmentation into bound clumps, while feedback-induced spurs are transient, being readily destroyed by subsequent feedback events. Nonlinear perturbations provided
by feedback wipe out linear modes of MJI that might otherwise have grown in our simulations. Observations of small-scale magnetic and velocity fields around spurs/feathers may provide tell-tale signs regarding whether they are produced by MJI or SN feedback.

In our simulations, we find no evidence for the growth of WI in arms downstream from spiral shocks, a mechanism that was proposed for inducing spur/feather formation by \citet{wad04}. As shown by \citet{kim14}, the WI develops as PV generated from deformed spiral shocks accumulates successively as orbiting gas passes through shocks periodically. This requires that the arm and shock fronts have a steady pattern speed and that the PV remains coherent before and after the shock fronts. In the present simulations, however, shock fronts oscillate around the mean position in the $x$-direction, making the density profile broader than the isothermal counterpart (see Figure \ref{fig:SigVxVy}). More importantly, strong turbulence driven by SN feedback rapidly mixes neighboring vortices with different polarity produced at a deformed shock front. This mixing is hostile to the growth of WI. The presence of magnetic fields and vertical shear also help to suppress the WI in our models (see also, \citealt{kim06,kim15,sor17}).

The number of spurs in our magnetized models is typically 2 or 3, corresponding to a mean spacing of $\lambda \sim2$--$3\kpc$. There are many factors including arm pitch angle and strength, angular velocity, and gas surface density that may affect the spur spacing significantly. In particular, the spacing of feedback-induced spurs should be affected by characteristic gravitational scales, such that the spacing of correlated star formation varies inversely with $\Sigma$ or $\SFRf$.
This is supported by recent observational results. By analyzing CO data for M51, for instance, \citet{sch17} identified nine molecular spurs with $\lambda \sim 0.33\kpc$ in the northern spiral arm located $\sim2.3\kpc$ away from the galaxy center.  This small value of $\lambda$ in M51 is likely due to high gas surface density $\Sigma\sim 160\Surf$ and high SFR surface density $\SFRf\sim 0.15\SFRunits$ \citep{ler17}, which are about an order of magnitude larger than the values in model {\tt F20B10}. On the other hand, Figure 1 of \citet{kre18} shows that the spiral arms in M74 have prominent molecular spurs with $\lambda\sim1$--$2\kpc$ in the regions at $R\sim5$--$6\kpc$ from the galaxy center, where both $\Sigma$ and $\SFRf$ are somewhat smaller than in the arms of M51.

That the total SFR in our simulations is insensitive to the arm strength is consistent with the observational results that the SFR is not much different in flocculent and grand-design spiral galaxies \citep{elm86} and that arm and interarm regions have similar gas depletion time \citep{foy10,ede13}.  This is also entirely consistent with the recent work of \citet{tre20}, who ran three-dimensional simulations of an interacting galaxy similar to the M51 system, and found that tidal interactions redistribute gas across the disk, and concentrate star formation within the tidally-induced  arms.
However, the total SFR is affected by less than a factor two (see their Figure 15), as we have also found.
Theoretically, the insensitivity of the total SFR to the arm strength is related to the fact that the SFR surface density is quasi-linearly proportional to the gas surface density. The quasi-linear relationship between $\SFRf$ and $\Sigma$ indicates a roughly constant gas depletion time, consistent with recent resolved observations of external galaxies (e.g., \citealt{ler17, kre18}; see also \citealt{won02,big08,sch11}, but note that star formation is as strongly correlated with stellar content as gas content -- e.g. \citealt{ler08,bol17}). This in turn implies that the arm star formation in our models corresponds to the regime where the external gravity from stars is more important than gaseous self-gravity in establishing the vertical force balance \citep{ost10,cgkim11,cgkim13}.  In gas-rich galaxies at high redshift or starbursts in the local Universe, there may be regions where self-gravity dominates the weight, such that
the equilibrium model for regulation of star formation yields $\SFRf\propto \Sigma^2$ \citep{ost11,she12}. Any physical process (including a spiral forcing, tidal interactions, or mergers) that concentrates gas enough to become self-gravitating is expected to enhance the total SFR significantly.

The fairly well-defined age gradient of young star clusters seen in our simulations may be a consequence of adopting a fixed gravitational potential for spiral arms.  The presence of the age gradient appears to depend on the nature of the spiral potential. Using numerical simulations, \citet{dob10} showed that galaxies with quasi-stationary spiral density waves exhibit a clear monotonic age gradient, while those with tidally-induced, transient arms do not. They proposed that the presence or absence of the age gradient can be used as a potential discriminant for the nature of spiral arms. \citet{sha18} found that NGC 1566, a grand-design barred-spiral galaxy with bisymmetric arms, has a noticeable age gradient across the spiral arms, suggesting that the arms may represent quasi-stationary density waves with a constant pattern speed, as envisaged by \citet{lin64,lin66}. For M51, on the other hand, they found star clusters with different ages are peaked almost at the same locations, although older clusters spread more widely (see also \citealt{kal10,cha17}). The lack of age gradient for the young star clusters in M51 is presumably  because its tidally-driven arms have not yet reached a steady state, and are still changing their amplitudes and pattern speed \citep{sha18}.

Radio synchrotron observations of external disk galaxies reveal that the pitch angles of large-scale magnetic fields are correlated with those of the gaseous arms, with the former systematically larger by $\sim5^\circ$--$10^\circ$, on average, than the latter (\citealt{van15,fri16}; see also the review by \citealt{bec16}). This is consistent with our numerical results that magnetic fields in the arm are inclined relative to the gaseous arm. In our models, the inclination angle $\theta_B$ of magnetic fields is determined by the competition among three agents:  spiral arm compression, SN feedback, and background shear.  Spiral compression and shear tend to decrease $\theta_B$, while SN feedback increases $\theta_B$ by creating  quasi-radial fields from  quasi-azimuthal magnetic fields. SN feedback and shear yield quasi-equilibrium regular fields with $\theta_B\sim 13^\circ$ in models without a spiral forcing, and the spiral compression (decompression) decreases (increases) $\theta_B$ to $\sim7^\circ$ ($\sim20^\circ$) in the arm (interarm) regions.

Our simulations show that the strength of magnetic fields are correlated with the gas surface density and SFR surface density via Equations \eqref{eq:SFRSigma} and \eqref{eq:mag}. Under the assumption of equipartition between the energy densities of magnetic fields and cosmic rays, \citet{tab13} used synchrotron emission to estimate the field strength across the disk of NGC 6946, finding that $B_\text{tot}\propto \SFRf^{0.14}$ and $B_\text{tot}\propto \Sigma^{0.23}$. These are shallower than our numerical results $B_\text{tot}\propto \SFRf^{0.54}$ and $B_\text{tot}\propto \Sigma^{0.61}$ (see Equation \eqref{eq:mag}). The shallow relations reported in observations suggest that the arm-to-interarm contrast of magnetic fields is very low ($\sim1.3$ in M51; see \citealt{fle11}). However, we caution that this shallow relation might be a consequence of the assumption of the energy equipartition.  There is no fundamental physical reason for equipartition to hold, especially at small scale \citep[e.g.][]{ste14}.  If instead comic rays are in fact more uniform than magnetic fields, then the true $B$--$\Sigma$ relationship should be steeper; if magnetic fields and cosmic rays are anticorrelated, then the relation would be much steeper.

Finally, we remark on a few important caveats of our simulations. First, our current models adopt local, spiral-arm coordinates which assume quite tightly-wound arms and neglect curvature terms.
The local models also cannot capture self-gravitating modes with wavelength longer than the $x$-width of the simulation box. Although we do not expect that qualititative results would change, to capture missing effects
it is desirable to run global simulations with radially-varying surface density, adopting more realistic arm pitch angles ($\sim20^\circ$--$30^\circ$). Second, by taking the diode-like vertical boundary conditions, our models do not allow for gas accretion from outside the simulation box and thus result in a secular decrease of the SFR over time. Observations indicate that the gas accretion rate to the Milky Way is  $\sim0.1$--$0.4\Aunit$ for the cold gas \citep{put12} and $\sim1\Aunit$ including the ionized gas \citep{leh11}.  Inclusion of accreted gas can offset the long-term decline of the gas mass via star formation and outflows in our simulations.
Third, in the current TIGRESS framework that we adopt, SN explosions are the only form of feedback that directly contributes to the turbulent pressure.  In reality, other forms of feedback such as stellar winds and ionizing radiation help to pressurize the ISM in star-forming regions before the onset of first SNe. Although the total (lifetime) momentum injection from these sources is small compared to that from SNe, the immediate onset may help to limit collapse of dense gas. Inclusion of additional ``early'' feedback represents an important direction for future high-resolution simulations of star formation and the ISM in spiral galaxies.

\acknowledgments
We appreciate a thoughtful report from the referee. W.-T.K.\ gratefully acknowledges the assistance and hospitality provided by the Department of Astrophysical Sciences at Princeton University during his sabbatical visit when this paper was prepared. The work of W.-T.K.\ was supported by the National Research Foundation of Korea (NRF) grant funded by the Korea government (MSIT) (2019R1A2C1004857), with partial sabbatical support from the Simons Foundation under grant 510940 to E.C.O..  The work of C.-G.K.\ and E.C.O.\ was partially supported by grant NNX17AG26G from NASA. C.-G.K.\ also acknowledges
support from the Simons Foundation Award No. 528307 (E.C.O). Computational resources for this project were provided by Princeton Research Computing, a consortium including PICSciE and OIT at Princeton University.

\textit{Software:} {\tt Athena} \citep{sto08},  {\tt numpy} \citep{van11}, {\tt matplotlib} \citep{hun07}, {\tt IPython} \citep{per07}, {\tt pandas} \citep{mck10}.

\appendix
\section{Equations of Motion for Star Particles}\label{sec:eom}

Here we derive the equations of motion for sink/star particles in spiral-arm coordinates. We start from Newton's force equation in the inertial  frame of reference
\begin{equation}\label{a:eom0}
  \ddot{ \mathbf{R}} =  -\boldsymbol\nabla \Phi_\text{tot}\,,
\end{equation}
where $\mathbf{R}=R\Rhat+z\zhat$ is the coordinate vector in cylindrical coordinates and $\Phi_\text{tot}=\Phi_\text{self} + \Phi_\text{ext}$ is the total gravitational potential. We assume that the axisymmetric part $\Phi_0$ of the total gravitational potential is separable from the remaining non-axisymmetric part $\Phi_1$ such that $\Phi_\text{tot}=\Phi_0(R) + \Phi_1 (R, \phi, z)$:
the axisymmetric part $\Phi_0$ is responsible for the galaxy rotation with angular frequency $\Omega=(R^{-1}d\Phi_0/dR)^{1/2}$, and $\Phi_1$ can be regarded as perturbations to $\Phi_0$. Equation \eqref{a:eom0} can then be rewritten as
\begin{equation}\label{a:eom1}
   (\ddot{R} - R\dot{\phi}^2 + R \Omega^2)\Rhat + (2\dot{R}\dot{\phi} + R\ddot{\phi}) \phat + \ddot{z} \zhat = -\vecnabla \Phi_1\,.
\end{equation}
Equation \eqref{a:eom1} states that the specific angular momentum $L_z=R^2\dot{\phi}$ is constant if $\Phi_\text{1}$ is axisymmetric.

\subsection{In the Frame Rotating at $\boldsymbol\Omega_p$}

We now consider a frame rotating at angular frequency $\boldsymbol\Omega_p=\Omega_p\zhat$ which is not necessarily same as the local angular frequency $\Omega_0=\Omega(R_0)$.  With a new angle variable $\phi_p \equiv  \phi-\Omega_p t$, Equation \eqref{a:eom1} becomes
 \begin{equation}\label{a:eom3}
    (\ddot{R} - R \dot{\phi_p}^2 )\Rhat + (2\dot{R} \dot{\phi}_p + R\ddot{\phi}_p )\phat + \ddot{z}\zhat = -\vecnabla \Phi_1 + R(\Omega_p^2-\Omega^2) \Rhat  - 2\Omg_p \times \dot{\mathbf{R}}_p \,,
\end{equation}
where $\dot{\mathbf{R}}_p \equiv  (\dot{R}, R\dot{\phi}_p, 0)$.

We set up a local Cartesian frame $(X, Y, Z)=(R-R_0, R_0\phi_p, z)$ centered at $(R_0, \Omega_pt, z)$, and make a local approximation such that $|X|,|Y-\dot{Y}_0t|,|Z|\ll R_0$ and $|\dot{X}|,|\dot{Y}-\dot{Y}_0|, |\dot{Z}| \ll R_0\Omega_0$, where $\dot{Y}_0 \equiv R_0(\Omega_0-\Omega_p)$ is the background rotational velocity seen in the rotating frame.
Then, Equation \eqref{a:eom3} is decomposed as
\begin{subequations}\label{a:Omegap}
\begin{align}
   \ddot{X} & = - \frac{\partial \Phi_1}{\partial X}  + 2\Omega_0 (\dot{Y} - \dot{Y}_0) + 2q \Omega_0^2 X\,, \label{a:Omega0a}  \\
   \ddot{Y} & = - \frac{\partial \Phi_1}{\partial Y} -2\Omega_0 \dot{X}
    \,,\label{a:Omega0b} \\
   \ddot{Z} & =  - \frac{\partial \Phi_1}{\partial Z}\,,    \label{a:Omega0c}
\end{align}
\end{subequations}
where $q=-d\ln\Omega/d\ln R|_{R_0}$ is the shear parameter. When $\Omega_p=\Omega_0$, that is, in the frame rotating at the local angular frequency $\Omega_0$, Equation \eqref{a:Omegap} becomes Hill's equations, which can be integrated to yield
\begin{equation}\label{a:eng0}
   \frac{1}{2}(\dot{X}^2 + \dot{Y}^2 + \dot{Z}^2)
    - q\Omega_0^2 X^2 + \Phi_1 = \text{constant}\,,
\end{equation}
corresponding to conservation of the total energy including
the tidal potential ($-q\Omega_0^2X^2$).

Equation \eqref{a:Omegap} possesses solutions for epicycle orbits in the $X$--$Y$ plane.  In the absence of the external forcing ($\Phi_1=0$), Equations \eqref{a:Omega0a} and \eqref{a:Omega0b} can be integrated to yield the orbit in the $z=0$ plane as
\begin{subequations}\label{a:episol}
\begin{align}
   X & = A\kappa_0 \sin(\kappa_0t + B) + X_0 \,, \label{a:episola}  \\
   Y & =  2A\Omega_0 \cos(\kappa_0t + B)- q\Omega_0X_0 t + \dot{Y}_0 t + Y_0\,,    \label{a:episolb}
\end{align}
\end{subequations}
where $A$, $B$, $X_0$, and $Y_0$ are constants to be determined subject to the initial conditions
and $\kappa_0=(4-2q)^{1/2}\Omega_0$ is the epicycle frequency. Note that the center of the epicycle $(X_0, Y_0)$ drifts at a constant speed $-q\Omega_0X_0 + \dot{Y}_0$ along the $Y$-direction due to the background rotation ($\dot{Y}_0$) and shear ($-q\Omega_0X_0$).

\subsection{In the Spiral-arm Coordinates}

We now tilt the local $(X, Y, Z)$ frame by an angle $i$ to construct another rectangular frame $(x, y, z)$, where $x$ and $y$ refer to the directions perpendicular and parallel to a local arm segment, respectively, while $z$ denotes the vertical direction.  This is achieved by the coordinate transformation
 \begin{gather}\label{a:rotmat}
 \begin{pmatrix}
   X \\ Y \\ Z
 \end{pmatrix} =
 \begin{pmatrix}
    \cos i & -\sin i & 0\\
    \sin i & \cos i  & 0\\
   0      &   0  & 1
 \end{pmatrix}
 \begin{pmatrix}
   x \\ y \\ z
 \end{pmatrix}.
\end{gather}

Assuming that the arm is tightly wound with $\sin i\ll1$, it is straightforward to transform Equation \eqref{a:Omegap} into the spiral-arm coordinates. Due to the constant drift of epicycle orbits,  however, one should be careful in handling the $-2q\Omega_0^2 y\sin i $ term originating from the last term of Equation \eqref{a:Omega0a}. Even if $\sin i$ is small under the tightly wound approximation,
\begin{equation}\label{a:epiapprox}
  y\sin i\approx Y\sin i \approx (\dot{Y}_0 \sin i)t
\end{equation}
can be of order unity for sufficiently large $t$,  representing the drift of a guiding center due to galaxy rotation (Equation \eqref{a:episolb}). Keeping all the terms of order unity, Equation \eqref{a:Omegap} is transformed to
\begin{subequations}\label{a:Rob0}
\begin{align}
    \ddot{x} &= -\frac{\partial \Phi_1}{\partial x}
   + 2\Omega_0 (\dot{y} -  \dot{y}_0 ) + 2q\Omega_0^2 (x- \dot{x}_0t),  \label{a:Rob0a}\\
    \ddot{y} &= -\frac{\partial \Phi_1}{\partial y }-  2\Omega_0
     (\dot{x} - \dot{x}_0)\,, \label{a:Rob0b} \\
     \ddot{z} &= - \frac{\partial \Phi_1}{\partial z}\,, \label{a:Rob0c}
\end{align}
\end{subequations}
where $\dot{x}_0=\dot{Y}_0 \sin i$ and $\dot{y}_0=\dot{Y}_0$. These are the desired set of equations for star particles in the spiral-arm coordinates.
The time $t$ in the last term of Equation \eqref{a:Rob0a} can be identified as the age of each star particle.
The corresponding equation for energy conservation reads
\begin{equation}\label{a:engcon}
   \frac{1}{2} \left|\dot{\bf x} -\dot{\bf x}_{0}\right|^2
   -q\Omega_0^2(x - \dot{x}_0t)^2 + \Phi_1 = \text{constant}\,,
\end{equation}
with $\dot{z}_0=0$.

When $\Phi_1=0$, the $x$ and $y$ coordinates are separable from the $z$ coordinate, and Equation \eqref{a:Rob0} can be integrated to give the epicycle orbit
\begin{subequations}\label{a:episol_sp}
\begin{align}
   x & = a\kappa_0 \sin(\kappa_0t + b)  + \dot{x}_0t + x_0 \,, \label{a:episol_spa}  \\
   y & =  2a\Omega_0 \cos(\kappa_0t + b) + (\dot{y}_0- q\Omega_0 x_0) t + y_0,    \label{a:episol_spb}
\end{align}
\end{subequations}
where $a$, $b$, $x_0$, and $y_0$ are integration constants determined by the initial conditions. It is apparent that the center $(x_0, y_0)$ of the epicycle drifts at a constant speed $(\dot{x}_0, \dot{y}-q\Omega_0x_0) = [ R_0(\Omega_0-\Omega_p)\sin i, R_0(\Omega_0-\Omega_p)-q\Omega_0x_0)]$ due to the background velocity ${\bf v}_0$ (Equation \eqref{eq:v0}) in the spiral-arm coordinates.

\begin{figure}
\epsscale{0.8} \plotone{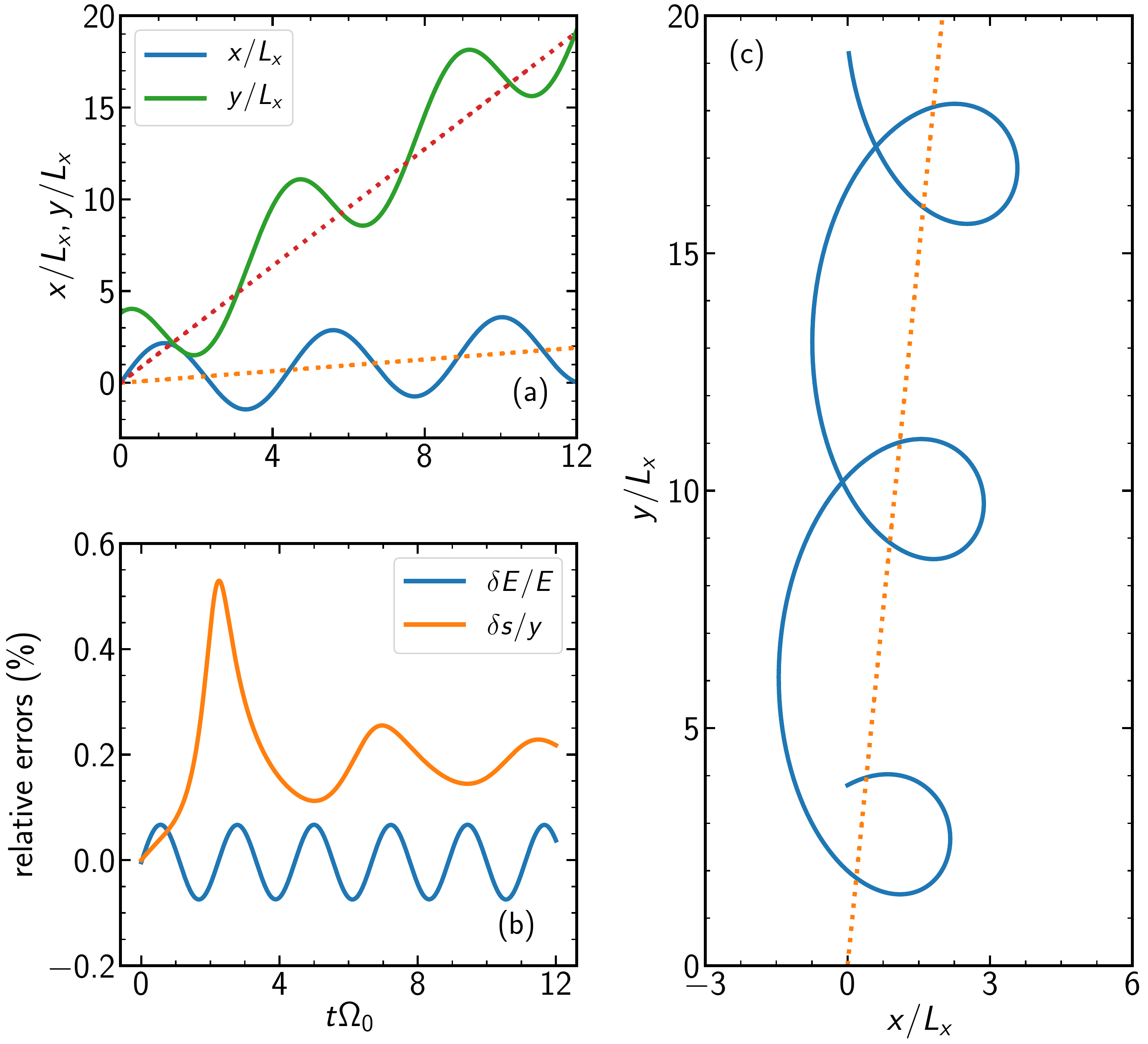}
\caption{Illustration of a particle orbit with the initial conditions $(x,y)/L_x=(0,3.80)$ and $(\dot{x},\dot{y})/(\Omega_0L_x)=(2.96, 1.59)$ at $t=0$, calculated from integrating Equation \eqref{a:Rob0} with $\Phi_1=0$.  Temporal changes of (a) the $x$- and $y$-coordinates of the particle and (b) the relative errors in the energy $\delta E/E$ and the position offset $\delta s$ between the numerical and analytic results. (c) The trajectory of the particle orbit in the $x$--$y$ plane. In (a) and (c), the dotted lines draw the movement of the guiding center of the epicycle.}\label{f:arm_epi}
\end{figure}

As an example, we take $\Omega_p=\Omega_0/2$, $\sin i=0.2$, $q=1$, and $L_x=\pi \sin i R_0$, and initially consider a star particle located at $(x,y)/L_x=(0,3.80)$  with velocity $(\dot{x},\dot{y})/(\Omega_0L_x)=(2.96, 1.59)$ at $t=0$, corresponding to $a\Omega_0/L_x=1.4$, $b=x_0=0$, and $y_0/L_x=1$. We integrate Equation \eqref{a:Rob0} with $\Phi_1=0$ using a kick-drift-kick scheme of the leap-frog integrator suggested by \citet{qui10}, with a time step of $\Delta t=10^{-3}/\Omega_0$. Figure \ref{f:arm_epi} plots the temporal changes of the orbit, the relative errors in the energy $\delta E/E$ and the position offset $\delta s/y$ between the numerical and analytic solutions (Equation \eqref{a:episol_sp}), and the trajectory in the $x$--$y$ plane. As expected, the stellar orbit consists of an epicycle motion in the clockwise direction and a constant drift of its guiding center indicated by the dotted lines. The energy is conserved within 0.1\%, while the position offset is less than 0.6\%.

\end{document}